\documentclass[aps,prc,reprint,showpacs,groupedaddress,onecolumn]{revtex4}
\usepackage{amsmath}
\usepackage{amssymb}
\usepackage{amsfonts}
\usepackage{graphics}
\usepackage[dvips]{color}
\def\beq{\begin{equation}}
\def\eeq{\end{equation}}
\def\eeqarray{\end{eqnarray}}
\def\bsig{\mbox{\boldmath$\sigma$}}

\begin{document}

\title{Representations of relativistic particles of arbitrary spin in
  Poincar\'e, Lorentz, and Euclidean covariant formulations of
  relativistic quantum mechanics}

\author{W.~N.~Polyzou}
\affiliation{Department of Physics and
Astronomy, The University of
Iowa, Iowa City, IA 52242}

\thanks{This work supported by the U.S. Department of Energy,
  Office of Science, Grant \#DE-SC16457}

\date{\today}


\begin{abstract}  

  {\bf Background:} Relativistic treatments of quantum mechanical
  systems are important for understanding hadronic structure and
  dynamics at sub-nucleon scales.  Relativistic invariance of a
  quantum system means that there is an underlying unitary
  representation of the Poincar\'e group.  This is equivalent to the
  requirement that the quantum observables (probabilities, expectation
  values and ensemble averages) for equivalent measurements performed in
  different inertial reference frames are identical.  Different
  representations are used in practice, including Poincar\'e covariant
  forms of dynamics, representations based on Lorentz covariant wave
  functions, Euclidean covariant representations and representations
  generated by Lorentz covariant fields.

  {\bf Purpose:} The purpose of this work is to illustrate the
  relation between the different equivalent representations of states
  in relativistic quantum mechanics.

  {\bf Method:} The starting point is a description of a particle of
  mass $m$ and spin $j$ using irreducible representations of the
  Poincar\'e group.  Since any unitary representation of the
  Poincar\'e group can be decomposed into a direct integral of
  irreducible representations, these are the basic building blocks of
  any relativistically invariant quantum theory.  The equivalence is
  established by constructing equivalent Lorentz covariant irreducible
  representations from Poincar\'e covariant irreducible
  representations and constructing equivalent Euclidean covariant
  irreducible representations from Lorentz covariant irreducible
  representations.

  {\bf Results:} Equivalent descriptions for positive mass
  representations of arbitrary spin are presented in each of these
  frameworks.  Dynamical realizations of the different representations
  are briefly discussed.

  {\bf Conclusion:} Poincar\'e covariant, Lorentz
  covariant and Euclidean covariant realizations of relativistic
  dynamics are shown to be equivalent by explicitly relating the
  positive-mass positive-energy irreducible representations of the
  Poincar\'e group that appear in the direct integral.

\end{abstract}

\maketitle


\section{Introduction}
Relativistic quantum mechanical models are important for modeling
hadronic structure.  Experiments using electromagnetic and weak probes
are designed to investigate the structure of hadronic targets.  The
relevant theoretical quantities are matrix elements of current
operators between initial and final hadronic states in different
inertial frames.  The finer the resolution of the probe, the larger
the momentum difference between the initial and final hadronic states.
First principles calculations of the initial and final hadronic wave
functions with quantifiable errors are challenging, especially when
they are needed in different inertial frames.  Relativistic models of
the hadronic states provide a consistent treatment of initial and
final states in different inertial reference frames.

There are many different formulations of relativistic quantum
mechanical models.  In this work the relation between different
quantum mechanical descriptions of relativistic particles is
systematically developed.  In order to take advantage of the relations
discussed in this work it is necessary to first have a dynamical
model.  While it is beyond the scope of this paper to give a detailed
discussion of dynamical models, typical relativistic wave functions are matrix
elements of a relativistic state in a dynamical model and a
non-interacting relativistic basis state.  For example, in describing
a nucleus as a system of constituent nucleons, the nuclear state is
the solution of a dynamical equation expressed in a basis of free
nucleon states.  In this work the focus is on deriving the relation
between different relativistic descriptions of these particle states.
This applies to both the interacting relativistic states and the
relativistic free-particle basis states.

In 1939 Wigner \cite{Wigner:1939cj} showed that the relativistic
invariance of a quantum system is equivalent to the requirement that
there is a unitary ray representation of the Poincar\'e group on the
Hilbert space of the quantum system.  This is the mathematical
formulation of the physical requirement that quantum observables
(probabilities, expectation values and ensemble averages) for
equivalent measurements performed in different inertial reference
frames are identical.  Physically this means that equivalent quantum
measurements in isolated systems cannot be used to distinguish
inertial frames.  This quantum mechanical formulation of relativistic
invariance focuses on the invariance of measurements, rather than the
transformation properties of equations, which is used in the classical
formulation of relativistic invariance.

Relativistically invariant quantum systems are represented using
Poincar\'e covariant methods, Lorentz covariant methods, Euclidean
covariant methods, and Lorentz covariant fields.  Each method provides
a different representation of the same physical system.  Each
representation has different advantages.  The purpose of these notes
is to exhibit the relation between these different representations.
While most of the content of this exposition can be found in
references, \cite{Wigner:1939cj} \cite{Dirac:1949cp}
\cite{Dirac:1945cm} \cite{Weinberg:1964ev} \cite{Weinberg:1964cn}
\cite{Weinberg:1969di} \cite{Weinberg:v1} \cite{Wightman:1980}
\cite{Osterwalder:1974tc} \cite{Osterwalder:1973dx} \cite{JacobWick},
it is difficult to find all of the relations in one place.

The starting point is the realization that any unitary representation
of the Poincar\'e group can be decomposed into a direct sum/integral
of irreducible representations.  These are the basic building blocks
of any relativistically invariant quantum theory.  The construction of
the direct integral is the dynamical problem, which is mathematically
equivalent to the simultaneous diagonalization of the Casimir
operators (mass and spin) of the Poincar\'e Lie algebra.  This is the
relativistic analog of diagonalizing a non-relativistic center of mass
Hamiltonian. It is a
non-trivial dynamical problem.  Wave functions of these irreducible
states are matrix elements of these states with free-particle
relativistic basis states.  The free particle states could be
irreducible basis states or tensor products of irreducible basis
states.  The relevant observation of this work is that once these wave
functions are found in one representation, the results of this work
can be applied to determine the corresponding relativistic wave
functions in different representations.

Because of this it is sufficient to understand the relation between
the different representations of the irreducible representations.  This
work considers only positive-mass positive-energy representations
of the Poincar\'e group
\cite{Bakamjian:1953kh}
\cite{Bargmann:1954gh}
\cite{Wigner:1957ep}
\cite{Weinberg:1964ev}
\cite{Weinberg:1964cn}
\cite{Weinberg:1969di}
\cite{Weinberg:v1}
\cite{Wightman}
\cite{Wightman:1980}
\cite{Osterwalder:1974tc}
\cite{Osterwalder:1973dx}.
These are the relevant representations for hadronic states.

The next section summarizes the notation used in the rest of this
paper and gives a brief description of the essential elements of the
Poincar\'e group.  Section three discusses the construction of
positive mass, positive energy unitary irreducible representations of
the Poincar\'e group for a particle of any (positive) mass and
spin. Single-particle states are represented by simultaneous
eigenstates of a complete set of commuting observables that are functions of the
infinitesimal generators of the Poincar\'e group.  These basis states
span a one-particle subspace, and the structure of the unitary
representation of the Poincar\'e group on that subspace is fixed by
the choice of commuting observables and group theory.  For a given
choice of commuting observables, there is a largest subgroup of the
Poincar\'e group where the transformations are independent of the
mass.  These subgroups are called {\it kinematic subgroups}.  Dirac
\cite{Dirac:1949cp} identified basis choices with the largest
kinematic subgroups.  He referred to them as defining ``forms of dynamics''.
Kinematic subgroups are useful because for transformations in this
subgroup, dynamical Poincar\'e transformations on interacting states
can be computed by applying the inverse kinematic transformation to
the free particle basis states.  This avoids the need to
explicitly compute the dynamical transformations.

Section four gives an introduction to $SL(2,\mathbb{C})$ which is
related to the Lorentz group like $SU(2)$ is related to $SO(3)$.
$SL(2,\mathbb{C})$ plays a central role in the construction of
Lorentz covariant descriptions of particles, Euclidean covariant
descriptions of particles and Lorentz covariant
fields.  This section includes a complete description of all of the
properties of $SL(2,\mathbb{C})$ that are needed in relativistic
quantum theories.

Section five discusses Lorentz covariant descriptions of particles.
In these representations the $SU(2)$ Wigner rotations are decomposed
into products of $SL(2,\mathbb{C})$ matrices.  The momentum-dependent
parts are absorbed into the definition of the wave functions.  The
result is a new wave function that transforms in a Lorentz covariant
way. In this representation the Hilbert space inner product acquires a
non-trivial kernel, which removes the momentum dependence that was
absorbed in the wave functions.  The resulting kernel is a 
free-particle Wightman function.  In addition, the $SU(2)$ identity,
$R=(R^{\dagger})^{-1}$ for the $SU(2)$ Wigner rotations leads to two
inequivalent decompositions of the Wigner rotation into products of
$SL(2,\mathbb{C})$ matrices.  The inequivalent representations are
related by space reflection.   The treatment of space reflection
in these representations is discussed.

Section six exhibits Euclidean covariant Green functions that lead to
all of the covariant representations constructed in section five.  The
interesting feature of this representation is that no analytic continuation is
needed to show equivalence with the Lorentz covariant representation.

Section seven discusses the construction of free Lorentz covariant
fields using the occupation number representation in the Lorentz
covariant description of particles.  In section eight the covariant fields
are used to construct local covariant fields.
Section nine discusses the role of dynamics in these representations.
Section ten contains a brief summary.

\section{The Poincar\'e group}   
\bigskip

The Poincar\'e group is the group of space-time coordinate
transformations that preserve the form of the source-free Maxwell's
equations.  It is also the group that relates different inertial
coordinate systems in special relativity.

In what follows the space and time coordinates of events are labeled
by components of a four vector
\beq
x^{\mu} =(ct,x^1,x^2,x^3).
\label{p.1}
\eeq
The convention for the Lorentz metric tensor is 
\beq
\eta^{\mu \nu} = \eta_{\mu \nu} :=
\left (
\begin{array}{cccc}
-1 & 0 & 0 &0 \\
0 & 1 &0 &0 \\
0 & 0 & 1 &0 \\
0 & 0 & 0 & 1 \\
\end{array}
\right)
\label{p.2}
\eeq
and repeated indices are assumed to be summed.  This choice of metric 
is natural for developing the relation with Euclidean representations.

The Poincar\'e group is the group of point transformations that preserve
the proper time between events:
\beq
\Delta \tau_{xy}^2 =  (x^0-y^0)^2-\vert \mathbf{x} - \mathbf{y} \vert^2 =
- \eta_{\mu \nu} (x^{\mu} - y^{\mu})(x^{\nu} - y^{\nu}).
\label{p.3}
\eeq
The general form of a point transformation, 
$x^{\prime \mu} = f^{\mu}(x)$, that preserves (\ref{p.3})
is
\beq
f^{\mu}(x) = 
x^{\prime \mu} = \Lambda^{\mu}{}_{\nu} x^{\nu} + a^{\mu}
\label{p.4}
\eeq
where $a^{\mu}$ and $\Lambda^{\mu}{}_{\nu}$ are constants and
the Lorentz transformation $\Lambda^{\mu}{}_{\nu}$ satisfies
\beq
\eta^{\mu\nu} =
\Lambda^{\mu}{}_{\alpha}\Lambda^{\nu}{}_{\beta}\eta^{\alpha \beta} .
\label{p.5}
\eeq
These relations can be derived by differentiating 
\beq
(f^{\mu}(x) -f^{\mu}(y)) (f^{\nu}(x) -f^{\nu}(y))\eta_{\mu\nu} =
(x^{\mu} -y^{\mu} ) (x^{\nu} -y^{\nu})\eta_{\mu\nu} 
\eeq
with respect to $x$, setting $x$ to $0$, and then doing the same with 
$y$.  In matrix form equation (\ref{p.5}) has the form
\beq
\eta = \Lambda \eta \Lambda^t
\label{p.6}
\eeq
which indicates that $\Lambda$ is a real orthogonal transformation with
respect to the Lorentz metric.  Equations (\ref{p.4}) and (\ref{p.6})
are relativistic generalizations of the fundamental theorem of rigid
body motion which asserts that any motion that preserves the distance
between points in a rigid-body is a composition of an orthogonal
transformation and a translation.

Equation (\ref{p.6}) implies that
\beq
\det (\Lambda)^2 =1 \qquad (\Lambda^0_{0})^2 = 1 + (\sum_i \Lambda^0_{i})^2 .
\label{p.6a}
\eeq
It follows from (\ref{p.6a}) that the Lorentz group has four topologically
disconnected components distinguished by
\beq
\mbox{det} (\Lambda) =  1 \qquad \Lambda^0{}_0 \geq 1
\label{p.7}
\eeq
\beq
\mbox{det} (\Lambda) =  1 \qquad \Lambda^0{}_0 \leq -1
\label{p.8}
\eeq
\beq
\mbox{det} (\Lambda) =  -1 \qquad \Lambda^0{}_0 \geq 1
\label{p.9}
\eeq
\beq
\mbox{det} (\Lambda) =  -1 \qquad \Lambda^0{}_0 \leq -1 .
\label{p.10}
\eeq
The component with $\det (\Lambda) = 1$ and $\Lambda^0{}_0 \geq 1$
contains the identity and is a subgroup.  These Lorentz
transformations are called proper Lorentz transformations.  This
subgroup is the symmetry group of special relativity.  The other three
components involve space and/or time reflections, which are not
symmetries of the weak interaction.  In what follows all Lorentz
transformations will be assumed to be proper transformations unless
otherwise specified.

The requirement that quantum observables are independent of inertial
coordinate system requires that equivalent states in different
inertial coordinate systems are related by a unitary (ray)
representation of the proper subgroup of the Poincar\'e group.  The
Poincar\'e group has ten infinitesimal generators that can be
expressed as components of operators that transform as a four vector
and an anti-symmetric rank-2 tensor under the unitary representation,
$U(\Lambda)$, of the Lorentz group:
\beq
P^{\mu} = (H,\mathbf{P})
\label{p.11}
\eeq
\beq
J^{\mu \nu} =
\left (
\begin{array}{cccc}
0 &-K^1& -K^2& -K^3\\
K^1 & 0 &J^3 & -J^2\\
K^2 &-J^3 & 0 &J^1 \\
K^3 & J^2 &-J^1 &0 \\
\end{array}
\right ) 
\label{p.12}
\eeq 
\beq
U(\Lambda) P^{\mu} U^{\dagger} (\Lambda)=
(\Lambda^{-1})^{\mu}{}_{\nu} P^{\nu} 
\label{p.13}
\eeq
\beq
U(\Lambda) J^{\mu\nu} U^{\dagger} (\Lambda)=
(\Lambda^{-1})^{\mu}{}_{\alpha}(\Lambda^{-1})^{\nu}{}_{\beta} J^{\alpha\beta} .
\label{p.14}
\eeq

The Pauli-Lubanski vector is the four-vector operator defined by
\beq
W^{\mu} = - {1 \over 2} 
\epsilon^{\mu \alpha \beta \gamma} P_{\alpha}J_{\beta\gamma}.
\label{p.15}
\eeq
The Lie algebra has two independent polynomial invariants
\beq
M^2 = -P^{\mu}P_{\mu} \qquad \mbox{and} \qquad
W^2 = W^{\mu}W_{\mu} = - M^2 \mathbf{j}^2 .
\label{p.16}
\eeq
When the spectrum of the mass operator, $\sigma ({M}) >0$, is positive 
spin operators are defined by 
\beq
(0,\mathbf{j}_x ) := -{1 \over M} B_x^{-1}(P/M)^{\mu}{}_{\nu}W^{\nu} 
\label{p.17}
\eeq
where $B_x^{-1}(P/M)^{\mu}{}_{\nu}$ is a matrix of operators that
transform $P^{\mu}$ to $(M,0,0,0)$:
\beq
B_x^{-1}(P/M)^{\mu}{}_{\nu} P^{\mu} = (M,\mathbf{0}) .
\label{p.18}
\eeq
A standard choice is the canonical (rotationless) boost $B_c(P/M)$  
defined by
\beq
B_c(V:=P/M)=
\left (
\begin{array}{cc}
V^0 & \mathbf{V} \\
\mathbf{V} & \delta_{ij} + {V^i V^j\over 1 + V^0}\\  
\end{array}
\right ). 
\eeq
The subscript $x$ indicates that both $B_x(P/M)$ and $\mathbf{j}_x$ are
not unique since for any $P$-dependent rotation $R_{xy}(P/M)$ 
\beq
B_y(P/M)^{\mu}{}_{\nu} := B_x (P/M)^{\mu}{}_{\rho}R_{xy}(P/M)
^{\rho}{}_{\nu}
\label{p.19}
\eeq
gives another matrix of operators with property (\ref{p.18}); however for any 
choice ($x$) the Poincar\'e commutation relations imply  
\beq
\mathbf{j}_x^2 = \mathbf{j}^2 = W^2/M^2
\label{p.20}
\eeq
\beq
[j_x^l,j_x^m]=i \sum_n \epsilon^{lmn}j_x^n
\label{p.21}
\eeq
\beq
[j_x^i,\mathbf ,P^{\mu}]=0 .
\label{p.22}
\eeq
It follows from (\ref{p.17}) that the different spin operators are related by 
\beq
(0,\mathbf{j}_x )^{\mu}  := B_x^{-1}(P/M)^{\mu}{}_{\rho}
B_y(P/M)^{\rho}{}_{\nu}(0,\mathbf{j}_y)^{\nu}.  
\label{p.23}
\eeq
The rotation
\beq
R_{xy}(P/M) := 
B_x^{-1}(P/M)
B_y (P/M)
\label{p.24}
\eeq
that relates different spin observables
is called a generalized Melosh rotation  \cite{Melosh:1974cu}.
The interpretation of
$\mathbf{j}_x$ is that it is the spin that would be measured in the
rest frame of a particle if it was Lorentz transformed to the rest
frame with the Lorentz transformation $B_x^{-1}(P/M)$.  This provides
a mechanism to compare spins in different inertial frames.  Different
kinds of spin arise because products of rotationless Lorentz boosts
can generate rotations.  This means that the spin measured in the rest
frame depends on the Lorentz transformation to the rest frame.  Note
that in spite of the 4 indices in (\ref{p.23}), the spin is not a
4-vector. This is because $B_x^{-1}(P/M)^{\mu}{}_{\rho}$ in equation
(\ref{p.17}) is a matrix of operators.

The spin can alternatively be expressed as
\beq
j_x^i = \epsilon_{ijk}B_x^{-1}(P/M)^{j}{}_{\mu}
B_x^{-1}(P/M)^{k}{}_{\nu}J^{\mu \nu}, 
\eeq
which can be interpreted as the angular momentum in the particle's rest frame,
which again depends on the Lorentz transformation used to get to the
rest frame.

Representations of the Poincar\'e group can be built up out of
irreducible representations.  The classification of the irreducible representations depends on the spectrum of invariant operators $M^2$ and $W^2$ and the
sign of $P^0$.  Wigner \cite{Wigner:1939cj} classified six classes of 
irreducible representations by the spectral properties of $P^2$ and
$P^0$: 
\begin{itemize}

\item[I.] $P^2 < 0 \qquad P^0>0$

\item[II.] $P^2 < 0 \qquad P^0<0$

\item[III.] $P^2 > 0 $ 

\item[IV.] $P^2 = 0 \qquad P^0>0$

\item[V.] $P^2 = 0 \qquad P^0<0$

\item[VI.] $P^{\mu}=0$ .

\end{itemize}
The physically interesting representations for particles are the ones
with $-P^2= M^2  > 0, \quad P^0>0$ (I) and $P^2 = 0, \quad P^0>0$  (IV)
which are associated with massive and massless particles respectively.
The irreducible representations are induced from a subgroup that
leaves a standard vector invariant in each of these classes.

\section{Poincar\'e covariant positive mass unitary irreducible representations}

For a particle of mass $m>0$ the mass, spin, and three components of
the linear momentum, and one component of $\mathbf{j}_x$ are a maximal
set of commuting self-adjoint functions of the infinitesimal generators
of the Poincar\'e group.  The standard vector can be taken as
$(m,0,0,0)$.  The rotation group is called the little group for these
representations because it leaves the standard vector invariant.  The
mass and spin$^2$ eigenvalues are fixed and label an irreducible
subspace.  Basis vectors can be taken as simultaneous eigenstates of
this maximal set of commuting operators
\beq
\vert (m,j)\mathbf{p} ,\mu \rangle . 
\label{p.29}
\eeq
In what follows the normalization convention
\beq
\langle (m,j)\mathbf{p}' ,\mu' \vert
(m,j)\mathbf{p} ,\mu \rangle = \delta (\mathbf{p}'-\mathbf{p}) 
\delta_{\mu' \mu} 
\label{p.30}
\eeq
is used.  The eigenvalue spectrum of both $\mathbf{p}$ and
$\mathbf{j}_x\cdot\hat{\mathbf{z}}$ is fixed by $j$ and group
properties ($\mathbf{p}$ can be boosted to any real value,
and the spin components satisfy SU(2) commutation relations (\ref{p.21})).

An irreducible unitary representation of the Poincar\'e group in this
basis can be constructed by considering the action of
elementary Poincar\'e transformations on the rest,
$(\mathbf{p}=\mathbf{0})$, eigenstates.  On these states rotations can
only affect the spin variables since they leave the rest four-momentum
(standard vector) unchanged.  The total spin constrains the structure
of the transformation - it must be a $2j+1$ dimensional irreducible 
unitary representation of $SU(2)$:
\beq
U(R,0) \vert (m,j) \mathbf{0},\mu \rangle =
\vert (m,j) \mathbf{0},\nu \rangle D^{j}_{\nu \mu}[R]
\label{p.31}
\eeq
where (see the Appendix) 
\[
D^{j}_{\nu \mu}[R] =
\langle j, \nu\vert   
U(R,0) \vert j, \mu \rangle =
\]
\beq
\sum_{k=0}^{j+\mu}
{\sqrt{(j+\nu)!(j-\nu)!(j+\mu)!(j-\mu)!}\over
k!(j+\nu-k)!(j+\mu-k)!(k-\nu-\mu)!}
R_{++}^k R_{+-}^{j+\nu-k}R_{-+}^{j+\mu -k} R_{--}^{k-\nu-\mu}  
\label{p.32}
\eeq
are the $2j+1$ dimensional unitary representations of $SU(2)$ 
in the $\vert j, \mu \rangle $ basis where
\beq
R = e^{{i\over 2} \pmb{\theta} \cdot \pmb{\sigma} }  =
\sigma_0 \cos (\theta /2) + i \hat{\pmb{\theta}}\cdot \pmb{\sigma} 
\sin (\theta /2) =
\left (
\begin{array}{cc}
R_{++} & R_{+-}\\
R_{-+} & R_{--}
\end{array}
\right ) .
\label{p.33}
\eeq
Here $\sigma_0$ is the $2 \times 2$ identity and $\pmb{\sigma}$ are
the Pauli spin matrices.  The Wigner function $D[R]$ is a
degree $2j$ polynomial in the components of $R$.  It
follows from (\ref{p.32}) that 
$D^{j}_{\nu \mu}[R^*]=(D^{j}_{\mu \nu}[R])^*$
and $D^{j}_{\nu \mu}[R^t]=D^{j}_{\mu \nu}[R]$.

Space-time translations of the rest state introduce a phase
\beq
U(I,a) \vert (m,j) \mathbf{0},\mu \rangle :=
e^{-i a^0 m} \vert (m,j) \mathbf{0},\mu \rangle , 
\label{p.34}
\eeq
while Lorentz boosts are unitary operators that change the rest vector to
$p^{\mu} = (\sqrt{m^2 +\mathbf{p}^2},\mathbf{p})$.  A different type of spin
is associated with each type of Lorentz boost.  The $x$-spin is the spin that
is unchanged when the basis vector is transformed to a rest vector
with the inverse boost $B^{-1}_x(p/m)$. The following definition is
consistent with the requirement that the $x$-spin is unchanged when
transformed to the rest frame with the inverse boost $B^{-1}_x(p/m)$:
\beq
U(B_x(p/m) ,0) \vert (m,j) \mathbf{0},\mu \rangle :=
\vert (m,j) \mathbf{p},\mu \rangle
\sqrt{{\omega_m(\mathbf{p}) \over m}} 
\label{p.35}
\eeq
where $\omega_m(\mathbf{p}):= \sqrt{m^2 +\mathbf{p}^2}$ is the energy
of the particle. 
The Jacobian is chosen to make the boost unitary for states with
the normalization (\ref{p.30}).  This can be seen by considering the
Lorentz invariant measure
\beq
\int d^4p \delta (p^2 +m^2) \theta (p^0) =
\int {d \mathbf{p} \over 2 \omega_m(\mathbf{p})}=
\int {d \mathbf{p}' \over 2 \omega_m(\mathbf{p}')}
\label{p.36}
\eeq
where $p'=\Lambda p$.
It follows that
\beq
I = \int \vert \mathbf{p} \rangle d\mathbf{p} \langle \mathbf{p} \vert
= \int \vert \mathbf{p}' \rangle d\mathbf{p}' \langle \mathbf{p}' \vert
=  \int \vert \mathbf{p} \rangle {d\mathbf{p}
  \over d\mathbf{p}'} d\mathbf{p}' \langle \mathbf{p} \vert
=\int \vert \mathbf{p} \rangle {2 \omega_m(\mathbf{p})
\over 2 \omega_m(\mathbf{p}')} d\mathbf{p}' \langle \mathbf{p} \vert
\label{p.37}
\eeq
which leads to the identification
\beq
\vert \mathbf{p}'(\mathbf{p}) \rangle =
\vert \mathbf{p} \rangle \sqrt{
{\omega_m(\mathbf{p})
\over \omega_m(\mathbf{p}')}}.
\label{p.38}
\eeq

A general unitary representation of the Poincar\'e group on any
basis state can be expressed as a product of these elementary
transformations on rest states
using the group representation property:
\[
U(\Lambda ,a) \vert (m,j) \mathbf{p},\nu \rangle =
U(I,a) U(\Lambda ,0) \vert (m,j) \mathbf{p},\nu \rangle =
\]
\[
U(I,a) U(\Lambda ,0) U(B_x (p/m),0) 
\vert (m,j) \mathbf{0},\nu \rangle \sqrt{{m \over \omega_m(\mathbf{p})}} =
\]
\[
U(B_x (\Lambda p/m),0) U(B^{-1}_x (\Lambda p/m),0)
U(I,a) U(\Lambda ,0) U(B_x (p/m),0)
\vert (m,j) \mathbf{0},\nu \rangle \sqrt{{m \over \omega_m(\mathbf{p})}} =
\]
\[
U(B_x (\Lambda p/m),0)
U(I,B_x^{-1} (\Lambda p/m) a) 
U(B^{-1}_x (\Lambda p/m),0)
U(\Lambda ,0) U(B_x (p/m),0) 
\vert (m,j) \mathbf{0},\nu \rangle \sqrt{{m \over \omega_m(\mathbf{p})}} =
\]
\beq 
e^{i \Lambda p \cdot a} 
\vert (m,j) \pmb{\Lambda}{p},\mu \rangle 
D^j_{\mu \nu} [B^{-1}_x (\Lambda p/m)\Lambda B_x (p/m)] 
\sqrt{{\omega_m(\pmb{\Lambda}{p}) \over \omega_m(\mathbf{p})}} .
\label{p.39}
\eeq
The rotation 
\beq
R_{wx}(\Lambda ,p) := 
B^{-1}_x (\Lambda p/m)\Lambda B_x (p/m)
\label{p.40}
\eeq
is called a {\it spin$_x$ Wigner rotation}.
The final result is the mass $m$ spin $j$ irreducible unitary
representation of the Poincar\'e group in the momentum-spin-x
basis:
\beq
\boxed{
U(\Lambda ,a) \vert (m,j) \mathbf{p},\mu \rangle =
e^{i \Lambda p \cdot a} 
\vert (m,j) \pmb{\Lambda}{p},\nu \rangle 
D^j_{\nu \mu} [R_{wx}(\Lambda,p)] 
\sqrt{{\omega_m(\pmb{\Lambda}{p}) \over \omega_m(\mathbf{p})}} .
} 
\label{p.41}
\eeq
Since  $U(\Lambda ,a)$ is defined as a product of unitary transformations,
it is unitary.

The momentum labels can be replaced by any three independent
functions, $\mathbf{f}(p)=
\mathbf{f}(\mathbf{p},m)$, 
of the four momentum $p^{\mu}$ and the spins can be replaced by any type of 
spin.  These replacements correspond to choosing a basis using a different
set of commuting observables.
Each replacement is just a unitary change of basis.  The
general form of the change of basis transformation is 
\[
\vert (m,j) \mathbf{f},\mu \rangle_y =
\]
\beq
\vert (m,j) \mathbf{p}(\mathbf{f},m),\nu \rangle_x 
D^{j}_{\nu\mu} [R_{xy}(p/m)]
\sqrt{\vert {\partial \mathbf{p}(\mathbf{f},m) \over \partial \mathbf{f}} \vert}. 
\label{p.42}
\eeq
Combining this with (\ref{p.39}) gives the
resulting unitary representation of the Poincar\'e group in the 
transformed basis
\[
U(\Lambda ,a) \vert (m,j) \mathbf{f},\mu \rangle_y =
\]
\beq
e^{i \Lambda p(f) \cdot a} 
\vert (m,j) \mathbf{f}(\Lambda p) ,\nu \rangle_y
D^j_{\nu \mu} [B^{-1}_y (\Lambda p(\mathbf{f})/m)\Lambda B_y (p(\mathbf{f})/m)] 
\sqrt{\vert {\partial \mathbf{f}(\Lambda p) \over \partial \mathbf{f}(p)} 
\vert} .
\label{p.43}
\eeq

There are four choices of commuting observables that are commonly used.
They involve a choice of continuous variables and a choice of spin
degrees of freedom.
They are distinguished by having some simplifying properties:
\beq
\mathbf{f} = \mathbf{p},\,  B_x(p/m) = B_c(p/m) \qquad
{\partial \mathbf{f}(\Lambda p) \over \partial \mathbf{f}(p)} =
{\omega_m (\pmb{\Lambda}p) \over \omega_m (\mathbf{p})}
\label{p.44a}
\eeq
\beq
\mathbf{f} = \mathbf{v} = \mathbf{p}/m, \,  B_x(p/m) = B_c(p/m) = B_c(v)
\qquad
{\partial \mathbf{f}(\Lambda p) \over \partial \mathbf{f}(p)} =
{\omega_1 (\pmb{\Lambda}v) \over \omega_1 (\mathbf{v})}
\label{p.44b}
\eeq
\beq
\mathbf{f} =\tilde{\mathbf{p}}:=  (p^+,\mathbf{p}_{\perp}) ,\,  B_x(p/m) = B_f(p/m) \qquad {\partial \mathbf{f}(\Lambda p) \over \partial \mathbf{f}(p)} = {(\Lambda p)^+ \over p^+} \qquad p^+ := p^0 + p^3;
\quad \mathbf{p}_{\perp} =(p^1,p^2)
\label{p.44c}
\eeq
\beq
\mathbf{f} = \mathbf{p},\,  B_x(p/m) = B_h(p/m) \qquad
{\partial \mathbf{f}(\Lambda p) \over \partial \mathbf{f}(p)} =
{\omega_m (\pmb{\Lambda}p) \over \omega_m (\mathbf{p})}
\label{p.45}
\eeq
these choices are associated with an instant, point, front-form, \cite{Dirac:1949cp}
or Jacob-Wick helicity dynamics \cite{JacobWick}.   The boost $B_c(p/m)$ is a rotationless boost,
$B_f(p/m)$ is a light-front preserving boost, and $B_h(p/m)$ is a 
helicity boost.  These choices lead to different spin observables.
The different types of boosts will be defined later.
The first three cases are distinguished by
the choice of a {\it kinematic subgroup}.  The kinematic subgroup 
is the subgroup of the Poincar\'e group where 
$\Lambda p(\mathbf{f})\cdot a $,
$B^{-1}_y (\Lambda p(\mathbf{f})/m)\Lambda B_y (p(\mathbf{f})/m) $
and  
${\partial \mathbf{f}(\Lambda p) \over \partial \mathbf{f}(p)}$
are all independent of $m$.  Since the transformations (\ref{p.42}) that relate
these representations involve the mass, they will generally have
different kinematic subgroups.
The choices (\ref{p.44a}-\ref{p.44c})
have the largest kinematic 
subgroups.  Kinematic subgroups are useful in dynamical theories
because transformations, $U(\Lambda,a)$, for $(\Lambda,a)$ 
in the kinematic subgroup can be computed exactly without having to  
diagonalize the mass and spin operators
using
\beq
\langle \phi_0 \vert U_I(\Lambda,a) \vert \phi_I\rangle =
\langle \phi_I \vert U_0^{\dagger} (\Lambda,a) \vert \phi_0\rangle^* .
\eeq

Explicit forms of the unitary irreducible representations of the
Poincar\'e group in each of these bases are given below
\beq
\boxed{
U(\Lambda ,a) \vert (m,j) \mathbf{p},\nu \rangle_c =
e^{i \Lambda p \cdot a} 
\vert (m,j) \pmb{\Lambda}{p},\mu \rangle_c 
D^j_{\mu \nu} [R_{wc}(\Lambda,p)] 
\sqrt{{\omega_m(\pmb{\Lambda}{p}) \over \omega_m(\mathbf{p})}}
}
\label{p.47}
\eeq
(instant form)

\beq
\boxed{
U(\Lambda ,a) \vert (m,j) \tilde{\mathbf{p}},\nu \rangle_f =
e^{i \Lambda p \cdot a} 
\vert (m,j) \tilde{\pmb{\Lambda}}{p},\mu \rangle_f 
D^j_{\mu \nu} [R_{wf}(\Lambda,p)] 
\sqrt{{ (\Lambda p)^{+} \over p^+}}
}
\label{p.48}
\eeq
(front from)

\beq
\boxed{
U(\Lambda ,a) \vert (m,j) \mathbf{v},\nu \rangle_v =
e^{i \Lambda v \cdot a} 
\vert (m,j) \pmb{\Lambda}{v},\mu \rangle_v 
D^j_{\mu \nu} [R_{wc}(\Lambda,v)]
\sqrt{{\omega_1(\pmb{\Lambda}{v}) \over \omega_1(\mathbf{v})}}
}
\label{p.49}
\eeq
(point form) 

\beq
\boxed{
U(\Lambda ,a) \vert (m,j) \mathbf{p},\nu \rangle_h =
e^{i \Lambda p \cdot a} 
\vert (m,j) \pmb{\Lambda}{p},\mu \rangle_h
D^j_{\mu \nu} [R_{jwh}(\Lambda,p)] 
\sqrt{{\omega_m(\pmb{\Lambda}{p}) \over \omega_m(\mathbf{p})}}
}
\label{p.50}
\eeq
(Jacob-Wick form).
These bases are called instant-form, front-form, point-form,
and Jacob Wick helicity bases.

In the instant-form case the kinematic subgroup is the six-parameter
three-dimensional Euclidean group.  In the point-form case, the
kinematic subgroup is the six parameter Lorentz group, and in the
light-front case the kinematic subgroup is the seven parameter
subgroup that leaves the plane $x^+=x^0+x^3= 0$ invariant.  

The light-front boosts have the distinguishing feature that they form
a subgroup - so light-front Winger rotations of light-front boosts are
the identity. The light-front representation has the largest kinematic
subgroup.  It is a natural representation for deep inelastic scattering.

The canonical boost has the distinguishing property that the Wigner
rotation of a rotation is the rotation.  This property is unique to
the canonical boost and is useful for adding angular momenta.
Both the point-form and instant-form representations use canonical
boosts to define the spins.

The helicity boost has the property that the Wigner rotation of any
Lorentz transformation is a phase.  The helicity spin is related to
the canonical spin by \cite{wpwg} 
$\mathbf{j}_h \cdot \hat{\mathbf{z}}=\mathbf{j}_c \cdot
\hat{\mathbf{p}}$.

These are the most commonly used Poincar\'e covariant representations
of single-particle states.  They are equivalent representations of a
free mass $m$ spin $j$ particle.  They are related by the unitary
transformations (\ref{p.42}).  These unitary equivalences
also apply to dynamical theories after the mass and spin are diagonalized.

These representations are the closest representations of single-particle
states to non-relativistic representations, but they 
are not the only representations used to describe relativistic
particles.  In addition to these there are representations that are
manifestly Lorentz covariant and representations that are also
Euclidean covariant.  In order to understand the relation of these
representations to the Poincar\'e covariant representations
constructed in this section it is useful to introduce the
group $SL(2,\mathbb{C})$, of complex $2\times 2$ matrices
with unit determinant, 
which is the covering group of the Lorentz
group.  The relation between $SL(2,\mathbb{C})$ and the Lorentz group
is analogous to the relation between $SU(2)$ and the rotation group
$SO(3)$.  It will be developed in the next section.
 
\section{$SL(2,\mathbb{C})$}

In order to motivate the connection of  $SL(2,\mathbb{C})$ 
with the Lorentz group it is useful to represent
space-time coordinates by $2 \times 2$ Hermitian matrices
\beq
X = x^{\mu}\sigma_{\mu} =
\left (
\begin{array}{cc}
x^0 + x^3 & x^1 -i x^2 \\
x^1+i x^2 & x^0-x^3
\end{array}
\right ) :=
\left (
\begin{array}{cc}
x^+ & x^*_{\perp} \\
x_{\perp} & x^-
\end{array}
\right ) .
\label{s.1}
\eeq
The inverse is
\beq
x^{\mu} = {1 \over 2} \mbox{Tr}(\sigma_{\mu}X) =
{1 \over 2} \mbox{Tr}(X\sigma_{\mu})
\label{s.2}
\eeq
which follows from properties of the Pauli matrices
\beq
\sigma_i \sigma_j = \delta_{ij} \sigma_0 + i \epsilon_{ijk} \sigma_k
\label{s.3}
\eeq
\beq
\mbox{Tr}(\sigma_i)=0 \qquad 
\mbox{Tr}(\sigma_0)=2 \qquad
\mbox{Tr}(AB)=\mbox{Tr}(BA) .
\label{s.4}
\eeq
The determinant of $X$ is the square of the proper time:
\beq
\mbox{det}(X) = (x^0)^2 - \mathbf{x} \cdot \mathbf{x} =
- \eta_{\mu\nu} x^{\mu}x^{\nu} =\tau^2 .
\label{s.5}
\eeq
Taking complex conjugates of (\ref{s.2}) gives
\beq
x^{\mu*} = {1 \over 2} \mbox{Tr}(\sigma_{\mu}^*X^*)=
{1 \over 2} \mbox{Tr}((\sigma_{\mu}^*X^*)^t)=
{1 \over 2} \mbox{Tr}(X^{\dagger}\sigma^{\dagger}_{\mu}) =
{1 \over 2} \mbox{Tr}(X^{\dagger}\sigma_{\mu}) =
{1 \over 2} \mbox{Tr}(\sigma_{\mu}X^{\dagger}).
\label{s.6}
\eeq
This will be equal to $x^{\mu}$ if and only if $X=X^{\dagger}$.  

It follows that any linear transformation that preserves both the 
Hermiticity and the determinant of $X$ must be a real Lorentz transformation.

A general linear transformation of the matrix $X$ has the form 
\beq
X' = A X B  .
\label{s.7}
\eeq
Hermiticity of $X'$ requires
\beq
AXB = B^{\dagger}X A^{\dagger}
\label{s.8}
\eeq
or 
\beq
A^{-1} B^{\dagger} X = X B A^{-1 \dagger}
\label{s.9}
\eeq
for any Hermitian $X$. 
If $X$ is set to the identity this becomes
\beq
C := B A^{-1 \dagger} =
A^{-1} B^{\dagger} =  C^{\dagger}.
\label{s.10} 
\eeq
Using (\ref{s.10}) in (\ref{s.9}) gives
\beq
CX = XC .
\label{s.11}
\eeq
This means that for any Hermitian $X$
\beq
[X,C]=0 .
\label{s.12}
\eeq
Since this must be true for $X=\sigma_{\mu}$ and any 
complex matrix can be expressed as 
$M=m^{\mu}\sigma_{\mu}$, it follows that 
$C$ commutes with every complex $2 \times 2$ matrix, so it must
be proportional to the identity, $C=cI$, with a real constant $c$
(by Hermiticity).
This leads to the relation
\beq
B= c A^{\dagger} .
\label{s.13}
\eeq
The condition on the determinant requires 
\beq
c^2 \vert \mbox{det} (A) \vert^2=1 .
\label{s.14}
\eeq
The magnitude of $c$ can be absorbed into the matrices
by redefining 
$A \to A'= {1 \over \sqrt{\vert c \vert}}A$ .  Then 
$c=\pm 1$ which gives
\beq
B = \pm A^{\dagger} .
\label{s.15}
\eeq
The (-1) changes the sign of all components of $X$
so it corresponds to a space-time reflection, 
which is not in the proper subgroup of the Lorentz group 
(the component connected to the identity).  
It follows that 
\beq
X'= A X A^{\dagger} \qquad \mbox{det} (A)=1 . 
\label{s.16}
\eeq
The determinant could be allowed to have a phase, but the
$\dagger$ will cause the phases to cancel, so there is no loss of
generality in choosing the determinant to be 1.

It follows that any $SL(2,\mathbb{C})$ matrix $A$ defines a real
proper Lorentz transformation by
\beq
\Lambda^{\mu}{}_{\nu} = {1 \over 2} \mbox{Tr}(\sigma_{\mu} A \sigma_\nu  
A^{\dagger}).
\label{s.16a}
\eeq

\begin{center}
{\bf
General form of $A$
}    
\end{center}

A general invertible complex $2\times 2$ matrix can always be expressed in
exponential form
\beq
A = e^M = e^{m^{\mu} \sigma_{\mu}}.
\label{s.17}
\eeq
The requirement that 
\beq
1 = \mbox{det}(A) =  e^{m^{\mu}\mbox{Tr}( \sigma_{\mu})} =
e^{2m^0} 
\label{s.18}
\eeq
holds for $m^0= n \pi i$.  This gives
\beq
A= \pm e^{\mathbf{z} \cdot \pmb{\sigma}}
\label{s.19a}
\eeq 
where $\mathbf{z}$ is a complex vector.  The minus sign can be absorbed
in $\mathbf{z}$ since 
\beq
-I = e^{i \pi \pmb{\sigma} \cdot \hat{\mathbf{a}}}
\label{s.19b}
\eeq
for any unit vector $\hat{\mathbf{a}}$, so a general $A \in SL(2,\mathbb{C})$
has the form 
\beq
\boxed{
A= e^{\mathbf{z} \cdot \pmb{\sigma}} .
}
\label{s.19c}
\eeq 
Note that both $A$ and $-A$ 
have determinant 1 and lead to the same
Lorentz transformation since the (-) signs cancel in
\beq
X' = A X A^{\dagger} . 
\label{s.20}
\eeq
This is the same behavior exhibited by $SU(2)$.

Finally note that $A(z)= e^{\mathbf{z} \cdot \mathbf{\sigma}}$ maps
the complex plane into $SL(2,\mathbb{C})$, so any path in $SL(2,\mathbb{C})$
is parameterized by a path in the complex plane that can be contracted to 
the identity, which implies that $SL(2,\mathbb{C})$ is simply connected.

\begin{center}
{\bf
Polar decomposition - generalized Melosh rotations and canonical boosts
}    
\end{center}  

$SL(2,\mathbb{C})$ matrices $A$ have polar decompositions
\beq
A = (AA^{\dagger})^{1/2}  (AA^{\dagger})^{-1/2}A =
A (A^{\dagger}A)^{-1/2}  (A^{\dagger}A)^{1/2}   
\label{s.21}
\eeq
where $(AA^{\dagger})^{1/2}$ and $(A^{\dagger}A)^{1/2}$ are positive 
Hermitian matrices and $(AA^{\dagger})^{-1/2}A$  and 
$A (A^{\dagger}A)^{-1/2}$ are $SU(2)$ matrices.
Define 
\beq
P_l := (AA^{\dagger})^{1/2} \qquad
U_r := (AA^{\dagger})^{-1/2}A
\label{s.22}
\eeq
\beq
P_r :=  (A^{\dagger}A)^{1/2} \qquad
U_l := A (A^{\dagger}A)^{-1/2} .
\label{s.23}
\eeq
Equation (\ref{s.21}) implies that a general $SL(2,\mathbb{C})$ matrix
$A$ has decompositions of the form
\beq
A = P_l U_r = U_l P_r  .
\label{s.24}
\eeq
The positive Hermitian $SL(2,\mathbb{C})$ matrices 
have the form 
\beq
P= e^{\pmb{\rho}  \cdot \pmb{\sigma} /2} = \cosh (\rho /2)\sigma_0
+ \hat{\pmb{\rho}}\cdot \pmb{\sigma} \sinh (\rho/2)
\label{s.25}
\eeq
while the unitary $SL(2,\mathbb{C})$ ones have the form
\beq
U= e^{i \pmb{\theta}  \cdot \pmb{\sigma} /2} =
\cos (\theta /2)\sigma_0
+i \hat{\pmb{\theta}}\cdot \pmb{\sigma} \sin (\theta/2) .
\label{s.26}
\eeq
The factor of ${1/2}$ is a convention motivated by the
$4\times 4$ matrix representations of the Lorentz group.

The Lorentz transformation $\Lambda^{\mu}{}_{\nu}$ is related to the
$SL(2,\mathbb{C})$ matrix $A$ by 
\beq
\Lambda^{\mu}{}_{\nu} = 
{1\over 2} \mbox{Tr}(\sigma_{\mu} A \sigma_{\nu}  A^{\dagger}). 
\label{s.27}
\eeq
It can be computed for both real and imaginary $\mathbf{z}$.  In the
positive case it is a rotationless or canonical boost.  In the unitary
case it is a rotation.

$SL(2,C)$ representatives of canonical boosts are given by:
\beq
A= e^{{1 \over 2}\pmb{\rho}\cdot \pmb{\sigma}} .
\label{s.28}
\eeq
This $A$ has the property that it transforms $(m,\mathbf{0})$ to
\beq
p^{\mu}\sigma_{\mu} = A m\sigma_0 A^{\dagger}
\qquad
\mbox{where}
\qquad
p^{\mu} = (\sqrt{m^2 + \mathbf{p}^2},\mathbf{p})
= {1 \over 2} \mbox{Tr} (\sigma^{\mu}A m\sigma_0 A^{\dagger}),  
\label{s.29}
\eeq
which represents a Lorentz boost with rapidity $\pmb{\rho}$ defined by 
\beq
\hat{\pmb{\rho}} = \hat{\mathbf{p}}=\hat{\mathbf{v}}
\label{s.30}
\eeq
and
\beq
\sinh (\rho) = {\vert \mathbf{p} \vert \over m} = \vert \mathbf{v} \vert
\label{s.31}
\eeq
\beq
\cosh (\rho) = {p^0  \over m} = v^0
\label{s.32}
\eeq
\beq
\sinh ({\rho \over 2} ) =  \sqrt{p^0 -m \over 2m}=\sqrt{v^0 -1 \over 2}
\label{s.33}
\eeq
\beq
\cosh ({\rho \over 2} ) =  \sqrt{p^0 +m \over 2m}=\sqrt{v^0 +1 \over 2} 
\label{s.34}
\eeq
with 
\[
A= B_c(v) := B_c(p/m)= 
\cosh (\rho/2) \sigma_0 + \sinh (\rho/2) \hat{\mathbf{v}} \cdot {\bsig} =
\]
\[
\sqrt{v^0 +1 \over 2}\sigma_0 + 
\sqrt{v^0 -1 \over 2}\hat{\mathbf{v}} \cdot {\bsig} =
\]
\[
{1 \over \sqrt{2(v^0+1)}} \left ( (v^0+1) \sigma_0 + \mathbf{v} 
\cdot {\bsig}\right ) =
\]
\beq
{1 \over \sqrt{2m(p^0+m)}} \left ( (p^0+m) \sigma_0 + \mathbf{p} 
\cdot {\bsig}\right ) 
\label{s.35}
\eeq
and
\beq
B^{\dagger}_c(v)= B_c(v).
\label{s.36}
\eeq
The inverse of a canonical boost can be computed by reversing the 
sign of $\mathbf{p}$ or
$\mathbf{v}$ or $\hat{\pmb{\rho}}$
\[
B^{-1}_c (v) = \sigma_2 {B}^*_c (v) \sigma_2  = 
\cosh (\omega/2) \sigma_0 - \sinh (\omega_2) \hat{\mathbf{v}} \cdot {\bsig} =
\]
\[
\sqrt{v^0 +1 \over 2}\sigma_0 -
\sqrt{v^0 -1 \over 2}\hat{\mathbf{v}} \cdot {\bsig} =
\]
\[
{1 \over \sqrt{2(v^0+1)}} \left ( (v^0+1) \sigma_0 - \mathbf{v} 
\cdot {\bsig}\right ) =
\]
\beq
{1 \over \sqrt{2m(p^0+m)}} \left ( (p^0+m) \sigma_0 - \mathbf{p} 
\cdot {\bsig}\right ) .
\label{s.37}
\eeq
This is not true for a general boost. 
Note that in all of the above expressions for the boosts, $v^0$ or $p^0$ 
represent ``on-shell'' quantities. 

Finally an important observation in what follows is  
\beq
B_c(\mathbf{p}/m)^2 = e^{\pmb{\rho}\cdot \pmb{\sigma}} = \cosh (\rho) \sigma_0 +
\hat{\mathbf{p}} \cdot \pmb{\sigma} \sinh (\rho) =
{1 \over m} p^{\mu}\sigma_{\mu}
\label{s.38}
\eeq
where $p^0=\sqrt{m^2 + \mathbf{p}^2}$.  This is a square of the
Hermitian matrix, $e^{\pmb{\rho}\cdot \pmb{\sigma}/2}$,
so it is a positive Hermitian matrix.

\begin{center}
{\bf
Inequivalence of conjugate representation: $A\not= SA^*S^{-1}$
}    
\end{center}
$SL(2,\mathbb{C})$ matrices have some important properties. 
Both  $SL(2,\mathbb{C})$ and the complex conjugate representation
are representations, but they are inequivalent.
This means that there is no single similarity transformation
$S$ that relates the two representations
\beq
A^* = S A S^{-1}
\label{s.39}
\eeq
for all $A$.  To show this note that if (\ref{s.39})
holds it follows that for $A= e^{{1 \over 2} \mathbf{z}\cdot \pmb{\sigma}}$ that  
\beq
\mathbf{z} \cdot S \pmb{\sigma} S^{-1} = \mathbf{z}^* \cdot \pmb{\sigma}^* 
\label{s.40}
\eeq
for all complex $\mathbf{z}$.  This can be rewritten 
\beq 
\mathbf{z} \cdot S \pmb{\sigma} S^{-1} = - \mathbf{z}^* \cdot  
\sigma_2 \pmb{\sigma} \sigma_2 . 
\label{s.41}
\eeq
For the special case that $\mathbf{z}=i\mathbf{y}$ is pure
imaginary this becomes
\beq
\mathbf{y}\cdot S \pmb{\sigma} S^{-1} = \mathbf{y}\cdot  
\sigma_2 \pmb{\sigma} \sigma_2 .
\label{s.42}
\eeq
This is because $\sigma_2$ is imaginary and anti-commutes
with $\sigma_1$ and $\sigma_2$.  
Thus for imaginary $\mathbf{z}$,  
$S= \sigma_2 C$ where $C$ is a matrix that commutes with 
$\pmb{\sigma}$. The only matrix 
commuting with all of the Pauli matrices is a constant
multiplied by the identity.  It follows that 
$S= c \sigma_2$ and $S^{-1} = c^{-1} \sigma_2$.  The constant factor
can be taken as 1 since it does not change the overall 
similarity transformation.  For real $\mathbf{z}$ 
this requires 
\beq 
\sigma_2 \pmb{\sigma} \sigma_2 =  \pmb{\sigma} 
\label{s.43}
\eeq
which is not true for $\sigma_1$ and $\sigma_3$.  This shows
that in general there is NO $S$ satisfying 
\beq
\boxed{
A^{*} = SAS^{-1}
}
\label{s.44}
\eeq
for all $A\in SL(2,\mathbb{C})$,
however it was demonstrated that
\beq
\boxed{
R^{*} = \sigma_2 R \sigma_2
}
\label{s.45}
\eeq
for all $A=R\in SU(2)$.

Equation (\ref{s.45}) is special case of the general property
of $SL(2,\mathbb{C})$ matrices
\beq
\boxed{
\sigma_2 A \sigma_2 = (A^{t})^{-1}
\qquad
\sigma_2 A^* \sigma_2 = (A^{\dagger})^{-1} .
}
\label{s.46}
\eeq
Equations (\ref{s.44}) and (\ref{s.45}) mean that while $SU(2)$
representations are equivalent to the
complex conjugate representations, this relation is not true for
$SL(2,\mathbb{C})$ representations.  This fact has implications for
structure of Lorentz covariant descriptions of free particles and
the treatment of space reflections in these representations.

\begin{center}
{\bf
Complex Lorentz transformations
}    
\end{center}

If both $A,B \in SL(2,\mathbb{C})$ then for
\beq
Y:= A X B^{t} 
\label{c.1}
\eeq
it still follows that 
\beq
\det{Y} = \det{X} \qquad \mbox{but} \qquad Y^{\dagger} \not = Y.
\label{c.2}
\eeq
This means that the pair $(A,B)$ represents a transformation
that preserves the proper time, $-x^2 =-y^2 $ with $y^{\mu *}\not= y^{\mu}$ 
i.e. it is a complex Lorentz transformation.

If $\sigma_0$ is replaced by $i \sigma_0$ and $\sigma_{e\mu}$ is defined by 
\beq
\sigma_{e\mu} : = (i \sigma_0, \pmb{\sigma})
\label{c.3}
\eeq
then
\beq
\mbox{det} (x_e^{\mu} \sigma_{e\mu}) = -(x^0_e)^2 - \mathbf{x}\cdot \mathbf{x}
\label{c.4}
\eeq
which is $(-)$ the square of the Euclidean length of $x_e^{\mu}$.
The Euclidean four vector
$x_e^{\mu}$ can also be represented by a $2 \times 2$ matrix:
\beq
X_e = x_e^{\mu} \sigma_{e\mu} 
\label{c.5}
\eeq
which can be inverted using
\beq
x_e^{\mu} = {1 \over 2} \mbox{Tr} (\sigma_{e\mu}^{\dagger}X_e). 
\label{c.6}
\eeq
It follows from (\ref{c.4}) that 
\beq
X_e' = A X_e B^t \qquad \mbox{det}(A) = \mbox{det}(B)=1
\label{c.7}
\eeq
also preserves the Euclidean distance.  This means that 
\beq
O^{\mu}{}_\nu (A,B) =
{1 \over 2} \mbox{Tr} (\sigma_{e\mu}^{\dagger}A\sigma_{e\nu}B^t) .
\label{c.8}
\eeq
is a complex four-dimensional orthogonal transformation.  The result
of these observations is that
$SL(2,\mathbb{C}) \times SL(2,\mathbb{C})$ represents both complex
Lorentz and complex orthogonal transformations.  The transpose is 
included in 
(\ref{c.7}) so the 
group multiplication property has the form
\beq
(A',B')(A,B) = (A'A,B'B)
\label{c.9}
\eeq
where each factor represents matrix multiplication.

If both $A$ and $B$ are $SU(2)$ matrices, then $(A,B)$ defines
a real four-dimensional orthogonal transformation.
To show reality when
$A$ and $B$ are $SU(2)$ matrices note that 
the transformed coordinates are 
\beq
y^\mu_e = {1 \over 2} \mbox{Tr}(\sigma_{e\mu}^{\dagger} A X_e B^{t} ).
\label{c.10}
\eeq
Taking complex conjugates (for $A,B\in SU(2)$) 
\beq
y^{\mu *}_e = {1 \over 2} \mbox{Tr}(\sigma_{e\mu}^{\dagger *} A^* X^*_e B^{t*} ).
\label{c.11}
\eeq
For $SU(2)$  matrices (\ref{s.45}) gives
\beq
A^* = \sigma_2 A \sigma_2 \qquad B^{t*} = \sigma_2 B^t \sigma_2 .
\label{c.12}
\eeq
Using (\ref{c.12}) in (\ref{c.11}) gives
\beq
y^{\mu *}_e = {1 \over 2} \mbox{Tr}(\sigma_{e\mu}^{\dagger *} \sigma_2
A \sigma_2 X^*_e \sigma_2 B^{t} \sigma_2  ). 
\label{c.13}
\eeq
For real $x_e^\mu$
\beq
\sigma_2 X^*_e \sigma_2 = - X_e
\label{c.14}
\eeq
so (\ref{c.13}) becomes
\beq
y^{\mu *}_e = {1 \over 2} \mbox{Tr}(-\sigma_{e\mu}^{\dagger *} \sigma_2
A  X_e  B^{t} \sigma_2  ) =
\label{c.15}
\eeq
\beq
{1 \over 2} \mbox{Tr}(-\sigma_2 \sigma_{e\mu}^{\dagger *} \sigma_2
A  X_e  B^{t} )= 
\label{c.16}
\eeq
\beq
{1 \over 2} \mbox{Tr}(\sigma_{e\mu}^{\dagger}
A  X_e  B^{t} )= y^{\mu}.
\label{c.17}
\eeq
This shows that pairs of $SU(2)$ matrices represent real
four-dimensional orthogonal transformations.

These considerations are relevant for Euclidean representations of
relativistic particles.

\begin{center}
{\bf
Rotations and canonical boosts
}
\end{center}  

$SU(2)$ rotations have the form
\beq
R = e^{i \pmb{\theta} \cdot \pmb{\sigma}/2} =
\cos (\theta/2) \sigma_0 + i \hat{\pmb{\theta}}
\sin (\theta/2)
\label{r.1}
\eeq
corresponding to a rotation about the $\hat{\pmb{\theta}}$ axis
by $\theta$.

The canonical boosts have the important property that the Wigner
rotation of a rotation is the rotation.  This is shown below.  The
following notation is used:
$R$ represents an $SU(2)$ rotation and $\mathbf{R}$ represents
the corresponding $SO(3)$ rotation:
\beq
R e^{{1 \over 2} \pmb{\rho}\cdot \pmb{\sigma}} R^{\dagger} =
e^{{1 \over 2} \pmb{\rho}\cdot R\pmb{\sigma}R^{\dagger} } =
e^{{1 \over 2} \pmb{\rho}\cdot (\mathbf{R}^T\pmb{\sigma}) }=
e^{{1 \over 2} (\mathbf{R}\pmb{\rho})\cdot\pmb{\sigma} }.
\label{r.2}
\eeq
This can be written as
\beq
R B_c(\mathbf{p}/m)R^{\dagger} = B_c(\mathbf{R}\mathbf{p}/m)
\label{r.3}
\eeq
or
\beq
\boxed{
R = B_c^{-1} (\mathbf{R}\mathbf{p}/m)R B_c(\mathbf{p}/m) =
R_{wc}(R,\mathbf{p}/m).
}
\label{r.4}
\eeq
This property is unique to canonical boosts.  The important property
is that the Wigner rotation of a rotation is the rotation, independent of
$\mathbf{p}$.  This means that if a rotation is applied to a 
many-particle system, where each particle has a different momentum, all of
the particles' spins will Wigner rotate the same way - independent of
their momenta.  This allows them to be coupled with ordinary
Clebsch-Gordan coefficients.  Adding angular momenta is most easily
preformed by transforming all of the spins to canonical spins.

\begin{center}
{\bf
Melosh Rotations
}
\end{center}  

In order to add spins it is necessary to first convert them to
canonical spins so they can be added.  After adding the spins they can
be converted back to their original spin representation.  The
matrices that transform the spins are generalized Melosh
rotations (the original Melosh transformation \cite{Melosh:1974cu}
relates light-front spins to canonical spins).

If a general boost is right multiplied by the inverse of a
canonical boost the result is a $SU(2)$ rotation, since it maps
zero momentum to zero momentum
\beq
R_{cx}(p/m)= B_c^{-1}(p/m) B_x(p/m)  .
\label{r.5}
\eeq
This can be expressed in the form
\beq
B_x(p/m) = B_c(p/m) R_{cx}(p/m)
\label{r.6}
\eeq
where $R_{cx}(p/m)$ is the $SU(2)$  (rotation) from the
polar decomposition (\ref{s.22}) of $B_x(p/m)$.
For
$A =B_x(p)$ the generalized Melosh rotation is given by
\beq
R_{cx}:= (AA^{\dagger})^{-1/2} A =
(B_x(p/m)B_x(p/m)^{\dagger})^{-1/2} B_x(p/m).
\label{r.7}
\eeq
while the associated canonical boost is
\beq
B_c(p/m)= (AA^{\dagger})^{1/2} .
\eeq

An important observation is that 
\beq
\boxed{
B_x(p/m)B^{\dagger} _x(p/m)= 
B_c(p/m) R_{cx}(p/m) R^{\dagger}_{cx}(p/m)B_c(p/m)
= B_c^2 (p/m) = {p^{\mu}\sigma_{\mu} \over m}
}
\label{r.8}
\eeq
{\it independent of} $x$.  This is a consequence of the
polar decomposition of the $SL(2,\mathbb{C})$ matrices.  It will be 
used to show that Dirac's forms of dynamics are irrelevant in Lorentz 
and Euclidean covariant representations of relativistic quantum mechanics.

The generalized Melosh rotations are used to change the type of spins
($y \to x)$: 
\[
\vert (m,j) \mathbf{p}, \mu \rangle_x =
U(B_x(p/m))\vert (m,j) \mathbf{0}, \mu \rangle_x \sqrt{{m \over\omega_m(p)}}=
U(B_y(p/m))U(B^{-1}_y(p/m) B_x(p/m))\vert (m,j) \mathbf{0}, \mu \rangle_x
\sqrt{{m \over \omega_m(p)}}=
\]
\[
U(B_y(p/m))\vert (m,j) \mathbf{0}, \nu \rangle_x
D^j_{\nu\mu} [B^{-1}_y(p/m) B_x(p/m)]
\sqrt{{m \over \omega_m(p)}}=
\vert (m,j) \mathbf{p}, \nu \rangle_y
D^j_{\nu\mu} [B^{-1}_y(p/m) B_x(p/m)]
\]

\begin{center}
{\bf  
$SL(2,\mathbb{C})$ representations of light-front boosts:
}
\end{center}

The light-front is the hyper-plane defined by points satisfying
$x^+=x^0+x^3=0$.  The kinematic subgroup of the light front
is the subgroup of Poincar\'e group that preserves $x^+=0$.

In $SL(2,\mathbb{C})$ the Lorentz transformations in this
subgroup are represented by lower triangular matrices.
$SL(2,\mathbb{C})$ representatives of light-front boosts are given by:
\beq
B_f (v) := 
\left (
\begin{array}{cc} 
\sqrt{v^+}  &0   \\
v_{\perp} /\sqrt{v^+} & 1/\sqrt{v^+}  
\end{array} 
\right ) =
\left (
\begin{array}{cc} 
\alpha  &0   \\
\beta/\alpha & 1/\alpha  
\end{array} 
\right ) 
\label{lf.1}
\eeq 

\beq
B_f^{-1} (v) := 
\left (
\begin{array}{cc} 
1/\sqrt{v^+}  &0   \\
-v_{\perp} /\sqrt{v^+} & \sqrt{v^+}  
\end{array} 
\right ) =
\left (
\begin{array}{cc} 
1/\alpha  &0   \\
-\beta/\alpha & \alpha  
\end{array} 
\right ) 
\label{lf.2}
\eeq 

\beq
B_f^{\dagger}  (v) := 
\left (
\begin{array}{cc} 
\sqrt{v^+}  &v^*_{\perp} /\sqrt{v^+}   \\
0 & 1/\sqrt{v^+}  
\end{array} 
\right ) =
\left (
\begin{array}{cc} 
\alpha  & \beta^*/\alpha  \\
0 & 1/\alpha  
\end{array} 
\right ) 
\label{lf.3}
\eeq 

\beq
\tilde{B}_f (v) := 
\left (
\begin{array}{cc} 
1/\sqrt{v^+}  &-v^*_{\perp} /\sqrt{v^+}  \\
0 & \sqrt{v^+}  
\end{array} 
\right ) =
\left (
\begin{array}{cc} 
1/\alpha  & -\beta^*/\alpha  \\
0 & \alpha  
\end{array} 
\right ) 
\label{lf.4}
\eeq 
where $\alpha := \sqrt{v^+}= \sqrt{p^+/m}$ and $\beta:= v_{\perp}:= (p_1+i p_2)/m$.  In (\ref{lf.4}) and in what follows the notation  $\tilde{A}:= (A^{\dagger})^{-1}$ is used. 
\bigskip

These lower triangular matrices with real quantities on the
diagonal form a group.   This is the subgroup of light-front 
boosts.  The light-front boost subgroup
can be expressed in terms of the light-front 
components of the four momentum and mass as:
\beq
B_f (p) := 
{1 \over \sqrt{mp^+} } \left (
\begin{array}{cc} 
p^+  &0   \\
p_{\perp}  & m  
\end{array} 
\right )
\label{lf.5}
\eeq

\beq
B_f^{-1} (p) := 
{1 \over \sqrt{mp^+} }
\left (
\begin{array}{cc} 
m  &0   \\
-p_{\perp}  & p^+  
\end{array} 
\right ) 
\label{lf.6}
\eeq 

\beq
B_f^{\dagger}  (p) := 
{1 \over \sqrt{mp^+} } \left (
\begin{array}{cc} 
p^+  &p^*_{\perp}   \\
0 & m   
\end{array} 
\right )
\label{lf.7}
\eeq 

\beq
\tilde{B}_f (p) := 
{1 \over \sqrt{mp^+} } \left (
\begin{array}{cc} 
m  &-p^*_{\perp}   \\
0 & p^+  
\end{array} 
\right ).
\label{lf.8}
\eeq 
These boosts are used to define light front spins.  Since these
boosts form a subgroup
the light-front boosts do not change the light-front spin.

\begin{center}
{\bf  
$SL(2,\mathbb{C})$ representations of Helicity boosts:
}
\end{center}

Helicity boosts are defined by

\beq
B_h(p/m) := B_c(p/m) R(\hat{\mathbf{z}} \to \hat{\mathbf{p}})=
R(\hat{\mathbf{z}} \to \hat{\mathbf{p}}) B_c(p_z/m).
\label{h.1}
\eeq
where the rotation 
\beq
R(\hat{\mathbf{z}} \to \hat{\mathbf{p}}) 
\label{h.3}
\eeq
is a rotation about the $\mathbf{z} \times \hat{\mathbf{p}}$ axis
through an angle $\theta = \cos^{-1}
(\mathbf{z} \cdot \hat{\mathbf{p}})$ given by
\beq
R(\hat{\mathbf{z}} \to \hat{\mathbf{p}})=
\sqrt{{1+ \hat{\mathbf{z}} \cdot \hat{\mathbf{p}} \over 2}}\sigma_0 +
\sqrt{{1- \hat{\mathbf{z}} \cdot \hat{\mathbf{p}} \over 2}}
{(\hat{\mathbf{z}} \times \mathbf{p})\cdot \pmb{\sigma} \over 
\vert \hat{\mathbf{z}} \times \mathbf{p}\vert} .
\label{h.4}
\eeq
The associated helicity-spin Wigner rotation is
\beq
R_{wh}(\Lambda,p) = R^{-1}(\hat{\mathbf{z}} \to \hat{\pmb{\Lambda}}p)
B_c^{-1}(\Lambda p/m) \Lambda B_c(p/m) R(\hat{\mathbf{z}} \to \hat{\mathbf{p}})
\label{h.2}
\eeq
which is always a rotation about the $z$ axis.  Because of this
property the Wigner D-function (\ref{p.32}) of the Jacob-Wick helicity
Wigner rotation is always a phase. 

The helicity spin and canonical spin are related by \cite{wpwg}
\beq
\mathbf{j}_h \cdot \hat{\mathbf{z}} =
\mathbf{j}_c \cdot \hat{\mathbf{p}}
\eeq
so the $z$ component of the helicity spin is the 
canonical spin projected in the direction of the momentum.  This
projection is the better known definition of the Jacob-Wick helicity.

\begin{center}
{\bf  
Lorentz Spinors:
}
\end{center}
The transformation property of a four vector represented by a
$2 \times 2$ Hermitian matrix can be expressed in tensor form as
\beq
X^{a\dot{a}} \to X^{\prime a\dot{a}}:=A^{ab}A^{*\dot{a}\dot{b}} X^{b\dot{b}}
\label{sp.1}
\eeq
where repeated matrix indices are assumed to be summed over two values. 
This looks like a rank two tensor with one index transforming under 
$SL(2,\mathbb{C})$ and one under the inequivalent complex conjugate
representation.  

This motivates the definition of Lorentz spinors.  These are
two-component vectors that transform under either of these
representations.  

The two-component spinors are characterized by their transformation
properties
\beq
\xi^a \to \xi^{a\prime} = A^{a b} \xi^b
\qquad 
\xi^{\dot a} \to \xi^{\dot a \prime} = A^{*\dot a \dot b} \xi^{\dot
b}
\label{sp.2}
\eeq
where a sum over repeated spinor indices is assumed.
These transformation properties define two different types of two
spinors that transform under the regular and complex conjugate
representations of $SL(2,\mathbb{C})$.  The upper un-dotted or dotted
indices identify the transformation properties.  These are referred
to as right- and left-handed spinors respectively.  The reason for this
designation will be discussed later.
  
It is possible to construct Lorentz invariant quadratic forms with either of
these types of spinors.  This follows from the general property of
$SL(2,C)$ matrices
(\ref{s.46}):
\beq
\sigma_2 A \sigma_2 = (A^{-1})^t .
\eeq
This leads to the definition of the metric spinor  
\label{sp.2}
\beq
\epsilon_{ab} = -\epsilon^{ab} = i(\sigma_2)_{ab}
\qquad 
\epsilon_{\dot a\dot b} = -\epsilon^{\dot a \dot b} = i(\sigma_2)_{\dot a \dot b}
\label{sp.3}
\eeq
and lower indexed spinors
\beq
\xi_a := \epsilon_{ab} \xi^b \qquad
\xi_{\dot a} := \epsilon_{\dot a \dot b} \xi^{\dot b}. 
\label{sp.4}
\eeq
The transformation properties of the lower index spinors are
\beq
\xi_a \to \xi_{a\prime} = \epsilon_{ab} A^{bc}
\epsilon^{cd} \epsilon_{de} \xi^e =
(A^{t})^{-1 ab} \xi_b 
\label{sp.5}
\eeq
and
\beq
\qquad 
\xi_{\dot a} \to \xi_{\dot a \prime} =
\epsilon_{\dot a \dot b}
A^{*\dot b \dot c} \epsilon^{\dot c \dot d} \epsilon_{\dot d \dot e} 
\xi^{\dot e}= (A^{\dagger})^{-1 \dot a \dot b} \xi_{\dot b}.
\label{sp.6}
\eeq

The metric spinor, $\epsilon_{ab}$ could also be taken to be
$(\sigma_{2})_{ab}$. It has the advantage that there are no sign
changes on raising and lowering indices, but the disadvantage is that
it is not real.  Equations (\ref{sp.5})-(\ref{sp.6})
show that the lower undotted and dotted indices
have different transformation properties than the corresponding
upper indices.

The metric spinor can be used to construct Lorentz invariant
scalars by contracting upper and lower indexed spinors of the
same type (dotted or undotted)
\beq
\chi'_a \xi^{\prime a} =
(A^{t})^{-1 ab} \chi_b A^{ac} \xi^c =
\chi_b (A)^{-1 ba}  A^{ac} \xi^c =
\chi_a \xi^{a}
\label{sp.7}
\eeq
and
\beq
\chi'_{\dot{a}} \xi^{\prime \dot{a}} =
(A^{\dagger})^{-1 \dot{a}\dot{b}} \chi_{\dot{b}} A^{*\dot{a}\dot{c}} \xi^{\dot{c}} =
\chi_{\dot{b}} (A)^{*-1 \dot{b}\dot{a}}  A^{*\dot{a} \dot{c}} \xi^{\dot{c}} =
\chi_{\dot{a}} \xi^{\dot{a}}.
\label{sp.8}
\eeq
It follows from the anti-symmetry of $\epsilon_{ab}$ that
\beq
\xi^a \xi_a = \epsilon_{ab} \xi^a \xi^b = 0 \qquad
\xi^{\dot a}\xi_{\dot a} =  
\epsilon_{ab} \xi^{\dot a}\xi^{\dot b} = 0
 .
\label{sp.9}
\eeq

The tensor product of a 2-spinor with its complex conjugate,
\beq
X^{a\dot{b}}:= \xi^a \xi^{*\dot{b}},
\label{sp.10}
\eeq
defines a real four vector; since it is Hermitian and the determinant
vanishes this defines a light-like four vector.

It follows from (\ref{sp.7}) and (\ref{sp.8}) that 
\beq
\xi^a \chi_a \qquad \xi^{\dot a}\chi_{\dot a}
\label{sp.11}
\eeq
are both invariant quadratic forms under $SL(2,C)$.
These forms are neither positive nor
sesquilinear.  Thus they cannot be used to construct a positive
invariant scalar product.  However in terms of the spinor indices it
is useful to define the following 4-momentum dependent $2 \times 2$
Hermitian matrices that transform like products of right and left handed
spinors
\beq
P^{a \dot a}  := (p^{\mu}\sigma_{\mu})^{a \dot a}
\label{sp.15}
\eeq
\beq
P_{a \dot a} := p^{\mu}(\sigma_2 \sigma_{\mu} \sigma_2)_{a \dot a} 
\label{sp.13}
\eeq
\beq
P^{\dot a a}  = (p^{\mu}\sigma^*_{\mu})^{a \dot a}
\label{sp.15a}
\eeq
\beq
P_{\dot a a} = (p^{\mu}\sigma_2 \sigma^*_{\mu}\sigma_2)_{\dot a a} 
\label{sp.16}
\eeq
The matrices
(\ref{sp.15},\ref{sp.13},\ref{sp.15a},\ref{sp.16})
are all positive definite (see (\ref{s.38}))
if $p$ is a time-like positive energy 
four vector.  They satisfy the following covariance properties
\beq
A^{a b} P^{b \dot c} A^{\dagger \dot c \dot d}   := (\Lambda p)^{\mu}\sigma_{\mu}^{a \dot d}
\label{sp.15c}
\eeq
\beq
(A^{t})^{-1}_{a b}  P_{b \dot c}(A^*)^{-1}_{\dot c \dot d} := (\Lambda p)^{\mu}(\sigma_2 \sigma_{\mu} \sigma_2)_{a \dot d} 
\label{sp.13c}
\eeq
\beq
A^{* \dot a \dot b} P^{\dot b c} A^{t c d}  = (\Lambda p)^{\mu}\sigma_{\mu}^{*
\dot a  d}
\label{sp.15c}
\eeq
\beq
(A^{\dagger})^{-1}_{\dot a \dot b} P_{\dot b c}A^{-1}_{c d} =
(\Lambda p)^{\mu}(\sigma_2 \sigma^*_{\mu}\sigma_2)_{\dot a d} 
\label{sp.16c}
\eeq

Because they are positive they can be used as kernels of the invariant
positive sesquilinear forms:
\beq
\xi_{a} \xi^*_{\dot a} P^{a \dot a}= 
\xi^{a} \xi^{* \dot a} P_{a\dot a} \geq 0
\eeq
\beq
\xi^{\dot a*} \xi^{a} P_{\dot a a} =
\xi_{\dot a}^* \xi_{a} P^{\dot a a} \geq  0.
\label{sp.18}
\eeq
The following identity is important in what follows, 
\beq
(p^{\mu}\sigma_2\sigma^*_{\mu}\sigma_2)^{a \dot a} =
(Pp)^{\mu}\sigma_{\mu a \dot a}
\label{sp.18}
\eeq
where $P$ represents a space reflection.

The matrices $P^{a\dot a}/m$,
$P_{a\dot a}/m$, $P^{\dot a a}/m$, $P_{\dot a a}/m$ are all
$SL(2,\mathbb{C})$ matrices. 
The $SL(2,\mathbb{C})$ spins can be added like $SU(2)$ spins
with $SU(2)$ Clebsch-Gordan coefficients.  This is 
because the $SU(2)$ identities 
\beq
\sum \langle j, \mu \vert j_1, \mu_1, j_2, \mu_2 \rangle
D^{j_1}_{\mu_1 \mu_1'}[R]
D^{j_2}_{\mu_2 \mu_2'}[R]
\langle j, \mu' \vert j_1, \mu_1', j_2, \mu_2' \rangle -
D^j_{\mu \mu'}[R] =0,
\label{sp.19}
\eeq
\beq
\sum \langle j, \mu \vert j_1, \mu_1, j_2, \mu_2 \rangle
D^j_{\mu \mu'}[R]
\langle j, \mu' \vert j_1, \mu_1', j_2, \mu_2' \rangle -
D^{j_1}_{\mu_1 \mu_1'}[R]
D^{j_2}_{\mu_2 \mu_2'}[R] =0 
\label{sp.20}
\eeq
also hold when $R$ is replaced by a $SL(2,\mathbb{C})$ matrix $A$.
This follows because both sides of these equations are finite degree
polynomials in the four components of $R$ which are entire analytic
functions of real angles.  This means that the left side of these
equations are entire functions of three complex angles that vanish
when all three angles are real.  It follows by analytic continuation
that they vanish for complex angles.  Thus they hold when $R\to A$ for
$A\in SL(2,\mathbb{C})$.  This means that there are higher spin versions
of the positive kernels (\ref{sp.15}-\ref{sp.16}).
In the next section the same method will be
used to show that $D^j_{\mu \mu'}[A]$ is a $2j+1$ dimensional
representation of $SL(2,\mathbb{C})$.

These relations can be used use to construct
$2j+1$ dimensional representations of $SL(2,\mathbb{C})$
that transform under
\beq D^j_{\mu \nu}[A],D^j_{\mu \nu}[A^*],D^j_{\mu \nu}[(A^t)^{-1}],
\mbox{ or } D^j_{\mu \nu}[(A^\dagger)^{-1}]
\label{sp.21}
\eeq
from the corresponding $2$-component $j=1/2$ spinors.  In these
expression the notation using the upper and lower dotted and undotted
indices is not used.

\section{Lorentz Covariant representations}

The unitary representation of the Poincar\'e group
for a particle of mass $m$ and spin $j$ has the form (\ref{p.41}) 
\beq
U(\Lambda ,a) \vert (m,j) \mathbf{p},\nu \rangle =
e^{i \Lambda p \cdot a} 
\vert (m,j) \pmb{\Lambda}{p},\mu \rangle 
D^j_{\mu \nu} [R_{wx}(\Lambda,p)] 
\sqrt{{\omega_m(\pmb{\Lambda}{p}) \over \omega_m(\mathbf{p})}} .
\label{cov.1}
\eeq
or one of the related forms (\ref{p.47}-\ref{p.50}).

In what follows the notation for $SL(2,\mathbb{C})$
matrices
\beq
\tilde{A} := (A^{\dagger})^{-1} = \sigma_2 A^* \sigma_2 
\label{cov.2}
\eeq
is used.  The spin$_x$ Wigner rotation can be written in either of
two equivalent ways
\beq
D^j_{\nu \mu} [B_x^{-1}(\Lambda p/m) A B_x(p/m)]
=
D^j_{\nu \mu} [\tilde{B}_x^{-1}(\Lambda p/m) \tilde{A} \tilde{B}_x(p/m)],
\label{cov.3}
\eeq
where $A$ and $\Lambda$ are related by (\ref{s.27}).
This is because $\tilde{R}:=(R^{\dagger})^{-1}= R$ for $R \in SU(2)$.

The Wigner function (\ref{p.32}) 
\beq
D^j_{\nu \mu} [e^{{i\over 2} \pmb{\theta}\cdot \pmb{\sigma}}], 
\label{cov.4}
\eeq
is a finite degree polynomial of entire analytic functions of the
three components of $\pmb{\theta}$.  It satisfies the group representation
property (the matrix indices are suppressed): 
\beq
D^j[R_2]D^j[R_1]-D^j[R_2R_1]=0 
\label{cov.5}
\eeq
for $R_1,R_2 \in SU(2)$.
Since the left side is an entire function of all 6 angle variables,
($\pmb{\theta}_1$, $\pmb{\theta}_2$),
that is 0 for all real variables, 
by analytic continuation the group representation property 
holds for complex angles
$i\pmb{\theta} \to \mathbf{z}= \pmb{\rho} + i\pmb{\theta}$.
It follows that $D^j[A]$ is a also $2j+1$ dimensional representation of
$SL(2,\mathbb{C})$.
 
This means that the Wigner rotation can be factored.
There are two possible factorizations that arise because
while $\tilde{R} =R$, this is not true for the $SL(2,\mathbb{C})$
transformations that are used to define the Wigner rotation.  This is due to 
the inequivalence of the two conjugate representations of
$SL(2,\mathbb{C})$.  This leads to the following two factorizations
of the Wigner rotation
\beq
D^j_{\nu \mu} [B_x^{-1}(\Lambda p/m) \Lambda B_x(p/m)]=
\left (
D^j [B_x^{-1}(\Lambda p/m)]
D^j [A ]
D^j [B_x(p/m)]
\right )_{\nu \mu}
\label{cov.6}
\eeq
and
\beq
D^j_{\nu \mu} [B_x^{-1}(\Lambda p/m) \Lambda B_x(p/m)]=
\left (
D^j [\tilde{B}_x^{-1}(\Lambda p/m)]
D^j [\tilde{A} ]
D^j [\tilde{B}_x(p/m)]
\right )_{\nu \mu}.
\label{cov.7}
\eeq
Using these factorizations and the group representation property, 
(\ref{cov.1}) can be equivalently written as
\[
U(\Lambda ,a) \vert (m,j) \mathbf{p},\nu \rangle_x
D^j_{\nu \mu} [B_x^{-1}(p/m)] \sqrt{\omega_m (\mathbf{p})}
=
\]
\beq
e^{i \Lambda p \cdot a} 
\vert (m,j) \pmb{\Lambda}{p},\nu \rangle_x 
D^j_{\nu \alpha} [B_x^{-1} (\Lambda p/m) ]
\sqrt{\omega_m(\pmb{\Lambda}{p})} 
D^j_{\alpha \mu} [A] 
\label{cov.8}
\eeq
or
\[
U(\Lambda ,a) \vert (m,j) \mathbf{p},\nu \rangle_x
D^s_{\nu \mu} [\tilde{B}_x^{-1}(p/m)] \sqrt{\omega_m (\mathbf{p})}
=
\]
\beq
e^{i \Lambda p \cdot a} 
\vert (m,j) \pmb{\Lambda}{p},\nu \rangle_x 
D^j_{\nu \alpha} [\tilde{B}_x^{-1} (\Lambda p/m) ]
\sqrt{\omega_m(\pmb{\Lambda}{p})} 
D^j_{\alpha \mu} [\tilde{A}] .
\label{cov.9}
\eeq

This leads to the definition of two types of Lorentz covariant states
\[
\vert (m,j) \mathbf{p},\mu \rangle_{cov} := 
\]
\beq
\vert (m,j) \mathbf{p},\nu \rangle_x
D^j_{\nu \mu} [B_x^{-1}(p/m)] \sqrt{\omega_m (\mathbf{p})}
\label{cov.10}
\eeq
and
\[
\vert (m,j) \mathbf{p},\nu \rangle_{cov*}:=
\]
\beq
\vert (m,j) \mathbf{p},\nu \rangle_x
D^j_{\nu \mu} [\tilde{B}_x^{-1}(p/m)] \sqrt{\omega_m (\mathbf{p})}.
\label{cov.11}
\eeq
These are called right and left handed Lorentz covariant states.

For these states equations (\ref{cov.8}) and (\ref{cov.9})
have the form
\beq
U(\Lambda ) \vert (m,j) \mathbf{p},\mu \rangle_{cov} =
\vert (m,j) \pmb{\Lambda}{p},\nu \rangle_{cov} D^j_{\nu \mu}[A]
\label{cov.12}
\eeq
\beq
U(\Lambda ) \vert (m,j) \mathbf{p},\mu \rangle_{cov*} =
\vert (m,j) \pmb{\Lambda}{p},\nu \rangle_{cov*} D^j_{\nu \mu}[\tilde{A}].
\label{cov.13}
\eeq
This appears to violate the condition that there are no finite
dimensional unitary representations of the Lorentz group.  The reason
that it does not is because the Hilbert space inner product in this
representation has a non-trivial momentum-dependent kernel.  To see
this it is instructive to write out the inner product of two vectors
in these representations.  Starting with the Poincar\'e covariant representation
\[
\langle \psi \vert \phi \rangle =
\int
\langle \psi \vert (m,j)\mathbf{p}, \mu \rangle_x d\mathbf{p}
_x\langle (m,j) \mathbf{p}, \mu \vert \phi \rangle =
\]
\[
\int
\langle \psi \vert (m,j)\mathbf{p}, \nu \rangle_{cov}
D^j_{\nu \mu} [B_x(p/m)B_x^{\dagger}(p/m) ]
{d\mathbf{p} \over \omega_m (\mathbf{p}) }
{}_{cov}\langle (m,j) \mathbf{p}, \mu \vert \phi \rangle =
\]
\[
\int
\langle \psi \vert (m,j)\mathbf{p}, \nu \rangle_{cov}
D^j_{\nu \mu} [B_c(p/m)B_c(p/m) ]
2 \delta (p^2+m^2)d^4 {p} \theta (p^0)
_{cov}\langle (m,j) \mathbf{p}, \mu \vert \phi \rangle =
\]
\beq
\int 
\langle \psi \vert (m,j) \mathbf{p}, \nu \rangle_{cov}
D^j_{\nu \mu} [p\cdot \sigma/m ]
2 \delta (p^2+m^2)d^4 {p} \theta (p^0)
_{cov}\langle (m,j) \mathbf{p} \mu \vert \phi \rangle .
\label{cov.14}
\eeq
Similarly for the left handed covariant representation
\[
\langle \psi \vert \phi \rangle =
\]
\beq
\int
\langle \psi \vert (m,j) \mathbf{p}, \nu \rangle_{cov*}
D^j_{\nu \mu} [p\cdot \sigma_2\sigma^*\sigma_2/m ]
2 \delta (p^2+m^2)d^4 {p} \theta (p^0)
_{cov*}\langle (m,j) \mathbf{p}, \mu \vert \phi \rangle .
\label{cov.15}
\eeq
Here (\ref{r.8}) was used to replace the $x$-boosts by
canonical boosts.
The Wigner functions in (\ref{cov.14}) have the form
(suppressing the spin indices) 
\beq
D^j[p\cdot \sigma /m] =D^j [B^2_c] =D^j[B_c]^{\dagger} D^j[B_c]>0  
\label{cov.16}
\eeq
and in (\ref{cov.15})
\beq
D^j[p\cdot \sigma_2\sigma^*\sigma_2/m]= D^j [B^{-2}_c] =D^j[B^{-1}_c]^{\dagger}D^j[B^{-1}_c]>0  
\label{cov.16}
\eeq
so they are positive kernels (note that these kernels are Hermitian
since $D^j_{\mu\nu}[A^t] =
D^j_{\nu\mu}[A]= (D^j_{\nu\mu}[A^*])^*$ follows from (\ref{p.32})).

The covariant kernels
\beq
D^j_{\nu \mu} [p\cdot \sigma /m]
2 \delta (p^2+m^2)d^4 {p} \theta (p^0)
\label{cov.17}
\eeq
and
\beq
D^j_{\nu \mu} [p\cdot \sigma_2\sigma^*\sigma_2/m ]
2 \delta (p^2+m^2)d^4 {p} \theta (p^0)
\label{cov.18}
\eeq
are (up to normalization) spin $j$-Wightman functions for right and left handed
free spin-$j$ particles.  They are $2j+1$ dimensional representations of the
positive forms (\ref{sp.15}) and (\ref{sp.16}).

Because 
\beq
p\cdot \sigma_2\sigma^*\sigma_2/m = (Pp)\cdot \sigma/m ,
\label{cov.19}
\eeq
where $P$ changes the sign of the spatial components of $p$, {\it the right
and left handed representations are related by space reflection}.  All
of these transformations are invertible so starting from any one of
them it is possible to return to any standard Poincar\'e covariant
description.  As long as space reflection is not needed, these are 
all equivalent descriptions of a mass $m$ spin $j$ particle.

To understand the role of space reflections
note that taking the complex conjugate of
\beq
X' = A X A^{\dagger} 
\label{cov.20}
\eeq
implies
\beq
X^{\prime *} = A^* X^* A^t .
\label{cov.21}
\eeq 
It follows that $X$ and $X^*$ transform under inequivalent representations
of $SL(2,\mathbb{C})$.  The operation $X \to X^*$ changes
the sign of $y$ which is equivalent to a space reflection in the $x-z$
plane.
This shows that
space reflection maps right handed to left handed
representations of the
Hilbert space.  

In the $2 \times 2$ matrix representation the space reflection, $\mathbf{x}\to -\mathbf{x}$, is represented by
\beq
X\to X' = \sigma_2 X^* \sigma_2.
\label{cov.22}
\eeq
This operation changes $A$ to $\tilde{A}:= \sigma_2 A^* \sigma_2
= (A^{\dagger})^{-1}$.
The problem with space reflections in Lorentz covariant representations is
that the kernel of the Hilbert space representation changes, to the
kernel for an inequivalent representation, so space reflection 
cannot be represented in the Hilbert space with the original Lorentz covariant
kernel because it will not transform
correctly with respect to Lorentz transformations.

The way to remedy this is to use a direct sum, where both kernels
appear on the diagonal.  Then space reflection can be realized on the
direct sum space by changing the sign of $\mathbf{p}$ and
interchanging the components of the direct sum.

In this case the representation of the Lorentz group is the 
chiral representation 
\beq
S[A] =
\left (
\begin{array}{cc} 
D^j[A] & 0\\
0 & D^j [\tilde{A}] \\
\end{array}
\right )
\label{cov.21}
\eeq
and the kernel of  the Hilbert space inner product is 
\beq
\delta (p^2 + m^2) \theta (p^0) 
\left (
\begin{array}{cc} 
D^j[ p \cdot \sigma/m ] & 0 \\
0 & D^j [p \cdot \sigma_2 \sigma^* \sigma_2/m]\\
\end{array}
\right ) =
\delta (p^2 + m^2) \theta (p^0) 
\left (
\begin{array}{cc} 
D^j[p \cdot \sigma/m] & 0 \\
0 & D^j [(Pp) \cdot \sigma /m]\\
\end{array}
\right )
.
\label{cov.22}
\eeq

The operation of space reflection on wave functions in
the doubled space becomes
\beq
P \left (
\begin{array}{c}
_{cov}\langle (m,j) \mathbf{p}, \mu \vert \phi_1 \rangle\\
_{cov*}\langle (m,j) \mathbf{p}, \mu \vert \phi_2 \rangle
\end{array}
\right ) =
\left (
\begin{array}{c}
_{cov*}\langle (m,j) -\mathbf{p}, \mu \vert \phi_2 \rangle\\
_{cov}\langle (m,j) -\mathbf{p}, \mu \vert \phi_1 \rangle
\end{array}
\right ) .
\label{cov.23}
\eeq
The kernels appearing in (\ref{cov.22}) arise naturally because they
come from the $SU(2)$ equivalence of $R$ and $\tilde{R}$, however the spin
kernel $D^j[p \cdot \sigma/m]$ could be replaced by $D^j[p \cdot
\sigma_2 \sigma\sigma_2 /m]$ and $D^j [p \cdot \sigma_2 \sigma^*
\sigma_2/m]$ could be replaced by $D^j [p \cdot \sigma^* /m]$ which
involve different equivalent representations of the right and left
handed spinor degrees of freedom.

An important observation is that the choice of kinematic variables
replacing $\mathbf{p}$ and the choice of boost in the spin
representation that characterize the Poincar\'e covariant forms of the
dynamics has disappeared in the Lorentz covariant representations.
The spins transform under a $2j+1$ dimensional representation of
$SL(2,\mathbb{C})$.  This means that there are ``no forms of
dynamics'' in Lorentz covariant representations.

Another observation is that in the Lorentz covariant representations
the Hilbert space kernels (\ref{cov.17}) and (\ref{cov.18}) have
a mass dependence, which for free particles defines the
dynamics.  In a dynamical Lorentz covariant model the kernel of the
Hilbert space inner product carries the dynamical content of the
theory.

\begin{center}
{\bf
$SL(2,\mathbb{C})\times SL(2,\mathbb{C})$ spinors
}
\end{center}

In order to understand the role played by the spinor degrees of freedom
in Euclidean representations of relativistic quantum mechanics
it is useful to define $SL(2,\mathbb{C})\times SL(2,\mathbb{C})$ spinors.

Let $Z:= z^{\mu} \sigma_{\mu} $ denote a complex 4 vector represented as
a $2 \times 2$ matrix.  Complex Lorentz
transformations are given by
\beq
Z \to Z' = A Z B^t
\label{er.2}
\eeq
where both $A$ and $B$ are $SL(2,\mathbb{C})$ matrices.

In this representation complex space reflection, which transforms
$(z^0,z^1,z^2,z^3)$ to $(z^0,-z^1,-z^2,-z^3)$  can be expressed
in matrix form as
\beq
Z \to Z' = PZ =\sigma_2 Z^t \sigma_2 .
\label{er.3}
\eeq
The transformation properties of $Z$ imply the transformation
properties of $Z':= PZ$:
\beq
PZ \to PZ' = \sigma_2 (B Z^t A^t) \sigma_2 =
(B^{-1})^t PZ A^{-1} .
\label{er.4}
\eeq
This means the under space reflection the complex spinor
transformation properties are replaced by
\beq
A \to (B^t)^{-1}  \qquad 
B \to (A^t)^{-1}.
\label{er.5}
\eeq

This suggests defining right and left handed 
$SL(2,\mathbb{C})\times SL(2,\mathbb{C})$ spinors
by their transformation properties
\beq
\xi^a \to A^{ab} \xi^b 
\label{er.6}
\eeq
\beq
\chi^{\dot{a}} \to B^{\dot{a}\dot{b}} \chi_{\dot{b}} 
\label{er.7}
\eeq
\beq
\xi_a \to ((A^{t})^{-1})_{ab}  \xi_b 
\label{er.8}
\eeq
\beq
\chi_{\dot{a}} \to ((B^t)^{-1})_{\dot{a}\dot{b}} \chi_{\dot{b}} 
\label{er.9}
\eeq
These definitions recover the $SL(2,\mathbb{C})$ transformation
properties of right and left handed spinors when $B=A^*$.
When $(A,B) \in SU(2)\times SU(2)$
these relations define the transformation properties of
right and left handed Euclidean spinors.

The definitions (\ref{er.6}-\ref{er.9}) are consistent with the the upper and
lower index spinors being related by $\epsilon_{ab}$ and
$\epsilon^{ab}$:
\beq
\xi_a = \epsilon_{ab} \xi^b
\qquad
\chi_{\dot{a}} = \epsilon_{\dot{a}\dot{b}} \chi^{\dot{b}}
\label{er.10}
\eeq
and the contraction of an upper and lower index spinor of the same
type (un-dotted or dotted) being invariant under
$SL(2,\mathbb{C})\times SL(2,\mathbb{C})$

The $SL(2,\mathbb{C})$ relations
\beq
D^{j}_{\mu \alpha}[A_2]D^{j}_{\alpha \nu}[A_1] - D^{j}_{\mu \nu}[A_2A_1] = 0
\label{er.11}
\eeq
\beq
\langle j, \mu \vert j_1, \mu_1, j_2, \mu_2 \rangle 
D^{j_1}_{\mu_1 \nu_1}[A]D^{j_2}_{\mu_2 \nu_2}[A] 
\langle j, \nu \vert j_1, \nu_1, j_2, \nu_2 \rangle 
- D^{j}_{\mu \nu}[A ] = 0
\label{er.12}
\eeq
\beq
\langle j, \mu \vert j_1, \mu_1, j_2, \mu_2 \rangle 
D^{j}_{\mu \nu}[R ]
\langle j, \nu \vert j_1, \nu_1, j_2, \nu_2 \rangle - 
D^{j_1}_{\mu_1 \nu_1}[R]D^{j_2}_{\mu_2 \nu_2}[R_]
= 0
\label{er.13}
\eeq
mean that both the group representation property and addition of
``spins'' extend unchanged to $SL(2,\mathbf{C})$.

\section{Euclidean covariant representations of relativistic quantum mechanics}

In the same way that Poincar\'e covariant representations were used to
construct equivalent Lorentz covariant representations of any spin,
the Lorentz covariant representations can be used to construct
equivalent Euclidean covariant representations.

Euclidean formulations of relativistic quantum mechanics are used in
path-integral representations, lattice calculations and with
Schwinger-Dyson equations.

While the transformation from a Euclidean covariant formalism
to a Lorentz covariant formalism normally requires an analytic
continuation, a fully relativistic form of quantum mechanics can be
formulated without explicit analytic continuation.  It requires that
the Euclidean analogs of the kernel of the inner product satisfies a
condition called reflection positivity
\cite{Osterwalder:1974tc} \cite{Osterwalder:1973dx}.  For irreducible
representations this condition can be satisfied for any spin.

The Euclidean representation of relativistic quantum mechanics has a
Hilbert space inner product that is defined by a kernel that is a
Euclidean covariant distribution left multiplied by a Euclidean time
reflection.  Both the initial and final states have to vanish for
negative Euclidean times.  The requirement that the resulting
quadratic form is non-negative is called reflection positivity.

In order to make contact with the Lorentz covariant representations
discussed above consider vectors represented by Euclidean covariant
spinor valued functions
$\langle \tau, \mathbf{x},\mu \vert \psi \rangle $ of four Euclidean
space-time variables with support for positive Euclidean time.  The
transformation properties of the spinor degrees of freedom will be
discussed in the next section.

In the Euclidean representation of relativistic quantum mechanics
of a particle of mass $m$ and spin $j$ the quantum mechanical inner
product is defined by
\[
\langle \phi \vert \psi \rangle :=   
{1 \over \pi}\int \sum
\langle \psi \vert -\tau_x, \mathbf{x},\mu \rangle  {e^{i p \cdot (x-y)} \over p^2 + m^2}
D^j_{\mu \nu}[p\cdot \sigma_e/m] \langle
\tau_y, \mathbf{y},\nu \vert \psi \rangle 
d^4 x d^4y d^4p =
\]
\beq
{1 \over \pi} \int \sum
\langle \phi \vert \tau_x, \mathbf{x},\mu \rangle {e^{-i p^0(\tau_x+\tau_y) + i \mathbf{p}
\cdot (\mathbf{x}-\mathbf{y})}
\over (p^0 - i \omega_m (\mathbf{p}))
(p^0 + i \omega_m (\mathbf{p}))}
D^j_{\mu \nu}[p\cdot \sigma_e/m]
\langle \tau_y, \mathbf{y}, \nu \vert \psi \rangle   
d^4 x d^4y d^4p 
\label{e.1}
\eeq
where $\omega_m (\mathbf{p}) = \sqrt{\mathbf{p}^2 + m^2}$ is the energy of
a particle of mass $m$ and momentum $\mathbf{p}$, and
all of the integration variables are Euclidean. 
The $-$ sign on
$\tau_x$ in the first term represents the Euclidean time
reflection discussed above.  In the second term the substitution
$\tau_x\to -\tau_x$ was made.  This, along with the Euclidean time
support condition of the wave functions, ensures that $\tau_x+\tau_y$
in the exponent of the second term is positive. 

To evaluate the $p^0$ integral, the $p^0$'s appearing in
$D^j_{\mu \nu} [p \cdot \sigma_e/m]$ can be replaced by
$-i {\partial \over \partial \tau_y}$ acting on the initial wave
function.  The $p^0$ integral can then be evaluated by the residue
theorem.  The $\tau_y$ derivatives can then moved back to
$D^j_{\mu \nu} [p \cdot \sigma_e/m]$ by a finite number of integrations by
parts,  since it is a polynomial in the
components of $p$.  This gives 
\beq
(\ref{e.1}) =\int
\langle \phi \vert \tau_x, \mathbf{x},\mu \rangle 
e^{- \omega_m (\mathbf{p})\tau_x+ i\mathbf{p} \cdot \mathbf{x}}
d^4x 
{ d\mathbf{p} \over \omega_m (\mathbf{p})}
D^j_{\mu \nu}[p_m \cdot \sigma/m]
e^{- \omega_m (\mathbf{p})\tau_y- i\mathbf{p} \cdot \mathbf{y}}
\langle \tau_y, \mathbf{y}, \nu \vert  \psi \rangle d^4y 
\label{e.2}
\eeq
where $p_m = (\omega_m (\mathbf{p}), \mathbf{p})$ and
\beq
D^j_{\mu \nu}[(-i \omega_m (\mathbf{p}),\mathbf{p}) \cdot \sigma_e/m] =
D^j_{\mu \nu}[p_m \cdot \sigma/m ].
\eeq
The resulting kernel 
\beq
{d\mathbf{p} \over \omega_m (\mathbf{p})}
D^j_{\mu \nu}[p_m \cdot \sigma/m]
\label{e.3}
\eeq
is {\it exactly} the Lorentz covariant measure appearing in (\ref{cov.14}) .
It follows that the Euclidean covariant distribution
\beq
{1 \over \pi} {D^j_{\mu \nu}[p\cdot \sigma_e/m] \over p^2 + m^2}   
\label{e.4}
\eeq
is reflection positive  because $D^s_{\mu \nu}[p \cdot \sigma_e/m]$
becomes a positive definite matrix after 
$p^0$ is set equal to $-i \omega_m (\mathbf{p})$.

The measure for the left handed (space reflected) representation is obtained by
replacing
\beq
\sigma_e \to \sigma_2 \sigma_e^t \sigma_2
\label{e.5}
\eeq
which changes the sign of the space components of $\sigma_{e\mu}$.
In this case 
\[
{1 \over \pi}\int 
\langle \phi \vert -\tau_x, \mathbf{x}\mu \rangle
{e^{i p \cdot (x-y)} \over p^2 + m^2}
D^j_{\mu \nu}[p\cdot \sigma_2 \sigma^t_e \sigma_2/m ]\langle \tau_y, \mathbf{y},
\nu \vert \psi \rangle 
d^4 x d^4y d^4p =
\]
\[
\int
\langle \phi \vert \tau_x, \mathbf{x},\mu \rangle 
e^{- \omega_m (\mathbf{p})\tau_x+ i\mathbf{p} \cdot \mathbf{x}}
d^4x 
{ d\mathbf{p} \over \omega_m (\mathbf{p})}
D^j_{\mu \nu}[p_m \cdot \sigma_2\sigma^*\sigma_2/m]
e^{- \omega_m (\mathbf{p})\tau_y- i\mathbf{p} \cdot \mathbf{y}}
\langle \tau_y, \mathbf{y} ,\nu \vert \psi \rangle  d^4y =
\]
\beq
\int
\langle \phi \vert \tau_x, \mathbf{x},\mu \rangle  
e^{- \omega_m (\mathbf{p})\tau_x+ i\mathbf{p} \cdot \mathbf{x}}
d^4x 
{ d\mathbf{p} \over \omega_m (\mathbf{p})}
D^j_{\mu \nu}[P p_m \cdot \sigma/m ]
e^{- \omega_m (\mathbf{p})\tau_y- i\mathbf{p} \cdot \mathbf{y}}
\langle \tau_y, \mathbf{y}, \nu \vert \psi \rangle  d^4y
\label{e.6}
\eeq
where
\beq
{ d\mathbf{p} \over \omega_m (\mathbf{p})}
D^j_{\mu \nu}[P p_m \cdot \sigma/m ]
\label{e.7}
\eeq
which is the Lorentz covariant kernel (\ref{cov.15}) for left handed 
spinors.  The positivity of the matrix $D^j_{\mu \nu}[P p_m \cdot \sigma/m ]$
implies that the Euclidean covariant distribution 
\beq
{1 \over \pi}{D^j_{\mu \nu}[p\cdot \sigma_2 \sigma^t_e \sigma_2/m ] \over p^2 + m^2}
=
{1 \over \pi} {D^j_{\mu \nu}[Pp\cdot \sigma_e /m ] \over p^2 + m^2}
\label{e.8}
\eeq
is also reflection positive.
By defining
\beq
\langle \mathbf{p},\nu \vert \chi \rangle:= \int e^{- \omega_m (\mathbf{p})\tau_y- i\mathbf{p} \cdot \mathbf{y}}
\langle \tau_y, \mathbf{y}, \nu \vert \psi \rangle  d^4y
\label{e.9}
\eeq
the norms can be expressed in the form
\beq
\langle \psi \vert \psi \rangle =
\int \langle \chi \vert \mathbf{p}, \mu \rangle 
{ d\mathbf{p} \over \omega_m (\mathbf{p})}
D^j_{\mu \nu}[p_m \cdot \sigma/m ]
\langle \mathbf{p},\nu \vert \chi \rangle
\label{e.10}
\eeq
and
\beq
\langle \psi \vert \psi \rangle =
\int \langle \chi \vert \mathbf{p}, \mu \rangle 
{ d\mathbf{p} \over \omega_m (\mathbf{p})}
D^j_{\mu \nu}[P p_m \cdot \sigma/m )]
\langle \mathbf{p}, \nu \vert \chi \rangle , 
\label{e.11}
\eeq
for the right and left handed representations respectively.  These
expressions have the same form as the Lorentz covariant (\ref{cov.14})
inner products with respect to the functions, $\langle \mathbf{p},
\nu \vert \chi \rangle$, up to a multiplicative constant.

As in the Lorentz covariant case, in the Euclidean case the
Euclidean covariant kernels are different for the right- and
left-handed representations:
\beq
{1 \over \pi}{1 \over p^2 + m^2 }
D^j_{\mu \nu}[p \cdot \sigma_e/m]
\label{e.12}
\eeq
\beq
{1 \over \pi} {1 \over p^2 + m^2 }
D^j_{\mu \nu}[p \cdot \sigma_2 \sigma_e^t \sigma_2/m].
\label{e.13}
\eeq

The $SU(2)\times SU(2)$ covariance property  
of the kernel (\ref{e.12}) is
\[
D^j [A] {D^j[p \cdot \sigma_e/m]\over p^2 + m^2 } D[B^t] =
{D^j[p \cdot A \sigma_e B^t/m] \over p^2 + m^2 } =
{D^j[p \cdot O^t(A,B) \sigma_e /m] \over p^2 + m^2 } =
\]
\beq
{D^j[O(A,B) p \cdot \sigma_e/m ]  \over p^2 + m^2 }
=
{D^j[O(A,B) p \cdot \sigma_e/m ] \over (O(A,B)p)^2 + m^2 } 
\label{e.14}
\eeq
Here $p^2 = (O(A,B) p)^2$ was used.  The corresponding covariance
property for the space reflected kernel,
(\ref{e.13}), can be obtained by taking the
transpose of (\ref{e.14}) and left and right multiplying by
$D^j_{\mu\nu} [\sigma_2] = (i)^{2\nu}\delta_{\mu -\nu}$ which gives
\[ 
D^j [\sigma_2 B \sigma_2]
{D^j [p \cdot \sigma_2 \sigma_e^t \sigma_2/m]
\over p^2 + m^2 }
D[\sigma_2 A^t \sigma_2] =
D^j [B^*]
{D^j [p \cdot \sigma_2 \sigma_e^t \sigma_2/m] \over p^2 + m^2 }
D^j[A^{\dagger}] =
\]
\beq
{D^j [O(A,B)p \cdot \sigma_2 \sigma_e^t \sigma_2 /m ] \over (O(A,B)p)^2 + m^2 }
\label{e.15}
\eeq

These results are abbreviated by
\beq
D[A] K_r(p) D[B^t] = K_r (O(A,B)p)
\label{e.16}
\eeq
\beq
D[B^*] K_l(p) D[A^{\dagger}] = K_l(O(A,B)p)
\label{e.17}
\eeq
where $K_r(p)$ and $K_l(p)$ are the right and left handed
reflection positive kernels (\ref{e.4}) and (\ref{e.8}).

While most treatments of Euclidean formulations of relativistic
quantum theories involve an analytic continuation in time, the
construction above shows how the right and left handed Lorentz
covariant irreducible representations (\ref{cov.12},\ref{cov.13}) are
recovered in the Euclidean formulation without any analytic
continuation.  This reason for this is that reflection positivity, the
spectral condition ($m>0$), and the assumption that the Euclidean
kernel is a tempered distribution ensures the existence of the
analytic continuation, however for the purpose of formulating
relativistic quantum mechanics the analytic continuation is not
needed.

\begin{center}
{\bf
Relativistic invariance in the Euclidean case
}  
\end{center}   

Relativistic invariance in the Euclidean case is a consequence of the
identities relating the Euclidean covariant inner product to the
Lorentz covariant inner product and the Poincar\'e covariant inner
product.


The relativistic transformation properties in the Euclidean
representation can be understood from the observation that the complex
orthogonal and complex Lorentz transformations have the same covering
group, $ SL(2,\mathbb{C})\times SL(2,\mathbb{C})$. This means that
the group of real Euclidean transformations can be identified with a
subgroup of the complex Lorentz group.  The real Euclidean group is a
10 parameter group.  Each generator can be thought of generating a
one-parameter subgroup of complex Poincar\'e transformations.  This
leads to a relation between the generators of real Euclidean
transformations and real Poincar\'e transformations that can be
realized in the Euclidean framework.

This relationship implies that the Poincar\'e Lie algebra is related to
the Euclidean Lie algebra by multiplying the generators of real Euclidean
transformations involving the Euclidean time by factors of $i$.  The
Euclidean generators involving the Euclidean time are the generator of
Euclidean time translations and the generators of rotations in
space-Euclidean time planes.  The resulting Poincar\'e generators for
time translation and canonical boosts are related to
the generators of Euclidean time translation and rotations in Euclidean
space-time planes by
\beq
H_m = i H_e  \qquad \mbox{and} \qquad \mathbf{K}\cdot \hat{\mathbf{n}}
= - i \mathbf{J}_{\hat{\mathbf{n}},\tau}.
\label{er.1}
\eeq
Both $H_m$ and $\mathbf{K}$ become Hermitian operators with respect to
the physical Hilbert space inner product (\ref{e.1}) that includes the
Euclidean time reflection.  On the physical Hilbert space real
Euclidean-time translations are represented by a contractive Hermitian
semi-group \cite{glimm} and the real rotations in space-Euclidean time planes are
represented by local symmetric semi-groups \cite{Klein:1981}\cite{Klein:1983}\cite{Frohlich:1983kp}.
The generators of these transformations are self-adjoint and are
exactly the Poincar\'e generators discussed above.

The $2 \times 2$ matrix representation of ordinary rotations in both
the Euclidean and Lorentz case can be represented by
\beq
X \to X' = A X B^{t}  \qquad X \to X_e' = A X_e B^{t}
\label{er.14}
\eeq
where $(A,B) = (A,A^*)$ for $A\in SU(2)$.

Euclidean rotations in space-Euclidean-time planes can be represented by
\beq
X_e \to X_e' = A X_e B^{t}
\label{er.15}
\eeq
where $(A,B) = (A,A^t)$ for $A\in SU(2)$, while rotationless Lorentz boosts
can be represented by a transformation of the same form,
\beq
X \to X' = A X_e B^{t}
\label{er.16}
\eeq
where $(A,B) = (A,A^t)$ and $A=A^{\dagger}$.

For a given $SU(2) \times SU(2)$ transformations $(A,B)$ there are
four types of Euclidean spinor wave functions that are identified by
their spinor transformation properties 
\beq
\psi^\mu (j,p) 
\to
\psi^{\mu\prime} (j,p)=
\psi^{\nu} (j,O(A,B) p) D^j_{\nu \mu} (A)
\label{er.19}
\eeq
\beq
\psi_\mu (j,p)
\to
\psi'_\mu (j,p)
=
\psi_\nu(j, O(A,B) p)D^j_{\nu \mu} (A^*)
\label{er.20}
\eeq
\beq
\psi^{\dot{\mu}} (j,p)
\to
\psi^{\dot{\mu}\prime} (j,p) =
\psi^{\dot{\nu}\prime} (j,O(A,B) p)
D^j_{\dot{\nu} \dot{\mu}} (B)
\label{er.21}
\eeq
\beq
\psi_{\dot{\mu}}(j,p)
\to
\psi_{\dot{\mu}}'(j,p)
=
\psi_{\dot{\nu}}(j,O(A,B) p)
D^j_{\dot{\nu} \dot{\mu}} (B^*)
\label{er.22}
\eeq
In equations (\ref{er.19}-\ref{er.22}) the bra-ket notation is not used
in order to differentiate the different types of spinor wave
functions.  The first two are right-handed wave functions; the last
two are left handed.

Representations of the Lorentz generators on each of these spinor
wave functions are obtained by first constructing the finite transformations
in 
(\ref{er.19}-\ref{er.22}) using
\beq
(A(\lambda),B(\lambda))_r=
(e^{i {\lambda \over 2} \hat{\mathbf{n}}\cdot \pmb{\sigma} },
(e^{i {\lambda \over 2} \hat{\mathbf{n}}\cdot \pmb{\sigma} })^*)
\eeq
for rotations about the $\hat{\mathbf{n}}$ axis and
\beq
(A(\lambda),B(\lambda)_b=
(e^{i {\lambda \over 2} \hat{\mathbf{n}}\cdot \pmb{\sigma} },
(e^{i {\lambda \over 2} \hat{\mathbf{n}}\cdot \pmb{\sigma} })^t)
\eeq
for rotations in the $\hat{\mathbf{n}}-\tau$ plane.

Representations for the generator of ordinary rotations about the
$\hat{\mathbf{n}}$ axis are obtained by using
$(A,B)=(A(\lambda),B(\lambda))_r$ in each of (\ref{er.19}-\ref{er.22}),
differentiating with respect to $\lambda$, setting $\lambda$ to 0 and
multiplying the result by $-i$.

Representations for the generator of rotationless boosts in the
$\hat{\mathbf{n}}$ direction are obtained by using
$(A,B)=(A(\lambda),B(\lambda))_b$ in each of (\ref{er.19}-\ref{er.22}),
differentiating each of (\ref{er.19}-\ref{er.22}) with respect to
$\lambda$, setting $\lambda$ to 0 and multiplying the result by $-1$.

The Hamiltonian and linear momentum operators in the Euclidean
representation are obtained by Fourier transforming
each of (\ref{er.19}-\ref{er.22}) followed by 
\beq
\mathbf{P} = -i \pmb{\nabla} \qquad H = {\partial \over \partial \tau}.
\eeq

The resulting operators satisfy the Poincar\'e  commutation relations
and are Hermitian when they are used in the inner product (\ref{e.1}).
As in the Lorentz covariant case, the dynamics enters through the
Euclidean kernel, which has all of the dynamics (mass dependence).

\section{Lorentz covariant fields}

Covariant fields are useful for treating systems of many identical
particles.  In many-body quantum mechanics fields are associated with
the occupation number representation.  They are constructed from a
single-particle basis $\{ \vert n\rangle \}$, and operators,
$a_n^{\dagger}$, that add and, $a_n$, that remove a particle in the
$n$-th single-particle state.  In this section the same methods are
used to develop Lorentz covariant fields for systems of
non-interacting particles of any spin.  Locality of the
fields is not assumed.  Local fields will be discussed in the next
section.

Field operators are defined in terms of a single particle basis by
\beq \Psi({x}) :=
\sum_n \langle \mathbf{x} \vert n \rangle a_n
\qquad
\Psi^{\dagger} (x) := \sum_n a_n^{\dagger}
\langle n \vert \mathbf{x} \rangle . 
\label{f.1}
\eeq
The field is independent of the choice of single-particle basis. 
In a plane-wave basis equations (\ref{f.1}) become
\beq
\Psi({x}) :=
\int d \mathbf{p} \langle \mathbf{x} \vert
\mathbf{p} \rangle 
a (\mathbf{p})
\qquad
\Psi^{\dagger} ({x}) := \int d\mathbf{p}
a^{\dagger}(\mathbf{p})  \langle \mathbf{p} \vert \mathbf{x} \rangle .
\label{f.3}
\eeq
The time dependence is determined by solving the Heisenberg equations
of motion
\beq
{d \Psi (\mathbf{x},t) \over dt} = i [H,\Psi (\mathbf{x},t)] .
\eeq
If $H$ is the free Hamiltonian the solution of the Heisenberg equations is
\beq
\Psi({x},t) :=
\int d \mathbf{p} \langle \mathbf{x} \vert
\mathbf{p} \rangle e^{-iE(\mathbf{p}) t}
a (\mathbf{p})
\qquad
\Psi^{\dagger} ({x},t) := \int d\mathbf{p}
a^{\dagger}(\mathbf{p}) e^{iE(\mathbf{p}) t}  \langle \mathbf{p} \vert \mathbf{x} \rangle 
\label{f.2}
\eeq
where $E(\mathbf{p})$ is the energy of a particle with momentum $\mathbf{p}$.

The vector $\vert 0 \rangle$, represents the no particle state.  It is
defined by the conditions
\beq
a_n \vert 0 \rangle = 0 \qquad \forall n \qquad \langle 0 \vert 0
\rangle =1 .
\label{f.3}
\eeq
The creation and annihilation operators satisfy the
commutation (anti-commutation) relations
\beq
[a_n,a_m^{\dagger}]_{\pm}
= \delta_{mn}
\qquad
\mbox{or}
\qquad
[a (\mathbf{p}),a_m^{\dagger}(\mathbf{p}') ]_{\pm}
= \delta (\mathbf{p}- \mathbf{p}') 
\label{f.4}
\eeq
depending on whether the particles are Bosons or Fermions.  

Free Lorentz covariant fields that transform under a
finite-dimensional representation of $SL(2,\mathbb{C})$ can be
constructed using the same method.  In this case the plane wave states
$\langle \mathbf{x} \vert \mathbf{p} \rangle $ are replaced by Lorentz
covariant plane wave states, and measure is replaced by the Lorentz
invariant measure.

Because the Lorentz covariant states can transform under
right or left-handed representations of $SL(2,\mathbb{C})$, the
corresponding covariant fields will also have a handedness.

In this section right and left handed spin-$j$ fields are constructed
with the following Poincar\'e covariance properties:
\beq
U(\Lambda ,a) \Psi_{r\mu} (x) U^{\dagger}(\Lambda, a) =
D^j_{\mu\nu}[A^{-1} ] \Psi_{r\nu} (\Lambda x+a) 
\label{f.5}
\eeq
\beq
U(\Lambda ,a) \Psi_{l\mu} (x) U^{\dagger}(\Lambda, a) =
D^j_{\mu\nu}[\tilde{A}^{-1}]
\Psi_{l\nu} (\Lambda x+a) 
\label{f.6}
\eeq
where $A$ and $\Lambda$ are related by (\ref{s.16a}).

The starting point is to define creation and annihilation operators
that transform like single-particle irreducible states.  These
create or destroy particles with a momentum $\mathbf{p}$ and
a magnetic quantum number associated with the $x$-type of spin,
as discussed in section 3.

The creation operators are assumed to have the following
transformation properties
\[
U(\Lambda ,a) a^{\dagger}_x(\mathbf{p},\mu)
U^{\dagger}(\Lambda, a) = e^{-i \Lambda p \cdot a} a^{\dagger}_x
(\pmb{\Lambda}p,\nu) D^j_{\nu\mu}[B_x^{-1}(\Lambda p/m)A
B_x( p/m)] \sqrt{{\omega_m (\pmb{\Lambda}p) \over \omega_m
(\mathbf{p})}} =
\]
\beq
e^{-i \Lambda p \cdot a} a^{\dagger}_x
(\pmb{\Lambda}p,\nu) D^j_{\nu\mu}[\tilde{B}_x^{-1}(\Lambda p/m)\tilde{A}
\tilde{B}_x(p/m)] \sqrt{{\omega_m (\pmb{\Lambda}p) \over \omega_m
(\mathbf{p})}} .
\label{f.7}
\eeq
The two expressions above are identical because
\beq
B_x^{-1}(\Lambda p/m)A
B_x( p/m) = R_{wx}(\Lambda ,p/m)  = (R_{wx}^{\dagger})^{-1}(\Lambda ,p/m)  =
\tilde{B}_x^{-1}(\Lambda p/m)\tilde{A}
\tilde{B}_x(p/m) . 
\label{f.8}
\eeq
The transformation properties of the creation operator (\ref{f.7})
is the same as the transformation properties a particle (\ref{p.41}),
except the sign of the phase is reversed because the time
dependence of the operator is given by the Heisenberg equations of motion.

The corresponding transformation properties for the annihilation operators
can be obtained by taking the adjoint of (\ref{f.7}):
\[
U(\Lambda ,a) a_x(\mathbf{p},\mu)
U^{\dagger}(\Lambda, a) = e^{i \Lambda p \cdot a}
D^j_{\mu\nu}[
B_x^{\dagger}(p/m))A^{\dagger} 
\tilde{B}_x(\Lambda p/m)]
a_x (\pmb{\Lambda}p,\nu)  \sqrt{{\omega_m (\pmb{\Lambda}p) \over \omega_m
    (\mathbf{p})}} =
\]
\beq
e^{i \Lambda p \cdot a}
D^j_{\mu\nu}[
B_x(p/m))^{-1}A^{-1} 
B_x(\Lambda p/m)]
a_x (\pmb{\Lambda}p,\nu)  \sqrt{{\omega_m (\pmb{\Lambda}p) \over \omega_m
(\mathbf{p})}}.
\label{f.9}
\eeq
 
Local fields are linear combinations of fields with creation and
annihilation operators that have the same covariance properties.  The
normal convention is to have the $SL(2,\mathbb{C})$ representation
matrices to the left of the creation and annihilation operators as in
(\ref{f.5}) and (\ref{f.6}).

This can be realized in (\ref{f.7}) by using the
$SU(2)$ identity (\ref{s.45}) 
\beq
R  = \sigma_2 (R^{t})^{-1} \sigma_2
\label{f.10}
\eeq
in the Wigner rotation
\beq
R:= B_x^{-1}(\Lambda p/m )A B_x(p/m). 
\label{f.11}
\eeq
which gives
\beq
D^j_{\mu \nu} (R) = 
D^j_{\mu \nu} (\sigma_2 (R^{t})^{-1} \sigma_2)=
D^j_{\nu\mu} ((-\sigma_2) R^{-1} (-\sigma_2))=
D^j_{\nu\mu} (\sigma_2 R^{-1}\sigma_2)
\label{f.12}
\eeq
the corresponding property of the Wigner functions
(note the reversal $\mu \leftrightarrow \nu$ of the spin indices).
Using this identity in (\ref{f.7}) gives
\[
U(\Lambda ,a) D^j_{\mu \nu} [\sigma_2] a^{\dagger}_x(\mathbf{p},\nu) 
U^{\dagger}(\Lambda, a) = e^{-i \Lambda p \cdot a}
D^j_{\mu\nu}[B_x^{-1}( p/m) A^{-1}
B_x(\Lambda p/m) \sigma_2 ]
a^{\dagger}_x
(\pmb{\Lambda}p,\nu)
\sqrt{{\omega_m (\pmb{\Lambda}p) \over \omega_m
(\mathbf{p})}} =
\]
\beq
U(\Lambda ,a) D^j_{\mu \nu} [\sigma_2] a^{\dagger}_x(\mathbf{p},\nu) 
U^{\dagger}(\Lambda, a) = e^{-i \Lambda p \cdot a}
D^j_{\mu\nu}[{B}^{\dagger}_x(p/m)
{A}^{\dagger} 
\tilde{B}_x(\Lambda p/m)\sigma_2]
a^{\dagger}_x
(\pmb{\Lambda}p,\nu)  \sqrt{{\omega_m (\pmb{\Lambda}p) \over \omega_m
(\mathbf{p})}} .
\label{f.13}
\eeq
Introducing the $\sigma_2$ factor gives the creation fields
the same covariance properties as the annihilation fields.

These operators determine the Poincar\'e transformation properties of 
the covariant spinor fields.
Different types of spinor fields 
are distinguished by their covariance properties.
General covariant fields of a given spin are built up out of four
types of elementary covariant fields, that are classified as right (r) or
left (l) handed and creation (c) or annihilation (a) fields.  The subscripts
$rc,lc,ra,la$ are used to distinguish the four different types of fields:
\[
\Psi^{\dagger}_{rc\mu}  (x) := \int {e^{-ip \cdot x} \over (2 \pi)^{3/2}}
{d \mathbf{p} \over \omega_m (\mathbf{p})}
D^j_{\mu \nu}
[{B}_x(p/m )\sigma_2 ]
a^{\dagger}_x(\mathbf{p},\nu) 
\sqrt{\omega_m (\mathbf{p})}  =
\]
\beq
\int {e^{-ip \cdot x} \over (2 \pi)^{3/2}}
{d \mathbf{p} \over \sqrt{\omega_m (\mathbf{p})}}
D^j_{\mu \nu}
[{B}_x(p/m )\sigma_2 ]
a^{\dagger}_x(\mathbf{p},\nu) 
\label{f.14}
\eeq

\[
\Psi_{ra\mu} (x) := \int {e^{ip \cdot x} \over (2 \pi)^{3/2}}
{d \mathbf{p} \over \omega_m (\mathbf{p})}
D^j_{\mu \nu}
[B_x(p/m )]
a^{}_x(\mathbf{p},\nu)
\sqrt{\omega_m (\mathbf{p})}  =
\]
\beq
\int {e^{ip \cdot x} \over (2 \pi)^{3/2}}
{d \mathbf{p} \over \sqrt{\omega_m (\mathbf{p})}}
D^j_{\mu \nu}
[B_x(p/m )]
a^{}_x(\mathbf{p},\nu)
\label{f.15}
\eeq

\[
{\Psi}^{\dagger}_{lc\mu}  (x) = \int e^{-ip \cdot x}
{d \mathbf{p} \over \omega_m (\mathbf{p})}
D^j_{\mu \nu}
[\tilde{B}_x(p/m )\sigma_2 ]
a^{\dagger}_x(\mathbf{p},\nu) \vert 0 \rangle
\sqrt{\omega_m (\mathbf{p})}=
\]
\beq
\int e^{-ip \cdot x}
{d \mathbf{p} \over \sqrt{\omega_m (\mathbf{p})}}
D^j_{\mu \nu}
[\tilde{B}_x(p/m )\sigma_2 ]
a^{\dagger}_x(\mathbf{p},\nu) \vert 0 \rangle
\label{f.16}
\eeq

\[
{\Psi}_{la\mu} (x) = \int e^{ip \cdot x}
{d \mathbf{p} \over \omega_m (\mathbf{p})}
D^j_{\mu \nu}
[\tilde{B}_x(p/m )]
a^{}_x(\mathbf{p},\nu) \vert 0 \rangle
\sqrt{\omega_m (\mathbf{p})} =
\]
\beq
\int e^{ip \cdot x}
{d \mathbf{p} \over \sqrt{\omega_m (\mathbf{p})}}
D^j_{\mu \nu}
[\tilde{B}_x(p/m )]
a^{}_x(\mathbf{p},\nu) \vert 0 \rangle
\label{f.17}
\eeq
where in all cases the 4-momenta are on shell:
\beq
p \cdot x = - \omega_m(\mathbf{p}^2) x^0 + \mathbf{p}\cdot \mathbf{x}.
\label{f.18}
\eeq

Note that $\Psi^{\dagger}_{xc\mu}(x)$ is not the
adjoint of $\Psi_{xa\mu}(x)$.  This is because of the factor
$\sigma_2$ that was introduced to  make both fields have the
same Lorentz covariance property.

The transformation properties of (\ref{f.14}-\ref{f.17})
follow directly from
the transformation properties of the creation and annihilation
operators (\ref{f.7}) and (\ref{f.9}) 
\beq
U(\Lambda ,b) {\Psi}^{\dagger}_{rc\mu} (x) U^{\dagger} (\Lambda ,b) =
D^j_{\mu \nu} [(A )^{-1}]
{\Psi}^{\dagger}_{rc\nu} (\Lambda x+b) 
\label{f.19}
\eeq
\beq
U(\Lambda ,b) {\Psi}_{ra\mu} (x) U^{\dagger} (\Lambda ,b) =
D^j_{\mu \nu} [(A )^{-1}]
{\Psi}_{ra\nu} (\Lambda x+b) 
\label{f.20}
\eeq
\beq
U(\Lambda ,b) {\Psi}^{\dagger}_{lc\mu} (x) U^{\dagger} (\Lambda ,b) =
D^j_{\mu \nu} [A^{\dagger}]
{\Psi}^{\dagger}_{lc\nu} (\Lambda x+b) 
\label{f.21}
\eeq
\beq
U(\Lambda ,b) {\Psi}_{la\mu} (x) U^{\dagger} (\Lambda ,b) =
D^j_{\mu \nu} [A^{\dagger}]
{\Psi}_{la\nu} (\Lambda x+b) .
\label{f.22}
\eeq
These fields can be
multiplied by any normalization constants.

These transformation properties can be used to construct invariant
operator densities.  Invariant products are constructed by taking the
product of a field of one handedness with the adjoint of a field of
the opposite handedness and summing over the spins.
Lorentz invariant Hermitian operators are
obtained by adding the Hermitian conjugate to each of the invariant pairs.
The following sums of products of left and right-handed
fields are Hermitian and transform like Lorentz scalars:
\beq
\sum_{\mu}\left (
\Psi_{lc\mu}(x) \Psi^{\dagger}_{rc\mu}(x) + \Psi_{rc\mu}(x) \Psi^{\dagger}_{lc\mu}(x) \right)
\label{f.23}
\eeq
\beq
\sum_{\mu}\left (
\Psi_{lc\mu}(x) \Psi_{ra\mu}(x) + \Psi^{\dagger}_{ra\mu}(x) \Psi^{\dagger}_{lc\mu}(x) \right)
\label{f.24}
\eeq
\beq
\sum_{\mu}\left (
\Psi_{rc\mu}(x) \Psi_{la\mu}(x)+\Psi^{\dagger}_{la\mu}(x) \Psi^{\dagger}_{rc\mu}(x) \right)
\label{f.25}
\eeq
\beq
\sum_{\mu}\left (
\Psi^{\dagger}_{la\mu}(x) \Psi_{ra\mu}(x)+ \Psi^{\dagger}_{ra\mu}(x) \Psi_{la\mu}(x) \right).
\label{f.26}
\eeq
Note that for free fields these expressions are normal ordered.
It is possible to make more complicated Lorentz invariant products of
field operators using $SU(2)$ Clebsch-Gordan coefficients and the
group representation properties of the Wigner functions.

The commutators or anti-commutators of the elementary fields with their true
adjoints are
\beq
[{\Psi}_{rc\mu} (x),{\Psi}^{\dagger}_{rc\nu} (y)]_\pm =
\int {d\mathbf{p} \over (2 \pi)^3 \omega_m(\mathbf{p})} e^{i p \cdot (x-y)} 
D^j_{\mu \nu} [\sigma^t \cdot p] = 2 \int {d^4 p\over (2 \pi)^3}
\delta (p^2+m^2) \theta(p^0) 
e^{i p \cdot (x-y)} 
D^j_{\mu \nu} [\sigma^t \cdot p]
\label{f.27}
\eeq
\beq
[{\Psi}_{ra\mu} (x),{\Psi}^{\dagger}_{ra\nu} (y)]_\pm =
2 \int {d^4 p\over (2 \pi)^3}
\delta (p^2+m^2) \theta(p^0) 
e^{i p \cdot (x-y)} 
D^j_{\mu \nu} [\sigma\cdot p]
\label{f.28}
\eeq
\beq
[{\Psi}_{lc\mu} (x),{\Psi}^{\dagger}_{lc\nu} (y)]_\pm =
2 \int {d^4 p\over (2 \pi)^3}
\delta (p^2+m^2) \theta(p^0) 
e^{i p \cdot (x-y)} 
D^j_{\mu \nu} [\sigma\cdot Pp]
\label{f.29}
\eeq
\beq
[{\Psi}_{la\mu} (x),{\Psi}^{\dagger}_{la\nu} (y)]_\pm =
2 \int {d^4 p\over (2 \pi)^3}
\delta (p^2+m^2) \theta(p^0) 
e^{i p \cdot (x-y)} 
D^j_{\mu \nu} [\sigma^t\cdot Pp]
\label{f.30}
\eeq
which are the spin-$j$ Wightman functions that define the kernels of
the Lorentz covariant inner products.  Note that 
$\sigma\cdot p$,  $\sigma^t\cdot p$, $\sigma\cdot Pp$ and
$\sigma^t\cdot Pp$ are all positive Hermitian matrices for
time-like $p$, so these kernels are all positive distributions.

These fields are analogous to the non-relativistic fields; they add 
or remove particles in the occupation number representation.
While they are not local, they are the basic building blocks of 
local free fields.   

The creation and annihilation operators can be extracted from the right or
left-handed fields using plane wave solutions of the Klein-Gordan equation
and spinor matrices
\beq
f_m (p,x) :=
\int {1 \over \sqrt{\omega_m (\mathbf{p})} (2 \pi)^{3/2}} e^{i p \cdot x}
\label{f.31}
\eeq
\[
a_x^{\dagger} (\mathbf{p},\mu) =
\]
\beq
{i \over 2}
D^j_{\mu \nu} [\sigma_2 B_x^{-1}(p/m)]
\int ({\partial f_m (p,x)\over \partial t} \Psi^{\dagger}_{rc\nu}(x) -
{\partial \Psi^{\dagger}_{rc\nu}(x) \over \partial t} f_m (x)))d\mathbf{x}
\label{f.32}
\eeq
\[
a_x (\mathbf{p},\mu) =
\]
\beq
-{i \over 2}
D^j_{\mu \nu} [B_x^{-1}(p/m)]
\int ( {\partial f_m^* (p,x)\over \partial t} \Psi_{ra\nu}(x) -
{\partial \Psi^{\dagger}_{ra\nu}(x) \over \partial t} f^*_m (x))d\mathbf{x}
\label{f.33}
\eeq
\[
a_x^{\dagger} (\mathbf{p},\mu) =
\]
\beq
{i \over 2} D^j_{\mu \nu} [\sigma_2 B_x^{\dagger}(p/m)]
\int ({\partial f_m (p,x)\over \partial t} \Psi^{\dagger}_{lc\nu}(x) -
{\partial \Psi^{\dagger}_{lc\nu}(x) \over \partial t} f_m (x))d\mathbf{x}
\label{f.34}
\eeq
\[
a_x (\mathbf{p},\mu) =
\]
\beq
-{i \over 2}D^j_{\mu \nu} [B_x^{\dagger}(p/m)]
\int ( {\partial f_m^* (p,x)\over \partial t} \Psi_{la\nu}(x) -
{\partial \Psi^{\dagger}_{la\nu}(x) \over \partial t} f^*_m (x))d\mathbf{x}
\label{f.35}
\eeq
where the integrals are evaluated at a common time.  

Fields that transform linearly under space reflection can be constructed
by taking a direct sum of a right and left handed field
\beq
\Psi_{c\mu}^{\dagger}(x) \to
\int e^{-i p\cdot x} {d \mathbf{p} \over \sqrt{\omega_m(\mathbf{p})}}
\left( 
\begin{array}{c}
D^j_{\mu_r \nu}[B_x(p/m)\sigma_2] \\
D^j_{\mu_l \nu}[\tilde{B}_x(p/m)\sigma_2]  
\end{array}
\right )
a_x^{\dagger}(\mathbf{p},\nu)
\label{f.36}
\eeq
where the matrix is a $2(2j+1)\times (2j+1)$ matrix.

The Poincar\'e transformation properties of these fields are
\beq
U(\Lambda, b) \left( 
\begin{array}{c}
\Psi_{rc\mu} (x)\\
{\Psi}_{lc\mu} (x) 
\end{array}
\right )    U^{\dagger}(\Lambda, b) =
\left( 
\begin{array}{cc}
D^j_{\mu_r\nu} [A^{-1} ]& 0 \\
0 & D^j_{\mu_l\nu} [A^{\dagger} ]
\end{array}
\right )
\left( 
\begin{array}{c}
\Psi_{rc\nu} (\Lambda x+b)\\
{\Psi}_{lc\nu} (\Lambda x+ b) 
\end{array}
\right ) 
\label{f.37}
\eeq
Space reflection changes the sign of the space component of 
$x$ and interchanges the right- and left-handed components
\beq
P \Psi^{\dagger}_{c\mu}(x)P^{-1}  = \Psi^{\dagger \prime}_{c\mu} (x) =  
P \left( 
\begin{array}{c}
\Psi^{\dagger}_{rc\mu} (x)\\
\Psi^{\dagger}_{lc\mu} (x) 
\end{array}
\right ) P =
\left( 
\begin{array}{c}
{\Psi}^{\dagger}_{lc\mu} (Px)\\
\Psi^{\dagger}_{rc\mu} (Px)\\
\end{array}
\right ) 
\label{f.38}
\eeq
The annihilation fields have the same structure
\beq
\Psi_{a\mu}(x) \to 
\left( 
\begin{array}{c}
\Psi_{ra\mu} (x)\\
{\Psi}_{la\mu} (x) 
\end{array}
\right ). 
\label{f.39}
\eeq
The Poincar\'e transformation properties of the annihilation fields are
\beq
U(\Lambda, b) \left( 
\begin{array}{c}
\Psi^{\dagger}_{ra\mu} (x)\\
{\Psi}_{la\mu} (x) 
\end{array}
\right )    U^{\dagger}(\Lambda, b) =
\left( 
\begin{array}{cc}
D^j_{\mu\nu} [A^{-1} ]& 0 \\
0 & D^j_{\mu\nu} [A^{\dagger} ]
\end{array}
\right )
\left( 
\begin{array}{c}
\Psi_{ra\nu} (\Lambda x+b)\\
{\Psi}_{la\nu} (\Lambda x+ b) 
\end{array}
\right ) .
\label{f.40}
\eeq
Space reflection changes with sign of the space component of 
$x$ and interchanges the right and left handed components
\beq
P \Psi_{a\mu}(x)P^{-1}  = \Psi'_{a\mu} (x) =  
P \left( 
\begin{array}{c}
\Psi_{ra\mu} (x)\\
\Psi_{la\mu} (x) 
\end{array}
\right ) P^{-1} =
\left( 
\begin{array}{c}
{\Psi}_{la\mu} (Px)\\
\Psi_{ra\mu} (Px)\\
\end{array}
\right ). 
\label{f.41}
\eeq
By analogy with the Dirac equation it is to useful to define
\beq
\Gamma^0 :=
\left( 
\begin{array}{cc}
0 & I\\
I & 0\\
\end{array}
\right )
\qquad
\Gamma^5 :=
\left( 
\begin{array}{cc}
I & 0 \\
0 & -I \\
\end{array}
\right )
\label{f.47}
\eeq
In this notation equations (\ref{f.38}) and (\ref{f.41}) can be 
written as
\beq
P \Psi^{\dagger}_{c\mu} (x)P^{-1}  = \Gamma^0 \Psi^{\dagger}_{c\mu} (Px)   
\label{f.48}
\eeq
and
\beq
P \Psi_{a\mu} (x)P^{-1}  = \Gamma^0 \Psi_{a\mu} (Px) .  
\label{f.49}
\eeq
In addition
\beq
\Pi_{l/r}= {I \pm \Gamma_5 \over 2}  
\label{f.50}
\eeq
projects on the right or left handed component of the field.

The matrices
\beq
D^j_{\mu \nu} [{B}_x(p/m )\sigma_2 ]
\qquad \mbox{and} \qquad
D^j_{\mu \nu} [{B}_x(p/m ) ]
\label{f.51}
\eeq
transform Wigner rotations 
(finite dimensional representations of the 
little group of positive-mass positive-energy representations of the 
Poincar\'e group) into finite dimensional representations of the 
Lorentz group
\beq
D^j[\Lambda  ]D[B_x(p/m)] 
=
D^j[B_x (\Lambda p/m)]
D^j[R_{wx}(\Lambda p/m)]
\label{f.52}
\eeq
and
\[
D^j[\Lambda]D^j[{B}_x(p/m )\sigma_2 ] =
D^j[{B}_x(\Lambda p/m) ]
D^j[{B}^{-1}_x(\Lambda p/m ] D^j[\Lambda]D^j[{B}_x(p/m )\sigma_2 ]=
\]
\beq
D^j[{B}_x(\Lambda p/m )\sigma_2 ]
D^j[R^*_{wx}(\Lambda p/m)]
%
\label{f.53}
\eeq
because they transform finite dimensional representations of the
Lorentz group into representations of a little group of the Poincar\'e
group.

\section{Local Lorentz (free) covariant fields}

The fields constructed in the previous section transform
covariantly, but they are not local.  While they are 
sufficient for use in many-body relativistic quantum mechanics,
they are not suitable for use in local relativistic quantum field
theory.  Local free fields are constructed from linear combinations of 
creation and annihilation fields:
\beq
\Psi_{loc \mu} (x) =   
\alpha\Psi_{ra\mu}(x) +\beta\Psi^{\dagger}_{rc\mu}(x)
\label{lft.1}
\eeq
\beq
\Psi_{loc \mu}^{\dagger}  (x) =  
\alpha^*\Psi^{\dagger}_{ra\mu}(y) +\beta^*\Psi_{rc\mu}(y)  
\label{lft.2}
\eeq
Normally the linear combinations involve  a particle creation operator with an
antiparticle annihilation operator.  

The coefficients of the linear combinations (\ref{lft.1}) 
are constrained by locality, but these
linear combinations transform covariantly for any constants $\alpha$ and $\beta$.

The commutator or anti-commutator of the linear combinations
(\ref{lft.1}-\ref{lft.2}) 
determines the constraints on the constants $\alpha$ and $\beta$
/imposed by locality
\[
[\alpha\Psi_{ra\mu}(x) +\beta\Psi^{\dagger}_{rc\mu}(x),
\alpha^*\Psi^{\dagger}_{ra\nu}(y) +\beta^*\Psi_{rc\nu}(y)]_{\pm} = 
\]
\[
\vert \alpha \vert^2 
[\Psi_{rc\mu}(x),
a^*\Psi^{\dagger}_{rc\nu}(y)]_{\pm} 
+\vert \beta \vert^2 
[\Psi^{\dagger}_{ra\mu}(x), 
\Psi_{ra\nu}(y)]_{\pm} = 
\]

\[
\int {d \mathbf{p} \over {\omega_m(\mathbf{p})}} 
(\vert \alpha \vert^2 e^{i p \cdot (x-y)}
D^j_{\mu \alpha}[B_x(p/m)] D^{j*}_{\nu \alpha}[B_x(p/m)]
\pm
\vert \beta \vert^2 e^{-i p \cdot (x-y)}
D^j_{\mu \alpha}[B_x(p/m)\sigma_2] D^{j*}_{\nu \alpha}[B_x(p/m)\sigma_2])=
\]

\beq
\int {d \mathbf{p} \over {\omega_m(\mathbf{p})}} 
(\vert \alpha \vert^2 e^{i p \cdot (x-y)} D^j_{\mu ,\nu}[\sigma\cdot p]
\pm
\vert \beta \vert^2 e^{-i p \cdot (x-y)}  D^j_{\mu ,\nu}[\sigma\cdot p]).
\label{lft.3}
\eeq
For $(x-y)^2>0$ the integral \cite{bogoliubov}
\beq
\int {d \mathbf{p} \over \omega_m(\mathbf{p})} 
e^{-i p \cdot (x-y)} =
- {4 \pi m \over \sqrt{(x-y)^2}}
K_1(m \sqrt{(x-y)^2})
\label{lft.4} 
\eeq
is an even function of $x-y$.  It follows that for $(x-y)^2>0$ this becomes 
\[
\left ( \vert \alpha \vert^2 D^j_{\mu ,\nu}(-\sigma\cdot i \partial_x)
\mp \vert \beta \vert^2 D^j_{\mu ,\nu}(\sigma\cdot i \partial_x)
\right )
\int {d \mathbf{p} \over \omega_m(\mathbf{p})} 
e^{i p \cdot (x-y)} =  
\]
\beq
\left (\vert \alpha \vert^2 (-)^{2j}
\mp \vert \beta \vert^2 \right )
D^j_{\mu ,\nu}(\sigma\cdot i \partial_x)
\int {d \mathbf{p} \over \omega_m (\mathbf{p})} 
e^{i p \cdot (x-y)}  
\label{lft.5}
\eeq
For this to vanish $\vert \alpha \vert^2$ = $\vert \beta\vert^2$ and
$(-)^{2j}= \pm 1$ this means that anti-commutation relations are
required for $j$ half integral, commutation relations for 
$j$ integer. 

Similar results are obtained for left handed spinors.
The only difference is that
$D^j(\sigma \cdot p)$ is replaced by 
$D^j(\sigma \cdot Pp)$.

Thus right and left handed spin $j$ free local fields have the form
\beq
\Psi_{r loc \mu} (x) = Z(\Psi_{ra\mu}(x)\pm \Psi^{\dagger}_{rc\mu}(x))
\label{lft.6}
\eeq
\beq
\Psi_{l loc \mu} (x) = Z(\Psi_{la\mu}(x)\pm \Psi^{\dagger}_{lc\mu}(x))
\label{lft.7}
\eeq
where $Z$ is a normalization constant.  Locality does not fix the
$\pm$ sign.  Local fields where space reflection acts linearly can be
constructed from these by taking the direct sum of a right and left
handed local field:
\beq
\Psi_{loc \mu}^{\dagger}(x) \to 
\left( 
\begin{array}{c}
\Psi_{rloc\mu}^{\dagger} (x)\\
{\Psi}_{lloc\mu}^{\dagger} (x) 
\end{array}
\right ) .
\label{lft.8}
\eeq

This structure will be used to construct a spin 1/2 field satisfying
the Dirac equation.  The structure of the gamma matrices follow from
the $SL(2,\mathbf{C})$ transformation properties of the Pauli matrices
and the $2\times 2$ identity.  The relevant representation of 
$SL(2,\mathbb{C})$  for a Dirac field is the direct sum of a right and left
handed representation of $SL(2,\mathbf{C})$:
\beq
S(A) =
\left (
\begin{array}{cc}
A & 0
\\
0 & \tilde{A}
\end{array}
\right ) =
\left (
\begin{array}{cc}
A & 0
\\
0 & ({A}^{\dagger})^{-1}
\end{array}
\right ) .
\label{lft.9}
\eeq
The representation of the $\gamma$-matrices follow from the transformation
properties of four vectors represented by $2 \times 2$
Hermitian matrices:
\beq
X:= x^{\mu} \sigma_{\mu} \qquad 
X' = A X A^{\dagger} \qquad A \in SL(2,\mathbb{C})
\label{lft.10}
\eeq
This can be expressed in terms of the components of $x$ as
\beq
\sigma_{\mu} \Lambda^{\mu}{}_{\nu}x^{\nu} =
A \sigma_{\nu} A^{\dagger}x^{\nu} . 
\label{lft.11}
\eeq
Equating the coefficients of $x^{\nu}$ gives
\beq
A \sigma_{\nu} A^{\dagger} =
\sigma_{\mu} \Lambda^{\mu}{}_{\nu}
\label{lft.12}
\eeq
Multiplying both sides of this equation by $\sigma_2$ and
taking complex conjugates gives
\beq
\tilde{A} \sigma_2\sigma^*_{\nu}\sigma_2 \tilde{A}^{\dagger} =
\sigma_2 \sigma*_{\mu}\sigma_2 \Lambda^{\mu}{}_{\nu} .  
\label{lft.13}
\eeq
Equations (\ref{lft.12}) and (\ref{lft.13}) can be combined into a
single equation
\beq
\left (
\begin{array}{cc}
A & 0
\\
0 & ({A}^{\dagger})^{-1}
\end{array}
\right )
\left (
\begin{array}{cc}
0 & \sigma_{\mu}
\\
\sigma_2 \sigma_{\mu}^* \sigma_2  & 0
\end{array}
\right )
\left (
\begin{array}{cc}
A^{-1} & 0
\\
0 & {A}^{\dagger}
\end{array}
\right ) =
\left (
\begin{array}{cc}
0 & \sigma_{\nu}
\\
\sigma_2 \sigma_{\nu}^* \sigma_2  & 0
\end{array}
\right )
\Lambda^{\nu}{}_{\mu}
\label{lft.14}
\eeq
which shows that the matrices 
\beq
\gamma_{\nu}:=
\left (
\begin{array}{cc}
0 & -\sigma_{\nu}
\\
-\sigma_2 \sigma_{\nu}^* \sigma_2  & 0
\end{array}
\right )
\label{lft.15}
\eeq
transform like four vectors with respect to the similarity
transformation
\beq
S(A) \gamma_{\mu} S(A^{-1}) = 
\gamma_{\nu} \Lambda_{\nu}{}^{\mu}. 
\label{lft.15}
\eeq
where the $-$ sign is a convention.   With this convention
\beq
\gamma^{0} =
\left (
\begin{array}{cc}
0 & \sigma_0 
\\
\sigma_{0}  & 0
\end{array}
\right )
\qquad
\gamma^{\mu} =
\left (
\begin{array}{cc}
0 & \sigma_2 \sigma^*_{\mu}\sigma_2
\\
-{\sigma}_{\mu}  & 0
\end{array}
\right )
\label{lft.16}
\eeq
and
\beq
\gamma^5 = \gamma_5 = i \gamma^0 \gamma^1 \gamma^2 \gamma^3 = 
\left (
\begin{array}{cc}
\sigma_0 & 0 \\
0 & -\sigma_0 
\end{array}
\right ).
\label{lft.17}
\eeq
In order to construct the Dirac field using matrix multiplication
it is useful to define the $4 \times 2$ matrix
\beq
u_{c \mu} :=
\left (
\begin{array}{c}
\sigma_0 \\
\sigma_0
\end{array}
\right ).
\label{lft.18}
\eeq
The Dirac field is a linear combination of the form (\ref{lft.6}-\ref{lft.7}):
\beq
\Psi_\alpha (x) =
\int
{d \mathbf{p} \over \sqrt{\omega_m(\mathbf{p})}}
\left (
e^{i p\cdot x} S(B_x (p/m))_{c d} u_{d \mu} a_x(\mathbf{p},\mu) +
e^{- p\cdot x} (\gamma_5 S(B_x (p/m))_{c c} u_{c \nu}
\sigma_{2\nu \mu} b^{\dagger}_x(\mathbf{p},\mu) 
\right )
\label{lft.19}
\eeq

The $\gamma_5$ commutes with $S(B_x (p/m))$, and
anti-commutes with $\gamma^{\mu}$.  While it changes the sign on the lower two
components, it is
consistent with the freedom to choose the sign of $\alpha$ and $\beta$ in
the locality constraint (\ref{lft.6}) and (\ref{lft.7}).

Multiplying (\ref{lft.18}) by the Dirac operator
$(-i \gamma^{\mu} {\partial \over \partial x^{\mu}} + mI)$ 
gives
\[
(-i \gamma^{\mu} {\partial \over \partial x^{\mu}} + mI)
\Psi (x)
=
\]
\[
\int
{d \mathbf{p} \over \sqrt{\omega_m(\mathbf{p})}}
\left (
e^{i p\cdot x} ( p\cdot \gamma +m) S(B_x (p/m)) u a_x(\mathbf{p}) +
e^{- p\cdot x} (-p\cdot \gamma +m) \gamma_5 S(B_x (p/m)) u \sigma_2  b^{\dagger}_x(\mathbf{p}) 
\right ) =
\]
\beq
\int
{d \mathbf{p} \over \sqrt{\omega_m(\mathbf{p})}}
\left (
e^{i p\cdot x} ( p\cdot \gamma +m) S(B_x (p/m))u  a_x(\mathbf{p}) +
e^{- p\cdot x} \gamma_5 (p\cdot \gamma +m) S(B_x (p/m)) u
\sigma_2  b^{\dagger}_x(\mathbf{p} ) 
\right ).
\label{lft.20}
\eeq
This vanishes because 
\beq
( p\cdot \gamma +m) S(B_x (p/m)) =
S(B_x (p/m))S^{-1} (B_x (p/m)) ( p\cdot \gamma +m) S(B_x (p/m)) =
S(B_x (p/m)) ( -m\gamma^0 + mI) 
\label{lft.21}
\eeq
which vanishes when applied to $u$ or $u \sigma_2$.

The quantities
\beq
u_{x c \mu}(p) := 
\sqrt{m}\left (S(B_x (p/m)) u \right )_{c \mu} 
\label{lft.22}
\eeq
\beq
v_{x c \mu}(p) := 
\sqrt{m} \left (\gamma_5 S(B_x (p/m)) u
\sigma_2 \right )_{c \mu}
\label{lft.23}
\eeq
are Dirac spinors.  Note that both the spinors and creation and
annihilation operators depend on the choice of boost, $B_x(p/m)$, but
the field itself is independent of this choice.

\section{Dynamics}

Dynamical relativistic models were not discussed in the previous
sections.  This is in part because there are many distinct
formulations of relativistic quantum mechanics that have been applied
to model few-hadron or few-quark systems.  Some examples are
\cite{Bakker:1979eg}
\cite{Coester:1982vt}
\cite{Crater:1983ew}
\cite{Crater:1983ni}
\cite{Grach:1983hd}
\cite{Polyzou:1986cv}
\cite{Glockle:1986zz}
\cite{Keister:1988zz}
\cite{Chung:1988my}
\cite{Chung:1988mu}
\cite{Glockle:1990}
\cite{Cardarelli:1995dc}
\cite{Polyzou:1996fb}
\cite{Krutov:1997wu}
\cite{Coester:1997ih}
\cite{Allen:2000xy}
\cite{Wagenbrunn:2000es}
\cite{Liu:2002cn}
\cite{Gross:2003qi} 
\cite{JuliaDiaz:2003gq}
\cite{Polyzou:2003dt}
\cite{Coester:2005cv}
\cite{Alba:2006hs}
\cite{Lin:2007kg}
\cite{deMelo:2008rj}
\cite{Huang:2008jd}
\cite{Lin:2008sy}
\cite{Biernat:2009my}
\cite{Desplanques:2009kj}
\cite{Frederico:2009fk}
\cite{Fuda:2009zz}
\cite{Whitney:2011aa}
\cite{Witala:2011yq}
\cite{Salme:2012tnh}
\cite{Carbonell:2014dwa}
\cite{Gross:2014zra}
\cite{Hadizadeh:2014yua}
\cite{Santopinto:2014opa}
\cite{Faustov:2015}
\cite{Arslanaliev:2018wkl}
\cite{Carbonell:2016udz}
\cite{dePaula:2016oct}
\cite{DelDotto:2016vkh}
\cite{Eichmann:2016yit}
\cite{Leitao:2017mlx}
.
An adequate discussion of
each one is beyond the intended scope of this work.  However, there
are common features in all formulations of relativistic quantum
theory.  The most important common feature is that the dynamics is
defined by an underlying unitary representation of the Poincar\'e group.
The dynamical representation can be decomposed into a direct integral of
irreducible unitary representations.  The structure of the direct
integral that defines the dynamics is the common element in all equivalent
formulations of relativistic quantum dynamics.  The dynamics
determines the spectrum and multiplicities of the masses and spins
that appear in the direct integral.

Another important common feature of relativistic quantum mechanical
models is the intended applications.  The physics goal of most
relativistic dynamical models is to understand the structure and
dynamics of hadronic systems at distance scales that are fractions of
a Fermi.  The cleanest way to study systems at this resolution is with
probes that interact weakly with these strongly interacting systems.
The basic observables are matrix elements of covariant current
operators that couple to weak and electromagnetic fields evaluated in
bound or scattering states of hadrons.  Since the probe must transfer
enough momentum to be sensitive to short-distance physics, the initial
and final hadronic states are needed in different Lorentz frames.
For an initial state in the laboratory frame the relevant matrix element is
\beq
_z\langle (m_f,j_f) \mathbf{p}_f,\mu_f ,\lambda_f \vert 
I^{\mu}(0) \vert (m,j) \mathbf{0}_i,\mu_i ,\lambda_i \rangle_z
\qquad
\mbox{where}
\qquad
\vert
(m_f,j_f) \mathbf{p}_f,\mu_f ,\lambda_f \rangle_z =
U(B_z(p),0 ) (m_f,j_f) \mathbf{0}_f,\mu_f ,\lambda_f \rangle_z .
\label{d.1}
\eeq
In this expression 
\beq
\vert (m,j) \mathbf{0}_i,\mu_i ,\lambda_i \rangle_z \qquad \mbox{and} \qquad 
\vert (m,j) \mathbf{0}_f,\mu_f ,\lambda_f \rangle_z
\label{d.2}
\eeq
are the initial and final dynamical mass and spin eigenstates in the
rest frame, $I^{\mu}(0)$ is a dynamical covariant current density at
$x=0$ and $U(B_x(p/m),0)$ is a dynamical Lorentz transformation from the
rest frame of the target to the frame of the recoiling hadronic system.

While QCD is assumed to be the theory of the strong interaction, there
are no known relativistically invariant approximations with
mathematically controlled errors at the interesting few-GeV energy
scale.  Relativistic quantum mechanical models provide a framework to
identify the important degrees of freedom and reaction mechanisms in a
manner that is consistent with the general principles of special
relativity and quantum mechanics.

Each dynamical formulation of relativistic quantum mechanics has it's
strengths and weaknesses for computing the matrix elements (\ref{d.1}).
The relation between different formulations of relativistic quantum
theory that arise from the common underlying Poincar\'e symmetry may
be used to take advantage of the strengths of different equivalent
formulations of the theory.  This section provides a brief summary of
the general structure of different dynamical formulations of
relativistic quantum mechanics, how they are related and how the dynamics
enters.

In order to understand the relation between different formulations of
relativistic quantum mechanics it is necessary to understand how the
direct integral of the dynamical irreducible representation of the
Poincar\'e group appears in the different formulations of relativistic
quantum mechanics.  The discussion that follows is limited to systems
of N-particles for the purpose of illustration, although the similar
considerations apply to systems that do not conserve particle
number.  The three different formulations of relativistic quantum
theory that were discussed are Poincar\'e covariant formulations,
Lorentz covariant formulations, and Euclidean covariant formulations.
These discussions were all in the context of a single particle or an
irreducible representation.

The dynamics in Poincar\'e covariant formulations of relativistic
quantum mechanics is defined by an explicit unitary representation of
the Poincar\'e group on the $N$-body Hilbert space
\cite{Sokolov:1977}\cite{Coester:1982vt}\cite{Keister:1991sb}.  The
mass and spin operators for this representation are dynamical
operators that act on this Hilbert space.  The direct integral results
from simultaneously diagonalizing both of these operators, and
introducing additional invariant degeneracy operators that separate
multiple copies of irreducible representations with same mass and
spin.

The dynamics in Lorentz covariant formulations of relativistic quantum
mechanics is defined by a N-particle Hilbert space with a Lorentz
covariant positive kernel \cite{Polyzou:1986cv}.  The need for a
dynamical kernel can be understood by observing that time evolution
cannot not be as trivial as shifting the time argument of a covariant
wave function.  For free particles the kernel was constructed from the
Poincar\'e covariant representation and was given by
(\ref{cov.14}-\ref{cov.15}).  The dynamics was given by the mass appearing
in these expressions.  In quantum field theory the covariant
Hilbert space kernels are the vacuum expectation values of products of
fields.  These are the Wightman functions \cite{Wightman:1980} of the
field theory.  The most direct way to understand the structure of the
Hilbert space kernels of Lorentz covariant formulations of
relativistic quantum mechanics is to compare them to field theoretic
kernels.

Vectors in quantum field theory can be constructed by applying 
polynomials of smeared Heisenberg fields to the physical vacuum  
\beq
\vert \psi \rangle := \int \sum \Psi_{\mu_N}(x_N) \cdots
\Psi_{\mu_1}(x_1)\vert 0 \rangle d^4x_1 \cdots d^4x_N f_{\mu_1}(x_1)
\cdots f_{\mu_N}(x_N)
\label{d.3}
\eeq
\beq
\vert \phi \rangle := \int \sum \Psi_{\mu_N}(x_N) \cdots
\Psi_{\mu_1}(x_1)\vert 0 \rangle d^4x_1 \cdots d^4x_N g_{\mu_1}(x_1)
\cdots g_{\mu_N}(x_N).
\label{d.4}
\eeq
The inner product of these vectors is an integral of
the product of covariant test functions with a covariant kernel
\[
\langle \phi \vert \psi \rangle = 
\]
\beq
\int \sum
g_{\mu_1}^*(y_1)
\cdots g_{\mu_N}^* (y_N) d^{4N}y
\langle 0 \vert \Psi^{\dagger}_{\mu_N}(y_N)  \cdots
\Psi^{\dagger}_{\mu_1}(y_1)
\Psi_{\nu_1}(x_1) \cdots
\Psi_{\nu_N}(x_N) \cdots
\vert 0 \rangle d^{4N}x
f_{\nu_1}(x_1)
\cdots f_{\nu_N}(x_N)   .
\label{d.5}
\eeq
The test functions represent the Lorentz covariant wave functions.
In this example the Lorentz covariant kernel is the vacuum expectation
value of the product of 2N fields:
\beq
W_{2N}(y_N, \mu_N, \cdots, y_1, \mu_1; x_1, \nu_1, \cdots, x_N, \nu_N) =    
\langle 0 \vert \Psi^{\dagger}_{\mu_N}(y_N)  \cdots
\Psi^{\dagger}_{\mu_1}(y_1)
\Psi_{\nu_1}(x_1) \cdots
\Psi_{\nu_N}(x_N)
\vert 0 \rangle .
\label{d.5}
\eeq
For fields that transform like (\ref{f.5}-\ref{f.6}) 
the kernel (\ref{d.5}) satisfies the covariance condition
\[
W_{2N}(y_N, \mu_N, \cdots, y_1, \mu_1; x_1, \nu_1, \cdots, x_N, \nu_N) =
\]
\beq
\prod_k D^{j_k}_{\mu_k\mu_k'}[A^{-1}]
W_{2N}(\Lambda y_N+a, \mu_N', \cdots, \Lambda y_1+a , \mu_1';
\Lambda x_1+a , \nu_1', \cdots, \Lambda x_N+a , \nu_N')
\prod_i D^{j_i}_{\nu_i'\nu_i}[\tilde{A}].
\label{d.6}
\eeq
This inner product is preserved for wave functions that transform like
\beq
f_{\nu_i}(x_i) \to \sum D^{j_i}_{\nu_i'\nu_i}[\tilde{A}]
f_{\nu_1}(\Lambda^{-1}(x_i-a))
\label{d.7}
\eeq
\beq
g^*_{\nu_i}(y_i) \to  \sum g^*_{\nu_i'}(\Lambda^{-1}(y_i-a))
D^{j_i}_{\nu_i'\nu_i}[{A}^{-1}].
\label{d.8}
\eeq
These equations are representative; there are similar relations for
fields that transform with different spinor representations of the
Lorentz group.  The invariance of the inner product with respect to
the Poincar\'e transformations means that the Poincar\'e transformations
in equations (\ref{d.7}-\ref{d.8}) are unitary.
 
In this example the dynamics is in the Heisenberg fields, which are
solutions of the field equations.  The direct integral of irreducible
representations enters the kernel by inserting a complete set of
intermediate Poincar\'e covariant states between these vectors
\[
\langle \phi \vert \psi \rangle = 
\]
\[
\int \sum g_{\mu_1}^*(y_1) \cdots g_{\mu_N}^* (y_N)
d^{4N}y  \langle 0 \vert
\Psi^{\dagger}_{\mu_N}(y_N) \cdots \Psi^{\dagger}_{\mu_1}(y_1) \vert
(m,j) \mathbf{p},\mu,\lambda \rangle_z
d\mathbf{p} dm d\lambda \times
\]
\beq
_z\langle (m,j)
\mathbf{p},\mu,\lambda \vert \Psi_{\nu_1}(x_1) \cdots
\Psi_{\nu_N}(x_N) \cdots \vert 0 \rangle d^{4N}x 
f_{\nu_1}(x_1) \cdots
f_{\nu_N}(x_N) 
\label{d.9}
\eeq
where $z$ indicates the type of spin as defined in (\ref{p.35}).
In this expression the matrix elements
\beq
 \langle 0 \vert
\Psi^{\dagger}_{\mu_N}(y_N) \cdots \Psi^{\dagger}_{\mu_1}(y_1) \vert
(m,j) \mathbf{p},\mu,\lambda \rangle_z
\label{d.10}
\eeq
and
\beq
_z\langle (m,j)
\mathbf{p},\mu_z,\lambda \vert \Psi_{\nu_1}(x_1) \cdots
\Psi_{\nu_N}(x_N) \vert 0 \rangle
\label{d.11}
\eeq
have mixed transformation properties.  The fields transform like
Lorentz covariant densities (\ref{f.5}-\ref{f.6}) while the states,
$\vert (m,j) \mathbf{p},\mu,\lambda \rangle_z$, transform like a mass
$m$ spin $j$ irreducible representation of the Poincar\'e group
(\ref{p.39}).  Following what was done in the single-particle case
(\ref{cov.14}), the Poincar\'e covariant intermediate states can be
replaced by equivalent Lorentz covariant intermediate states
\[
\int \sum g_{\mu_1}^*(y_1) \cdots g_{\mu_N}^* (y_N) d^{4N}y \langle 0 \vert
\Psi^{\dagger}_{\mu_N}(y_N) \cdots \Psi^{\dagger}_{\mu_1}(y_1) \vert
(m,j) p,\mu,\lambda \rangle_{cov} \delta (m^2 + p^2 ) \theta (p^0)
D^j_{\mu \nu} [p\cdot \sigma] d^4p dm d\lambda \times
\]
\beq
_{cov}\langle (m,j)
{p},\nu,\lambda \vert \Psi_{\nu_1}(x_1) \cdots
\Psi_{\nu_N}(x_N) \cdots \vert 0 \rangle d^{4N}x  f_{\nu_1}(x_1) \cdots
f_{\nu_N}(x_N)  
\label{d.12}
\eeq
where right-handed representations were used in (\ref{d.12}) for the
purpose of illustration.  In the dynamical case the masses, spins and
degeneracy quantum numbers are the masses, spins and degeneracy
quantum numbers that appear in the complete set of intermediate
states.  These states may include single-particle states, bound states or
scattering states.  This change of representation results in a
manifestly Lorentz covariant expression for the intermediate states
in the direct integral..

Matrix elements of normalizable vectors with the irreducible dynamical
eigenstates that appear in the direct integral in Poincar\'e
covariant formulations of relativistic quantum
mechanics are identified with the field theoretic amplitudes
by
\[
\langle \psi \vert (m,j)\mathbf{p},\mu ,\lambda \rangle_z =
\]
\beq
\int \sum g_{\mu_1}^*(y_1) \cdots g_{\mu_N}^* (y_N) d^{4N}y \langle 0 \vert
\Psi^{\dagger}_{\mu_N}(y_N) \cdots \Psi^{\dagger}_{\mu_1}(y_1) \vert
(m,j) \mathbf{p},\mu,\lambda \rangle_z
\label{d.13}
\eeq
and in the (right handed) Lorentz covariant formulations by
\[
\langle \psi \vert (m,j)\mathbf{p},\mu ,\lambda \rangle_{cov} =
\]
\beq
\int \sum g_{\mu_1}^*(y_1) \cdots g_{\mu_N}^* (y_N) d^{4N}y \langle 0 \vert
\Psi^{\dagger}_{\mu_N}(y_N) \cdots \Psi^{\dagger}_{\mu_1}(y_1) \vert
(m,j) p,\mu,\lambda \rangle_{cov}
\label{d.14}
\eeq
where (\ref{d.13}) and (\ref{d.14}) are related by 
\[
\int \sum g_{\mu_1}^*(y_1) \cdots g_{\mu_N}^* (y_N) d^{4N}y \langle 0 \vert
\Psi^{\dagger}_{\mu_N}(y_N) \cdots \Psi^{\dagger}_{\mu_1}(y_1) \vert
(m,j) p,\mu,\lambda \rangle_{cov} = 
\]
\beq
\int \sum g_{\mu_1}^*(y_1) \cdots g_{\mu_N}^* (y_N) d^{4N}y \langle 0 \vert
\Psi^{\dagger}_{\mu_N}(y_N) \cdots \Psi^{\dagger}_{\mu_1}(y_1) \vert
(m,j) \mathbf{p},\nu,\lambda \rangle_z
D^j_{\nu \mu}[B_z^{-1}(p/m)] \sqrt{\omega_m (\mathbf{p})} .
\label{d.15}
\eeq


A dynamical kernel for a general Lorentz covariant formulation of
relativistic $N$-body quantum mechanics has the same
structure as (\ref{d.12})
\[
W(x_N, \mu_N, \cdots x_1, \mu_1; y_1 , \nu_1, \cdots, y_N, \nu_N) =
\]
\[
\sum\int 
w^*_{cov}(x_N, \mu_N, \cdots , x_1, \mu_1 \vert (m,j) p, \mu, \lambda) 
\delta (p^2 +m^2) \theta (p^0)d^4 p dm d\lambda
D^j_{\mu \nu} [p\cdot \sigma/m] \rho (m, \lambda)
\times
\]
\beq
w_{cov}((m,j)p, \nu, \lambda \vert y_1, \nu_1, \cdots ,  y_N, \nu_N) .
\label{d.16}
\eeq
The Lorentz covariant transformation properties
and interpretation of the amplitudes 
\beq
w_{cov}((m,j)p, \nu, \lambda \vert y_1, \nu_1, \cdots,  y_N, \nu_N)
\qquad \mbox{and} \qquad 
w_{cov}^*( y_N, \nu_N, \cdots,  y_1, \nu_1 ;(m,j)   p, \nu, \lambda)
\label{d.17}.
\eeq
are identical to the corresponding properties of the field amplitudes
\beq
\langle 0 \vert
\Psi^{\dagger}_{\mu_N}(y_N) \cdots \Psi^{\dagger}_{\mu_1}(y_1) \vert
(m,j) p,\mu,\lambda \rangle_{cov}
\qquad \mbox{and} \qquad 
_{cov}\langle (m,j) p,\mu,\lambda \vert
\Psi_{\mu_1}(x_1) \cdots \Psi_{\mu_N}(x_N) \vert 0
 \rangle .
 \label{d.18}
 \eeq
The kernel (\ref{d.16}) also can be factored into
amplitudes involving Poincar\'e covariant states
\[
W(x_N, \mu_N, \cdots x_1, \mu_1; y_1 , \nu_1, \cdots, y_N, \nu_N) =
\]
\[
\sum\int 
w^*(x_N, \mu_N, \cdots , x_1, \mu_1 \vert (m,j) \mathbf{p}, \mu,
\lambda)_z 
d^3 p dm d\lambda
\times
\]
\beq
w_z((m,j)\mathbf{p}, \nu, \lambda \vert y_1, \nu_1 \cdots  y_N, \nu_N) .
\label{d.19}
\eeq
where the Lorentz covariant and Poincar\'e covariant
amplitudes are related by 
\[
w^*_{cov}(x_N, \mu_N, \cdots , x_1, \mu_1 \vert (m,j) p, \mu,
\lambda)
= 
\]
\beq
\sum w^*_z(x_N, \mu_N, \cdots , x_1, \mu_1 \vert (m,j) \mathbf{p}, \nu,
\lambda) 
D^j_{\nu \mu}[B_x^{-1}(p/m)] \sqrt{\omega_m (\mathbf{p})} .
\label{d.20}
\eeq
The sum and integral over the mass, spin and degeneracy parameters is
the direct integral that defines the dynamics.    
It is still necessary to specify the kernel in order to define the
dynamics.

The third class of formulations of relativistic quantum mechanics are
Euclidean covariant formulations.  Schwinger \cite{Schwinger:1958qau}
showed that time-ordered vacuum expectation values of products of
fields satisfying the spectral
condition could be analytically continued to imaginary time.  For free
particles the relation between the Euclidean and Lorentz covariant
representations of a single particle was given by (\ref{e.4}).  For
the interacting case the field theoretic example provides some
insight.  Since the intermediate states between each pair of fields in
the Wightman functions have positive energy (after recursively
subtracting vacuum contributions), they can be analytically continued
to regions of the complex plane where the imaginary parts of the
relative times are negative.  In the field theory case the domain of
analyticity can be extended using covariance with respect to complex
Lorentz transformations and locality \cite{Wightman:1980}.  Both the
time-ordered Green's functions and the Wightman functions can be
recovered from the Euclidean Green's functions using different limits.
Osterwalder and Schrader
\cite{Osterwalder:1973dx}\cite{Osterwalder:1974tc} considered the
inverse problem of identifying the conditions on a collection of
Euclidean covariant distributions that are needed to define a Lorentz
covariant quantum field theory or relativistic quantum theory.  Since
the group of real Euclidean transformations is a subgroup of the
complex Lorentz group, the generators of the two groups are formally
related by (\ref{er.1}) for the generators of transformations
involving the Euclidean time.  The generators of rotations and space
translations are the same in the Euclidean and Minkowski cases.
However the generators (\ref{er.1}) are not Hermitian on a space
defined by a Euclidean covariant kernel.  Reversing the signs of the
Euclidean times in the final state makes the corresponding Lorentz
generators formally Hermitian.  The Euclidean sesquilinear form with
the time reflection cannot be positive on arbitrary functions of
Euclidean variables, since functions that are odd or even under
Euclidean time reflection will lead to norms with opposite signs.
However for suitable Euclidean covariant distributions positivity can
hold on a subspace.  For systems of particles the relevant subspace is
the space of functions of Euclidean space time variables with positive
relative Euclidean time support.

The kernel of the physical Hilbert space scalar product is a Euclidean
covariant distribution, with Euclidean time reflection operators on
the final Euclidean times. It must be non-negative on the space of
Euclidean covariant functions with support for positive relative
times.  When this is satisfied the Euclidean covariant distribution is
called reflection positive.  Reflection positivity is responsible for
both the positivity of the physical Hilbert space norm and the
spectral condition \cite{glimm}.

The structure of the physical inner product in the Euclidean
framework is motivated by local field theory, where the kernels are
analytic continuations of real-time Green's functions:
\[
\langle \psi \vert \phi \rangle =
\]
\[
\int 
g_{\mu_1}^*(y_{e1})
\cdots g_{\mu_N}^* (y_{eN}) d^{4N}y_e
E(\theta y_{eN}, \mu_N, \cdots  \theta y_{e1}, \mu_1 ;
x_{e1}, \nu_1, \cdots x_{eN} ,\nu_N) d^{4N}x_e \times
\]
\beq
f_{\nu_1}(x_{e1})
\cdots f_{\nu_N}(x_{eN}). 
\label{d.21}
\eeq
In this expression 
$\theta$ is the Euclidean time reflection operator and the
test functions are non-zero for $0<\tau_{x1}<\tau_{x2} \cdots \tau_{xN}$
and $0<\tau_{y1}<\tau_{y2} \cdots \tau_{yN}$.  The direct integral enters
the Euclidean kernel in the form
\[
E(y_{eN}, \mu_N, \cdots  y_{e1}, \mu_1 ;
x_{e1}, \nu_1, \cdots x_{eN} ,\nu_N) =
\]
\beq
\sum\int 
e^*(y_{eN}, \mu_N, \cdots , y_{e1}, \mu_1 \vert (m,j) p_e, \mu, \lambda)
\delta (p_e^2 +m^2) dm d\lambda
{1 \over \pi} {D^j_{\mu \nu} [p_e\cdot \sigma_e/m]\over p_e^2 +m^2}
e((m,j) p_e, \nu, \lambda \vert x_{e1}, \nu_1 \cdots  x_{eN}, \nu_{N})
\label{d.22}
\eeq
where the individual factors
$e^*(y_{eN}, \mu_N, \cdots , y_{e1}, \mu_1 \vert (m,j) p_e, \mu, \lambda)$
are Euclidean covariant.  The matrix elements with the covariant states
are related by 
\beq
\langle \psi \vert (m,j) p_e, \mu, \lambda \rangle_{cov} = 
\sum\int
g_{\mu_1}^*(y_{e1})
\cdots g_{\mu_N}^* (y_{eN}) d^{4N}y_e
e^*(y_{eN}, \mu_N, \cdots , y_{e1}, \mu_1 \vert (m,j) (-i \omega_m(\mathbf{p}),
\mathbf{p}, \mu, \lambda) 
\label{d.23}
\eeq
where the amplitude is analytic in the lower half $p^0_e$ plane
provided the test functions satisfy the positive relative Euclidean
time support condition.

The resulting inner product is the physical inner product.  On this
space, with the Euclidean time reflection, the 10 generators of the
Euclidean group, with the modifications (\ref{er.1}), are formally
Hermitian and satisfy the Poincar\'e commutations relations.  These
generators can be exponentiated to construct the unitary
representation of the Poincar\'e group on the Euclidean representation
of the Hilbert space.  Calculations of inner products of physical
states and matrix elements of operators in these states can be
performed by integrating over the Euclidean variables without
performing an explicit analytic continuation.

While this general discussion explains where the direct integral that
defines the dynamics enters in each of the formulations of
relativistic quantum mechanics, it does not explain how to construct
dynamical models in each of the formulations of relativistic quantum
mechanics.  Several methods for constructing dynamical models in each
of these frameworks are discussed below.

The class of models that have the most in common with non-relativistic
quantum models are Poincar\'e covariant quantum models.  They are
defined by constructing an explicit dynamical unitary representation
of the Poincar\'e on the $N$-particle Hilbert space.  One strategy for
constructing the dynamical unitary representation of the Poincar\'e
group, due to Bakamjian and Thomas \cite{Bakamjian:1953kh},
is to
start with a system of non-interacting relativistic particles and
decompose the $N$-particle states into irreducible subspaces labeled
by mass and spin using Clebsch-Gordan \cite{Coester:1965zz}
\cite{Polyzou:19892b}\cite{Keister:1991sb} coefficients of the
Poincar\'e group. Interactions that commute with
the $N$-particle spin are added to the free invariant mass operator.
Diagonalizing this operator in the irreducible free-particle basis
gives a complete set of states labeled by the mass and spin
eigenvalues.  This is the relativistic analog of diagonalizing a
non-relativistic center of mass Hamiltonian.  The dynamical
eigenstates are complete and transform like irreducible
representations of the Poincar\'e group (\ref{p.39}) with the particle mass
replaced by the mass eigenvalue.  This defines the dynamical unitary
representation of the Poincar\'e group on a basis.  The choice of
basis used to compute the Clebsch-Gordan coefficients has dynamical
consequences in this framework.

The advantages of the Poincar\'e covariant framework are (1) bound state and
scattering state solutions can be calculated using the same methods
that are employed in non-relativistic few-body calculations (2)
standard high-precision nucleon-nucleon interactions
\cite{Wiringa:1994wb}\cite{Entem:2003ft} that are fit to experimental
data can be \cite{Coester:1975zz} reinterpreted and used directly in
these calculations (3) reactions with particle production can be
treated.  This is the most mature method in terms of the complexity of
models that have been treated.  Applications include
relativistic constituent quark models of mesons and baryons
\cite{Salme:2012tnh}\cite{Fuda:2009zz} 
two and three-nucleon bound state calculations
\cite{Bakker:1979eg}\cite{Glockle:1986zz}\cite{Witala:2011yq}
\cite{Hadizadeh:2014yua}\cite{Arslanaliev:2018wkl}
relativistic two and three-nucleon scattering calculations
\cite{Lin:2007kg}\cite{Lin:2008sy}\cite{Witala:2011yq}
and electromagnetic observables of hadrons 
\cite{Grach:1983hd}
\cite{Chung:1988my}
\cite{Keister:1988zz}
\cite{Chung:1988mu}
\cite{Chung:1991st}
\cite{Coester:1994wp}
\cite{Polyzou:1996fb}
\cite{Coester:1997ih}
\cite{Krutov:1997wu}
\cite{Wagenbrunn:2000es}
\cite{JuliaDiaz:2003gq}
\cite{Huang:2008jd}
\cite{deMelo:2008rj}
\cite{Frederico:2009fk}
\cite{Krutov:2009zz}
\cite{Desplanques:2009kj}
\cite{Biernat:2009my}
\cite{Santopinto:2014opa}
\cite{DelDotto:2016vkh}.

The fundamental challenge with Poincar\'e covariant models
is that there are an infinite
number of equivalent representations \cite{Polyzou:19892b}
\cite{Polyzou:2002id} associated with different irreducible basis
choices, none of which can be trivially derived from QCD.  The unitary
transformations that relate the different equivalent representations
generate many-body interactions and many-body current operators in
transforming from one representation to another \cite{Glockle:1990}.
This makes it difficult to construct equivalent interactions in
different representations or assign any special significance
to interactions in a given representation.
Cluster properties, for systems of
more than two particles, require an additional class of
momentum-dependent many-body interactions
\cite{Sokolov:1977}
\cite{Coester:1982vt}
\cite{Keister:1991sb}
\cite{Polyzou:2003dt}.
When the many-body operators are not included the equivalence of the
different forms of dynamics breaks down \cite{Desplanques:2009kj}.
While two-body interactions are constrained by experiment, it is more
difficult to constrain the three and four-body interactions and the
two, three and four-body exchange currents.  Comparison with
experiment suggests that these operators cannot be ignored.  Another
consequence of the non-trivial dynamical structure of representations
of the Poincar\'e group is that it is not possible to construct
Poincar\'e covariant one-body current operators.  This means that
typical impulse approximations that are used in hard scattering
calculations cannot be formulated in a fully Poincar\'e covariant
manner using these models.  Normally ``impulse approximations'' in this
framework are
defined by using the one-body parts of the current operator to compute
a set of preferred independent matrix elements, with the remaining
matrix elements generated by covariance and current conservation.  The
results, while covariant, depend on the choice of independent matrix
elements.  These are not true impulse approximations because
covariance condition implicitly generates exchange current
contributions in the remaining current matrix elements.  It is no
substitute for having an explicit covariant current operator, which is
needed to have a meaningful probe of these system.

The most useful Poincar\'e covariant representations are the ones
discovered by Dirac \cite{Dirac:1949cp}, which are characterized by
dynamical representations of the Poincar\'e group that have a 6 or 7
parameter (kinematic) subgroup that is free of interactions.  These
are called Dirac's forms of dynamics.  Each form of dynamics has
different advantages.  In the ``instant form'' rotations and
translations are kinematic.  Some representative calculations are
\cite{Glockle:1986zz}
\cite{Krutov:1997wu}
\cite{Lin:2007kg}
\cite{Lin:2008sy}
\cite{Fuda:2009zz} 
\cite{Witala:2011yq}
\cite{Hadizadeh:2014yua}
\cite{Arslanaliev:2018wkl}.
The
difficulties are that the Lorentz boosts that are needed to compute
current matrix elements are dynamical. This means that ``impulse
approximations'' are frame dependent in the sense that an impulse
approximation in the Breit frame is not an impulse approximation in
the laboratory frame.  In the ``point form'' of the dynamics Lorentz
transformations are kinematic, but the translations that transform the
current to $x=0$ are not.  The
kinematic Lorentz invariance implies that one-body operators remain
one-body operators in all frames related by Lorentz transformations.
However, because translations are dynamical, the momentum transferred
to a system in an ``impulse matrix element'' is not the same as the
momentum transferred to the constituents \cite{Allen:2000xy}. Some
representative point form
calculations are
\cite{Coester:1997ih}
\cite{Allen:2000xy}
\cite{Wagenbrunn:2000es}
\cite{JuliaDiaz:2003gq}
\cite{Biernat:2009my}
\cite{Santopinto:2014opa}.
Front-from or light-front dynamics has
the largest (7-parameter) kinematic subgroup, which is the subgroup of
the Poincar\'e group that leaves a plane tangent to the light-cone
invariant.  It has a three-parameter subgroup of boosts that are free
of interactions.  The light-front spins do not Winger rotate when
transformed with this subgroup.  It also has a three-parameter
subgroup of translations tangent to the light front that are free of
interactions.  Finally it has frame-independent ``impulse
approximations'' where the momentum transferred to the target is the
same as the momentum transferred to the constituents.  The difficulty
is that rotations are dynamical.  Rotational covariance of current
operators in this representation is difficult to realize at the
operator level. Representative front-form calculations are
\cite{Bakker:1979eg}
\cite{Grach:1983hd}
\cite{Chung:1988my}
\cite{Chung:1988mu}
\cite{Keister:1988zz}
\cite{Chung:1989zz}
\cite{Cardarelli:1995dc}
\cite{Coester:1997ih}
\cite{Coester:2004qu}
\cite{Coester:2005cv}
\cite{Huang:2008jd}
\cite{deMelo:2008rj}
\cite{Frederico:2009fk} 
\cite{Desplanques:2009kj}.
Equations (\ref{p.42}) can be used to relate the dynamical mass-spin
eigenstates and wave functions in each of these forms.  The rotations
$R_{xy}(p/m)$ and Jacobians $ \vert{\partial \mathbf{p} (\mathbf{f},m)
\over \partial \mathbf{f}} \vert$ in the transformations of the
eigenstates in (\ref{p.42}) depend on the mass eigenvalues, so they
are dynamical and become non-trivial operators when the mass is not
diagonalized.  For instance, equations (\ref{p.42}) could be used to
transform an instant-form calculation of a triton bound state wave
function to a
light-front representation of the same state with a light-front spin,
which is preferred in structure function calculations. In this case
the triton mass eigenvalue appears in the rotation and Jacobian.
If this accompanied by the corresponding transformation
of the variables of the wave functions,  the combination of these two
transformations, one dynamical and one kinematic, makes the
light-front boosts kinematic.

In Lorentz covariant formulations of relativistic quantum mechanics
the wave functions depend on $4N$ space-time variables and spins that
transform under a finite dimensional representation of the Lorentz
group.  Lorentz covariant quantum theories are closely related to the
Klein-Gordon or Dirac equations.  Quantum field equations are the
operator versions of these equations.  The difference is that the
Klein-Gordon and Dirac equations are for a single particle, while the
corresponding field equations are for systems of an infinite number of
degrees of freedom.  The interactions in quantum field theory are
expressed in terms of products of fields at the same point, which are
mathematically ill defined.  Lorentz covariant quantum theories, which
involve a finite number of degrees of freedom, fall between these two
extremes.  They are mathematically well-defined theories of a finite
number of degrees of freedom.

A feature of Lorentz covariant quantum theories is that the Hilbert
space inner product must have a non-trivial Poincar\'e covariant
kernel {\cite{Polyzou:1986cv} that defines the dynamics as in
(\ref{cov.14}-\ref{cov.15}).
The covariance ensures the invariance of the Hilbert space inner
product with respect to Poincar\'e transformations of the arguments
of the covariant wave functions.  This defines the dynamical unitary
representation of the Poincar\'e group.  In most
applications of Lorentz covariant quantum mechanics the inner
product exists in the background, and is not explicitly utilized in
calculations.

While the Dirac and Klein-Gordon equations are one-particle equations,
many-particle systems can be treated by coupled Dirac or Klein-Gordon
equations that satisfy first class \cite{Dirac:1951zz} constraints.
The first class condition is an integrability condition for the
coupled equations.  The first class condition also ensures that a
covariant ``quasi-Wightman kernel'' can be defined as a product of
delta functions in the constraints (the first class condition implies
that the product of these delta functions is independent of the order
of the products).  Using this kernel to calculate scalar products is
equivalent to solving the coupled Dirac or Klein-Gordon equations.
For free particles the constraints are products of positive energy
mass-shell constraints.  This is identical to products of free-field
two-point Wightman functions.  The mass shell constraints are first
class because they commute.  Dynamics is introduced by adding
covariant interactions to the kinematic constraints that preserve the
first-class condition.  The simplest way to do this is to replace the
constraints by an equivalent set of constraints that are the sum of
the free-particle constraints and an independent set of the
differences.  Interactions that commute with the difference
constraints can be added to the sum of the constraints. For example,
for two interacting scalar particles a dynamical kernel
is given by 
\beq
W(x_1,x_2;y_1,y_2) = \langle x_1 ,x_2 \vert \delta (p_1^2 + m_1^2
- p_2^2 - m_2^2)\theta (p_1^0) \theta (p_2^0) \delta ((p_1^2 + m_1^2 +
p_2^2 + m_2^2 + V) \vert y_1, y_2 \rangle
\label{d.24}
\eeq
where
\beq
[p_1^2 +
m_1^2 - p_2^2 - m_2^2,V]=0.
\label{d.25}
\eeq
Some applications of constraint dynamics to
meson and baryon spectra and nucleon-nucleon scattering are
\cite{Crater:1983ew}
\cite{Crater:1983ni}
\cite{Liu:2002cn}
\cite{Alba:2006hs}
\cite{Whitney:2011aa}.
While the construction described above can be applied
to systems of any number of particles, it does not satisfy cluster
properties for systems of more than two particles.  The challenge is
to add interactions to each mass-shell constraint that preserve the
first class condition.

An alternative to constraint dynamics is to build Lorentz covariant
models by making finite number of degree of freedom truncations to
the Heisenberg fields.  The N-quantum approximation
\cite{Greenberg:2010wx} starts by representing the Heisenberg fields
as an expansion in normal products of an irreducible set of asymptotic
fields called the Haag expansion \cite{Glaser:1957vrs}.  The field
equations give an infinite set of coupled equations for the
coefficients of the expansion.  A consistent treatment requires
a-priori knowledge of the asymptotic fields that appear in the
expansion.  This is equivalent to knowing the spectral content of the
direct integral that defines the dynamics.  Lorentz covariant models
can be constructed by truncating this expansion to a finite number of
experimentally relevant terms.  Since the asymptotic fields and
Heisenberg fields have the same vacuum, the resulting fields can be
used to calculate model Wightman functions.  Model Wightman functions
constructed by truncating the expansion will be covariant and can lead
to a limited kind of positivity that gives a Lorentz covariant
relativistic quantum theory.  Because the coefficients of the
expansion are related to observables, \cite{Glaser:1957vrs},
observables can be calculated directly from the coefficients of the
expansion without explicitly utilizing the Hilbert space inner
product.  The N-quantum approximation has the computational advantage
that the variables associated with the asymptotic fields remain on
shell.

The most common Lorentz covariant models are based on truncations of
the Schwinger-Dyson equations,
\cite{Dyson:1949ha}\cite{Schwinger:1951xk}\cite{Schwinger:1953tb},
which are an infinite set of coupled equations for vacuum expectation
values of time-ordered products of fields.  The advantage of this
method is that matrix elements of any observable can be calculated without
directly utilizing the Hilbert space representation
\cite{Huang:1974cd}.  To make relativistic quantum models the infinite
set of equations must be truncated, and the input that would have been
determined by the discarded equations has to be modeled.  The Lorentz
covariance is easily preserved under truncation.  The existence of an
underlying Hilbert space with positive norm and a dynamics satisfying
a spectral condition constrains the model input to the truncated
system. 

The model-time ordered Green's functions are constrained by assuming
the existence of an underlying relativistic quantum theory.
This assumption implies that there is a complete set of
positive mass (plus the vacuum) intermediate states (the direct
integral) between any pair of fields in the Green's function.
Choosing time orderings that separate the desired initial final states,
and inserting a complete set of Poincar\'e covariant intermediate
states leads to an expression with a pole in the intermediate
energy variable.  The residue is $-{1 \over 2 \pi i}$ times a product of
amplitudes of the form
\beq
\langle 0\vert T (\Psi_{\mu_N}(x_N) \cdots \Psi_{\mu_1}(x_1)) \vert  
(m,j) \mathbf{p},\mu,\lambda \rangle_z
\label{d.26}
\eeq
and
\beq
_z\langle (m,j) \mathbf{p},\mu,\lambda \vert T (\Psi_{\mu_1}(y_1) \cdots \Psi_{\mu_N}(y_N)) \vert 0 \rangle .
\label{d.27}
\eeq
where $T$ is the time-ordering operator.  When these are integrated
over test functions with support for a given time ordering, these
quantities are identical to (\ref{d.14}).

As in the N-quantum approximation, direct use of the underlying
Hilbert space representation can be avoided.  Matrix elements of
operators are obtained by inserting the operator in the product of
fields, choosing a time ordering so the products of fields on the left
and right of the operator are interpolating fields for chosen initial
and final states \cite{Huang:1974cd}.  The residue of the poles that
select the initial and final states is a product of the matrix
elements of the operators between these two states (\ref{d.1}) with
two ``covariant wave functions''.  To isolate the desired matrix
element the two covariant wave functions can be eliminated using a
quadrature associated with normalization of the fields.

The simplest Schwinger-Dyson equation is the Bethe-Salpeter equation,
which is used for both two-body bound and scattering states
\cite{Carbonell:2014dwa}\cite{Carbonell:2016udz}\cite{dePaula:2016oct}.
The input is a pair of dynamical time-ordered two-point functions and
a connected dynamical time-ordered four-point function.  Both of these
quantities are unknown, and have to be modeled by appealing to
experiment and general principles.  The equations are more complicated
to solve than the corresponding Poincar\'e covariant equations due to
the presence of additional relative time or energy variables.  The
calculation of matrix elements of operators requires using a dynamical
normalization condition.  The complications increase with the
three-body problem, especially for the scattering problems.  One
advantage that comes from the explicit covariance is the existence of
a relativistic impulse approximation when coupling to currents.  When
the additional relative energy or time variable is eliminated
\cite{Coester:1994wp} it generates an effective exchange current.  The
relation between the Lorentz covariant and Poincar\'e covariant
representations can be used to motivate the structure of exchange
currents \cite{Huang:2008jd} in Poincar\'e covariant quantum models
that arise impulse currents in Lorentz covariant theories.

A related class of models that are used in calculations of strongly
interacting systems are quasipotential methods.  These methods are
formally equivalent to the Bethe-Salpeter equation.  They are derived
by replacing the Bethe-Salpeter equation by an equivalent pair of
equations.  One involves a new kernel with a constraint that reduces
the number of integration variables.  The second relates the new
kernel to the original Bethe Salpeter kernel.  When the quasipotential
kernel is calculated, the equations are equivalent to the
Bethe-Salpeter equation.  However since both kernels are modeled in
practice, it is no reason to assume the one is more fundamental than
the other. Quasipotential methods have been used to compute
scattering, structure, electromagnetic observables
\cite{Gross:2003qi}
\cite{Gross:2014zra} 
\cite{Biernat:2015xya}
\cite{Faustov:2015}
\cite{Leitao:2017mlx}
.

The last class of theories are based on the Euclidean formulation of
relativistic quantum mechanics.  Euclidean Green's functions are used in
truncations of the Euclidean form of the Schwinger-Dyson equations as well
as in lattice truncations of QCD.  While lattice truncations are a
powerful computational tool, they break Poincar\'e (or Euclidean)
invariance.  As in the Minkowski case the input to the Euclidean
Schwinger-Dyson equation needs to be modeled.  In the Euclidean case
the Schwinger-Dyson equations are much simpler than the corresponding
Minkowski equations.


The challenges with the Euclidean approach arise because physical
observables involve real time.  This normally requires an
analytic continuation in the time variable of a quantity that is often
calculated either numerically or statistically.
In spite of these
challenges there have been many advances using these methods. 

Applications based on the Euclidean formulation of the Schwinger-Dyson
equations have been performed on the and two nucleon
\cite{Carbonell:2016udz} the three quark \cite{Eichmann:2016yit}
systems.  An aspect of the Osterwalder and Schrader
reconstruction theorem is that the physical Hilbert space and the
unitary representation of the Poincar\'e group can be constructed
directly from the Euclidean Green functions without analytic
continuation.  The advantage of the Hilbert space representation are
that the input involves solutions of relatively well-behaved
Euclidean Green functions and there are explicit expressions for the
Hamiltonian and the other nine self-adjoint Poincar\'e generators on
this space.  Because the Hilbert space representation is the physical
representation, direct calculations of scattering observables
\cite{Polyzou:2013nga} \cite{Aiello:2015jgc}\cite{Kopp:2011vv} can be
preformed without analytic continuation.  The challenges are to ensure
that the kernel is reflection positive.

To summarize, all formulations of relativistic quantum mechanics
involve a direct integral of irreducible representations.  Equivalent
models involve different representations of the same direct integral.
In all three formulations the matrix elements, $\langle \psi \vert
(m,j)\mathbf{p},\mu ,\lambda \rangle_z$, can be extracted, where
$\vert \psi \rangle$ is a normalizable vector in the Hilbert space.
How the vectors $\vert \psi \rangle$ are represented
depends on the representation of the Hilbert space.  
 
\section{Summary}

Relativistically invariant  treatments of quantum mechanics are needed
to understand physics on distance scales that are small compared to
the Compton wavelength of the relevant particles.  Of particular importance
is the need to consistently calculate matrix elements of hadronic currents
when the initial and final hadronic states are in different Lorentz frames.
The strength of the interaction precludes a perturbative treatment
to the hadronic structure or final state interactions in these matrix
elements.

Relativistic invariance in quantum mechanics means that measurements
of quantum observables - probabilities, expectation values and
ensemble averages cannot be used to distinguish inertial coordinate
systems.  This is equivalent to the requirement that equivalent
operators and states in different inertial coordinate systems are
related by a unitary ray representation of the Poincar\'e group on the
Hilbert space of the quantum theory.  Unitary representations of the
Poincar\'e group can always be decomposed into direct integrals of
irreducible representations.  This step is the relativistic analog of
diagonalizing the Hamiltonian in non-relativistic quantum theory.  The
structure of the invariant mass $m$ spin $j$ irreducible subspaces are
fixed by group theory.  Different treatments of relativistic quantum
theory use different ways of representing these elementary building
blocks of the theory.  Since each representation has is own advantages,
it is useful to understand how the different representations are related.

In this work mass $m$ spin $j$ irreducible representations of the
Poincar\'e group were constructed using a basis of simultaneous
eigenstates of independent commuting functions of the Poincar\'e
generators.  The relevant Hilbert space was the space of square
integrable functions of the eigenvalues of these operators.  The
eigenvalue spectrum of these commuting observables is fixed by properties
of the Poincar\'e group.  The transformation properties of the
Poincar\'e generators led to an explicit unitary representation of the
Poincar\'e group on this representation of the Hilbert space.
Different choices of the commuting observables lead to different
representations that are related by unitary transformations.

Factoring the Wigner rotations that appear in these irreducible
representations into products of Lorentz
$SL(2,\mathbb{C})$ transformations, and using group representation
properties of $SL(2,\mathbb{C})$, led to equivalent Lorentz
covariant representations, where the states transform under finite
dimensional representations of $SL(2,\mathbb{C})$.  In these
representations the Hilbert space inner product has a non-trivial
kernel, which was shown to be, up to normalization,
the two-point Wightman function of a free quantum field theory. 

These Lorentz covariant representations were shown to be
derivable from a representation of the Hilbert space with a Euclidean
covariant kernel and a Euclidean time reflection on the final states.
In this representation of the Hilbert space, the inner product involves an
integral over Euclidean variables; it does not require analytic
continuation.

Finally covariant fields were constructed from the Lorenz covariant
wave functions.  These fields have the property that the vacuum
expectation value of products of two fields recover the free-field
Wightman functions that appear in the kernel of the Lorentz covariant
representations.

While the Lorentz covariant, Euclidean covariant, and field
representations were constructed starting with irreducible
representations of the Poincar\'e group, the process could easily be
reversed by factoring the Wightman functions.

These relations indicate how the many different representations
that are used in applications are related.  The discussion in sections
(3-8) was limited to one-particle or Poincar\'e irreducible states. 
These same representations appear in dynamical models in the form
of direct integrals of irreducible eigenstates.  A discussion
of these same relations in the context of dynamical models was
given in section 9.

\appendix*

\section{Wigner D-functions} 

The $2j+1$ dimensional unitary representations of $SU(2)$ in the basis
of eigenstates of $j^2,j_z$ (Wigner functions) are used extensively in
these notes.  Most derivations in the literature are for the
expression in terms of Euler angles, rather than in terms of the
$SU(2)$ matrix elements.  The expression directly in terms of $SU(2)$
matrix elements was used to extend these to representations of $SL(2,C)$.

%
The most straightforward derivation of the formula (\ref{p.32}) for
the Wigner $D$-function in terms of the $SU(2)$ matrix elements uses
Schwinger's formulation of the angular momentum algebra
\cite{schwinger} using creation and annihilation operators.

The main elements of this formalism are pair of creation and
annihilation operators.  Angular momentum state are relabeled with 
\beq
n_{\pm} := j \pm m
\eeq
which are related to the standard angular momentum labels by
\beq
j: = {1 \over 2} (n_+ + n_-)
\qquad
m := {1 \over 2} (n_+ - n_-)
\eeq
\beq
\vert n_+, n_-\rangle := \vert j,m\rangle  
\eeq
The creation and annihilation operators are defined by 
\beq
a_{\pm}^{\dagger} \vert n_{\pm} \rangle = \sqrt{n_{\pm}+1} \vert
n_{\pm} +1 \rangle 
\eeq
\beq
a_{\pm} \vert n_{\pm} \rangle = \sqrt{n_{\pm}} \vert
n_{\pm} -1 \rangle 
\eeq
With these definitions
\beq
J_\pm = a_\pm^{\dagger} a_\mp \qquad
J_z = {1 \over 2} (a_+^{\dagger} a_+ - a_-^{\dagger} a_-)
\eeq
\beq
\mathbf{J} = {1 \over 2} 
\left ( 
\begin{array}{cc}
a^{\dagger}_+ 
a^{\dagger}_- \\  
\end{array} 
\right )
\pmb{\sigma} 
\left ( 
\begin{array}{c}
a_+ \\
a_- \\  
\end{array} 
\right ) 
\eeq
This can be used to show
\beq
e^{i \mathbf{J} \cdot \pmb{\theta}}a^{\dagger}_{\pm} 
e^{-i \mathbf{J} \cdot \pmb{\theta}} =
(a_+^{\dagger}R_{+\pm} + a_-^{\dagger}R_{-\pm})
\eeq
where
\beq
R= \cos ({\theta \over 2})I  + i \pmb{\hat{\theta}}\cdot
\pmb{\sigma} \sin ({\theta \over 2}) =
\left (
\begin{array}{cc}
R_{++} & R_{+-} \\
R_{-+} & R_{--}
\end{array}  
\right ).  
\eeq
Normalized angular momentum eigenstates have the form 
\beq
\vert n_+, n_- \rangle =
{(a_+^{\dagger})^{n_+} \over \sqrt{n_+!}}
{(a_-^{\dagger})^{n_-} \over \sqrt{n_-!}}\vert 0,0 \rangle .
\eeq
Combining these results gives
\beq
D^j_{m'm}[R]= 
\langle n_+', n_-' \vert
e^{i \mathbf{J} \cdot \pmb{\theta}} \vert
n_+, n_- \rangle =
\eeq
\beq
{1 \over \sqrt{n_+'! n_-'! n_+! n_-! } }
\langle 0,0 \vert (a_+)^{n_+'}(a_-)^{n_+'}
(a_+^{\dagger}R_{++} + a_-^{\dagger}R_{-+})^{n_+}
(a_+^{\dagger}R_{+-} + a_-^{\dagger}R_{--})^{n_-}
\vert 0,0 \rangle
\eeq
(this vanishes unless $j=j'$).
Expanding
$(a_+^{\dagger}R_{++} + a_-^{\dagger}R_{-+})^{n_+}$
and
$(a_+^{\dagger}R_{+-} + a_-^{\dagger}R_{--})^{n_-}$
using the binomial series and properties of the
creation and annihilation operators gives
the result (\ref{p.32}).


\begin{thebibliography}{93}
\expandafter\ifx\csname natexlab\endcsname\relax\def\natexlab#1{#1}\fi
\expandafter\ifx\csname bibnamefont\endcsname\relax
  \def\bibnamefont#1{#1}\fi
\expandafter\ifx\csname bibfnamefont\endcsname\relax
  \def\bibfnamefont#1{#1}\fi
\expandafter\ifx\csname citenamefont\endcsname\relax
  \def\citenamefont#1{#1}\fi
\expandafter\ifx\csname url\endcsname\relax
  \def\url#1{\texttt{#1}}\fi
\expandafter\ifx\csname urlprefix\endcsname\relax\def\urlprefix{URL }\fi
\providecommand{\bibinfo}[2]{#2}
\providecommand{\eprint}[2][]{\url{#2}}

\bibitem[{\citenamefont{Wigner}(1939)}]{Wigner:1939cj}
\bibinfo{author}{\bibfnamefont{E.~P.} \bibnamefont{Wigner}},
  \bibinfo{journal}{Annals Math.} \textbf{\bibinfo{volume}{40}},
  \bibinfo{pages}{149} (\bibinfo{year}{1939}).

\bibitem[{\citenamefont{Dirac}(1949)}]{Dirac:1949cp}
\bibinfo{author}{\bibfnamefont{P.~A.~M.} \bibnamefont{Dirac}},
  \bibinfo{journal}{Rev. Mod. Phys.} \textbf{\bibinfo{volume}{21}},
  \bibinfo{pages}{392} (\bibinfo{year}{1949}).

\bibitem[{\citenamefont{Dirac}(1945)}]{Dirac:1945cm}
\bibinfo{author}{\bibfnamefont{P.~A.~M.} \bibnamefont{Dirac}},
  \bibinfo{journal}{Proc. Roy. Soc. Lond.} \textbf{\bibinfo{volume}{A183}},
  \bibinfo{pages}{284} (\bibinfo{year}{1945}).

\bibitem[{\citenamefont{Weinberg}(1964{\natexlab{a}})}]{Weinberg:1964ev}
\bibinfo{author}{\bibfnamefont{S.}~\bibnamefont{Weinberg}},
  \bibinfo{journal}{Phys. Rev.} \textbf{\bibinfo{volume}{134}},
  \bibinfo{pages}{B882} (\bibinfo{year}{1964}{\natexlab{a}}).

\bibitem[{\citenamefont{Weinberg}(1964{\natexlab{b}})}]{Weinberg:1964cn}
\bibinfo{author}{\bibfnamefont{S.}~\bibnamefont{Weinberg}},
  \bibinfo{journal}{Phys. Rev.} \textbf{\bibinfo{volume}{133}},
  \bibinfo{pages}{B1318} (\bibinfo{year}{1964}{\natexlab{b}}).

\bibitem[{\citenamefont{Weinberg}(1969)}]{Weinberg:1969di}
\bibinfo{author}{\bibfnamefont{S.}~\bibnamefont{Weinberg}},
  \bibinfo{journal}{Phys. Rev.} \textbf{\bibinfo{volume}{181}},
  \bibinfo{pages}{1893} (\bibinfo{year}{1969}).

\bibitem[{\citenamefont{Weinberg}(1995)}]{Weinberg:v1}
\bibinfo{author}{\bibfnamefont{S.}~\bibnamefont{Weinberg}},
  \emph{\bibinfo{title}{The Quantum Theory of Fields}},
  vol.~\bibinfo{volume}{I} (\bibinfo{publisher}{Cambridge University Press,
  NY}, \bibinfo{year}{1995}).

\bibitem[{\citenamefont{Streater and Wightman}(1980)}]{Wightman:1980}
\bibinfo{author}{\bibfnamefont{R.~F.} \bibnamefont{Streater}} \bibnamefont{and}
  \bibinfo{author}{\bibfnamefont{A.~S.} \bibnamefont{Wightman}},
  \emph{\bibinfo{title}{PCT, Spin and Statistics, and All That}}
  (\bibinfo{publisher}{Princeton Landmarks in Physics}, \bibinfo{year}{1980}).

\bibitem[{\citenamefont{Osterwalder and Schrader}(1975)}]{Osterwalder:1974tc}
\bibinfo{author}{\bibfnamefont{K.}~\bibnamefont{Osterwalder}} \bibnamefont{and}
  \bibinfo{author}{\bibfnamefont{R.}~\bibnamefont{Schrader}},
  \bibinfo{journal}{Commun. Math. Phys.} \textbf{\bibinfo{volume}{42}},
  \bibinfo{pages}{281} (\bibinfo{year}{1975}).

\bibitem[{\citenamefont{Osterwalder and Schrader}(1973)}]{Osterwalder:1973dx}
\bibinfo{author}{\bibfnamefont{K.}~\bibnamefont{Osterwalder}} \bibnamefont{and}
  \bibinfo{author}{\bibfnamefont{R.}~\bibnamefont{Schrader}},
  \bibinfo{journal}{Commun. Math. Phys.} \textbf{\bibinfo{volume}{31}},
  \bibinfo{pages}{83} (\bibinfo{year}{1973}).

\bibitem[{\citenamefont{Jacob and Wick}(1959)}]{JacobWick}
\bibinfo{author}{\bibfnamefont{M.}~\bibnamefont{Jacob}} \bibnamefont{and}
  \bibinfo{author}{\bibfnamefont{G.~C.} \bibnamefont{Wick}},
  \bibinfo{journal}{Annals of Physics} \textbf{\bibinfo{volume}{7}},
  \bibinfo{pages}{404} (\bibinfo{year}{1959}).

\bibitem[{\citenamefont{Bakamjian and Thomas}(1953)}]{Bakamjian:1953kh}
\bibinfo{author}{\bibfnamefont{B.}~\bibnamefont{Bakamjian}} \bibnamefont{and}
  \bibinfo{author}{\bibfnamefont{L.~H.} \bibnamefont{Thomas}},
  \bibinfo{journal}{Phys. Rev.} \textbf{\bibinfo{volume}{92}},
  \bibinfo{pages}{1300} (\bibinfo{year}{1953}).

\bibitem[{\citenamefont{Bargmann}(1954)}]{Bargmann:1954gh}
\bibinfo{author}{\bibfnamefont{V.}~\bibnamefont{Bargmann}},
  \bibinfo{journal}{Annals Math.} \textbf{\bibinfo{volume}{59}},
  \bibinfo{pages}{1} (\bibinfo{year}{1954}).

\bibitem[{\citenamefont{Wigner}(1957)}]{Wigner:1957ep}
\bibinfo{author}{\bibfnamefont{E.~P.} \bibnamefont{Wigner}},
  \bibinfo{journal}{Rev. Mod. Phys.} \textbf{\bibinfo{volume}{29}},
  \bibinfo{pages}{255} (\bibinfo{year}{1957}).

\bibitem[{\citenamefont{Wightman}(1960)}]{Wightman}
\bibinfo{author}{\bibfnamefont{A.~S.} \bibnamefont{Wightman}},
  \emph{\bibinfo{title}{{L'Invariance Dans La Mecanique Quantique
  Relativiste}}}, vol.~\bibinfo{volume}{7} (\bibinfo{publisher}{Hermann,
  Paris}, \bibinfo{year}{1960}).

\bibitem[{\citenamefont{Melosh}(1974)}]{Melosh:1974cu}
\bibinfo{author}{\bibfnamefont{H.~J.} \bibnamefont{Melosh}},
  \bibinfo{journal}{Phys. Rev.} \textbf{\bibinfo{volume}{D9}},
  \bibinfo{pages}{1095} (\bibinfo{year}{1974}).

\bibitem[{\citenamefont{Polyzou et~al.}(2013)\citenamefont{Polyzou, Glockle,
  and Wita{\l}a}}]{wpwg}
\bibinfo{author}{\bibfnamefont{W.~N.} \bibnamefont{Polyzou}},
  \bibinfo{author}{\bibfnamefont{W.}~\bibnamefont{Glockle}}, \bibnamefont{and}
  \bibinfo{author}{\bibfnamefont{H.}~\bibnamefont{Wita{\l}a}},
  \bibinfo{journal}{Few Body Syst.} \textbf{\bibinfo{volume}{54}},
  \bibinfo{pages}{1667} (\bibinfo{year}{2013}).

\bibitem[{\citenamefont{Gilmm and Jaffe}(1981)}]{glimm}
\bibinfo{author}{\bibfnamefont{J.}~\bibnamefont{Gilmm}} \bibnamefont{and}
  \bibinfo{author}{\bibfnamefont{A.}~\bibnamefont{Jaffe}},
  \emph{\bibinfo{title}{Quantum Physics - A functional Integral Point of View}}
  (\bibinfo{publisher}{Springer}, \bibinfo{year}{1981}).

\bibitem[{\citenamefont{Klein and L.}(1981)}]{Klein:1981}
\bibinfo{author}{\bibfnamefont{A.}~\bibnamefont{Klein}} \bibnamefont{and}
  \bibinfo{author}{\bibfnamefont{L.}~\bibnamefont{L.}}, \bibinfo{journal}{J.
  Functional Anal.} \textbf{\bibinfo{volume}{44}}, \bibinfo{pages}{121}
  (\bibinfo{year}{1981}).

\bibitem[{\citenamefont{Klein and L.}(1983)}]{Klein:1983}
\bibinfo{author}{\bibfnamefont{A.}~\bibnamefont{Klein}} \bibnamefont{and}
  \bibinfo{author}{\bibfnamefont{L.}~\bibnamefont{L.}}, \bibinfo{journal}{Comm.
  Math. Phys} \textbf{\bibinfo{volume}{87}}, \bibinfo{pages}{469}
  (\bibinfo{year}{1983}).

\bibitem[{\citenamefont{Frohlich et~al.}(1983)\citenamefont{Frohlich,
  Osterwalder, and Seiler}}]{Frohlich:1983kp}
\bibinfo{author}{\bibfnamefont{J.}~\bibnamefont{Frohlich}},
  \bibinfo{author}{\bibfnamefont{K.}~\bibnamefont{Osterwalder}},
  \bibnamefont{and} \bibinfo{author}{\bibfnamefont{E.}~\bibnamefont{Seiler}},
  \bibinfo{journal}{Annals Math.} \textbf{\bibinfo{volume}{118}},
  \bibinfo{pages}{461} (\bibinfo{year}{1983}).

\bibitem[{\citenamefont{Bogoliubov and Shirkov}(1959)}]{bogoliubov}
\bibinfo{author}{\bibfnamefont{N.~N.} \bibnamefont{Bogoliubov}}
  \bibnamefont{and} \bibinfo{author}{\bibfnamefont{D.~V.}
  \bibnamefont{Shirkov}}, \emph{\bibinfo{title}{Introduction to the theory of
  quantized fields}} (\bibinfo{publisher}{Wiley-Interscience},
  \bibinfo{year}{1959}).

\bibitem[{\citenamefont{Bakker et~al.}(1979)\citenamefont{Bakker, Kondratyuk,
  and Terentev}}]{Bakker:1979eg}
\bibinfo{author}{\bibfnamefont{B.~L.~G.} \bibnamefont{Bakker}},
  \bibinfo{author}{\bibfnamefont{L.~A.} \bibnamefont{Kondratyuk}},
  \bibnamefont{and} \bibinfo{author}{\bibfnamefont{M.~V.}
  \bibnamefont{Terentev}}, \bibinfo{journal}{Nucl. Phys.}
  \textbf{\bibinfo{volume}{B158}}, \bibinfo{pages}{497} (\bibinfo{year}{1979}).

\bibitem[{\citenamefont{Coester and Polyzou}(1982)}]{Coester:1982vt}
\bibinfo{author}{\bibfnamefont{F.}~\bibnamefont{Coester}} \bibnamefont{and}
  \bibinfo{author}{\bibfnamefont{W.~N.} \bibnamefont{Polyzou}},
  \bibinfo{journal}{Phys. Rev.} \textbf{\bibinfo{volume}{D26}},
  \bibinfo{pages}{1348} (\bibinfo{year}{1982}).

\bibitem[{\citenamefont{Crater and Van~Alstine}(1983)}]{Crater:1983ew}
\bibinfo{author}{\bibfnamefont{H.~w.} \bibnamefont{Crater}} \bibnamefont{and}
  \bibinfo{author}{\bibfnamefont{P.}~\bibnamefont{Van~Alstine}},
  \bibinfo{journal}{Annals Phys.} \textbf{\bibinfo{volume}{148}},
  \bibinfo{pages}{57} (\bibinfo{year}{1983}).

\bibitem[{\citenamefont{Crater and Van~Alstine}(1988)}]{Crater:1983ni}
\bibinfo{author}{\bibfnamefont{H.~W.} \bibnamefont{Crater}} \bibnamefont{and}
  \bibinfo{author}{\bibfnamefont{P.}~\bibnamefont{Van~Alstine}},
  \bibinfo{journal}{Phys. Rev.} \textbf{\bibinfo{volume}{D37}},
  \bibinfo{pages}{1982} (\bibinfo{year}{1988}).

\bibitem[{\citenamefont{Grach and Kondratyuk}(1984)}]{Grach:1983hd}
\bibinfo{author}{\bibfnamefont{I.~L.} \bibnamefont{Grach}} \bibnamefont{and}
  \bibinfo{author}{\bibfnamefont{L.~A.} \bibnamefont{Kondratyuk}},
  \bibinfo{journal}{Sov. J. Nucl. Phys.} \textbf{\bibinfo{volume}{39}},
  \bibinfo{pages}{198} (\bibinfo{year}{1984}).

\bibitem[{\citenamefont{Polyzou}(1985)}]{Polyzou:1986cv}
\bibinfo{author}{\bibfnamefont{W.~N.} \bibnamefont{Polyzou}},
  \bibinfo{journal}{Phys. Rev.} \textbf{\bibinfo{volume}{D32}},
  \bibinfo{pages}{995} (\bibinfo{year}{1985}).

\bibitem[{\citenamefont{Glockle et~al.}(1986)\citenamefont{Glockle, Lee, and
  Coester}}]{Glockle:1986zz}
\bibinfo{author}{\bibfnamefont{W.}~\bibnamefont{Glockle}},
  \bibinfo{author}{\bibfnamefont{T.~S.~H.} \bibnamefont{Lee}},
  \bibnamefont{and} \bibinfo{author}{\bibfnamefont{F.}~\bibnamefont{Coester}},
  \bibinfo{journal}{Phys. Rev.} \textbf{\bibinfo{volume}{C33}},
  \bibinfo{pages}{709} (\bibinfo{year}{1986}).

\bibitem[{\citenamefont{Keister}(1988)}]{Keister:1988zz}
\bibinfo{author}{\bibfnamefont{B.~D.} \bibnamefont{Keister}},
  \bibinfo{journal}{Phys. Rev.} \textbf{\bibinfo{volume}{C37}},
  \bibinfo{pages}{1765} (\bibinfo{year}{1988}).

\bibitem[{\citenamefont{Chung et~al.}(1988{\natexlab{a}})\citenamefont{Chung,
  Polyzou, Coester, and Keister}}]{Chung:1988my}
\bibinfo{author}{\bibfnamefont{P.~L.} \bibnamefont{Chung}},
  \bibinfo{author}{\bibfnamefont{W.~N.} \bibnamefont{Polyzou}},
  \bibinfo{author}{\bibfnamefont{F.}~\bibnamefont{Coester}}, \bibnamefont{and}
  \bibinfo{author}{\bibfnamefont{B.~D.} \bibnamefont{Keister}},
  \bibinfo{journal}{Phys. Rev.} \textbf{\bibinfo{volume}{C37}},
  \bibinfo{pages}{2000} (\bibinfo{year}{1988}{\natexlab{a}}).

\bibitem[{\citenamefont{Chung et~al.}(1988{\natexlab{b}})\citenamefont{Chung,
  Coester, and Polyzou}}]{Chung:1988mu}
\bibinfo{author}{\bibfnamefont{P.~L.} \bibnamefont{Chung}},
  \bibinfo{author}{\bibfnamefont{F.}~\bibnamefont{Coester}}, \bibnamefont{and}
  \bibinfo{author}{\bibfnamefont{W.~N.} \bibnamefont{Polyzou}},
  \bibinfo{journal}{Phys. Lett.} \textbf{\bibinfo{volume}{B205}},
  \bibinfo{pages}{545} (\bibinfo{year}{1988}{\natexlab{b}}).

\bibitem[{\citenamefont{Gl{\"o}ckle and Polyzou}(1990)}]{Glockle:1990}
\bibinfo{author}{\bibfnamefont{W.}~\bibnamefont{Gl{\"o}ckle}} \bibnamefont{and}
  \bibinfo{author}{\bibfnamefont{W.~N.} \bibnamefont{Polyzou}},
  \bibinfo{journal}{Few-Body Systems} \textbf{\bibinfo{volume}{9}},
  \bibinfo{pages}{97} (\bibinfo{year}{1990}).

\bibitem[{\citenamefont{Cardarelli et~al.}(1995)\citenamefont{Cardarelli, Pace,
  Salme, and Simula}}]{Cardarelli:1995dc}
\bibinfo{author}{\bibfnamefont{F.}~\bibnamefont{Cardarelli}},
  \bibinfo{author}{\bibfnamefont{E.}~\bibnamefont{Pace}},
  \bibinfo{author}{\bibfnamefont{G.}~\bibnamefont{Salme}}, \bibnamefont{and}
  \bibinfo{author}{\bibfnamefont{S.}~\bibnamefont{Simula}},
  \bibinfo{journal}{Phys. Lett.} \textbf{\bibinfo{volume}{B357}},
  \bibinfo{pages}{267} (\bibinfo{year}{1995}).

\bibitem[{\citenamefont{Polyzou and Gl{\"o}ckle}(1996)}]{Polyzou:1996fb}
\bibinfo{author}{\bibfnamefont{W.~N.} \bibnamefont{Polyzou}} \bibnamefont{and}
  \bibinfo{author}{\bibfnamefont{W.}~\bibnamefont{Gl{\"o}ckle}},
  \bibinfo{journal}{Phys. Rev.} \textbf{\bibinfo{volume}{C53}},
  \bibinfo{pages}{3111} (\bibinfo{year}{1996}).

\bibitem[{\citenamefont{Krutov}(1997)}]{Krutov:1997wu}
\bibinfo{author}{\bibfnamefont{A.~F.} \bibnamefont{Krutov}},
  \bibinfo{journal}{Phys. Atom. Nucl.} \textbf{\bibinfo{volume}{60}},
  \bibinfo{pages}{1305} (\bibinfo{year}{1997}).

\bibitem[{\citenamefont{Coester and Riska}(1998)}]{Coester:1997ih}
\bibinfo{author}{\bibfnamefont{F.}~\bibnamefont{Coester}} \bibnamefont{and}
  \bibinfo{author}{\bibfnamefont{D.~O.} \bibnamefont{Riska}},
  \bibinfo{journal}{Few Body Syst.} \textbf{\bibinfo{volume}{25}},
  \bibinfo{pages}{29} (\bibinfo{year}{1998}).

\bibitem[{\citenamefont{Allen et~al.}(2000)\citenamefont{Allen, Payne, and
  Polyzou}}]{Allen:2000xy}
\bibinfo{author}{\bibfnamefont{T.~W.} \bibnamefont{Allen}},
  \bibinfo{author}{\bibfnamefont{G.~L.} \bibnamefont{Payne}}, \bibnamefont{and}
  \bibinfo{author}{\bibfnamefont{W.~N.} \bibnamefont{Polyzou}},
  \bibinfo{journal}{Phys. Rev.} \textbf{\bibinfo{volume}{C62}},
  \bibinfo{pages}{054002} (\bibinfo{year}{2000}).

\bibitem[{\citenamefont{Wagenbrunn et~al.}(2001)\citenamefont{Wagenbrunn,
  Boffi, Klink, Plessas, and Radici}}]{Wagenbrunn:2000es}
\bibinfo{author}{\bibfnamefont{R.~F.} \bibnamefont{Wagenbrunn}},
  \bibinfo{author}{\bibfnamefont{S.}~\bibnamefont{Boffi}},
  \bibinfo{author}{\bibfnamefont{W.}~\bibnamefont{Klink}},
  \bibinfo{author}{\bibfnamefont{W.}~\bibnamefont{Plessas}}, \bibnamefont{and}
  \bibinfo{author}{\bibfnamefont{M.}~\bibnamefont{Radici}},
  \bibinfo{journal}{Phys. Lett.} \textbf{\bibinfo{volume}{B511}},
  \bibinfo{pages}{33} (\bibinfo{year}{2001}).

\bibitem[{\citenamefont{Liu and Crater}(2003)}]{Liu:2002cn}
\bibinfo{author}{\bibfnamefont{B.}~\bibnamefont{Liu}} \bibnamefont{and}
  \bibinfo{author}{\bibfnamefont{H.}~\bibnamefont{Crater}},
  \bibinfo{journal}{Phys. Rev.} \textbf{\bibinfo{volume}{C67}},
  \bibinfo{pages}{024001} (\bibinfo{year}{2003}).

\bibitem[{\citenamefont{Gross et~al.}(2004)\citenamefont{Gross, Stadler, and
  Pena}}]{Gross:2003qi}
\bibinfo{author}{\bibfnamefont{F.}~\bibnamefont{Gross}},
  \bibinfo{author}{\bibfnamefont{A.}~\bibnamefont{Stadler}}, \bibnamefont{and}
  \bibinfo{author}{\bibfnamefont{M.~T.} \bibnamefont{Pena}},
  \bibinfo{journal}{Phys. Rev.} \textbf{\bibinfo{volume}{C69}},
  \bibinfo{pages}{034007} (\bibinfo{year}{2004}).

\bibitem[{\citenamefont{Julia-Diaz et~al.}(2004)\citenamefont{Julia-Diaz,
  Riska, and Coester}}]{JuliaDiaz:2003gq}
\bibinfo{author}{\bibfnamefont{B.}~\bibnamefont{Julia-Diaz}},
  \bibinfo{author}{\bibfnamefont{D.~O.} \bibnamefont{Riska}}, \bibnamefont{and}
  \bibinfo{author}{\bibfnamefont{F.}~\bibnamefont{Coester}},
  \bibinfo{journal}{Phys. Rev.} \textbf{\bibinfo{volume}{C69}},
  \bibinfo{pages}{035212} (\bibinfo{year}{2004}).

\bibitem[{\citenamefont{Polyzou}(2003)}]{Polyzou:2003dt}
\bibinfo{author}{\bibfnamefont{W.~N.} \bibnamefont{Polyzou}},
  \bibinfo{journal}{Phys. Rev.} \textbf{\bibinfo{volume}{C68}},
  \bibinfo{pages}{015202} (\bibinfo{year}{2003}).

\bibitem[{\citenamefont{Coester and Polyzou}(2005)}]{Coester:2005cv}
\bibinfo{author}{\bibfnamefont{F.}~\bibnamefont{Coester}} \bibnamefont{and}
  \bibinfo{author}{\bibfnamefont{W.~N.} \bibnamefont{Polyzou}},
  \bibinfo{journal}{Phys. Rev.} \textbf{\bibinfo{volume}{C71}},
  \bibinfo{pages}{028202} (\bibinfo{year}{2005}).

\bibitem[{\citenamefont{Alba et~al.}(2007)\citenamefont{Alba, Crater, and
  Lusanna}}]{Alba:2006hs}
\bibinfo{author}{\bibfnamefont{D.}~\bibnamefont{Alba}},
  \bibinfo{author}{\bibfnamefont{H.~W.} \bibnamefont{Crater}},
  \bibnamefont{and} \bibinfo{author}{\bibfnamefont{L.}~\bibnamefont{Lusanna}},
  \bibinfo{journal}{J. Phys.} \textbf{\bibinfo{volume}{A40}},
  \bibinfo{pages}{9585} (\bibinfo{year}{2007}).

\bibitem[{\citenamefont{Lin et~al.}(2008{\natexlab{a}})\citenamefont{Lin,
  Elster, Polyzou, and Glockle}}]{Lin:2007kg}
\bibinfo{author}{\bibfnamefont{T.}~\bibnamefont{Lin}},
  \bibinfo{author}{\bibfnamefont{C.}~\bibnamefont{Elster}},
  \bibinfo{author}{\bibfnamefont{W.~N.} \bibnamefont{Polyzou}},
  \bibnamefont{and} \bibinfo{author}{\bibfnamefont{W.}~\bibnamefont{Glockle}},
  \bibinfo{journal}{Phys. Lett.} \textbf{\bibinfo{volume}{B660}},
  \bibinfo{pages}{345} (\bibinfo{year}{2008}{\natexlab{a}}).

\bibitem[{\citenamefont{de~Melo et~al.}(2009)\citenamefont{de~Melo, Frederico,
  Pace, Pisano, and Salme}}]{deMelo:2008rj}
\bibinfo{author}{\bibfnamefont{J.~P. B.~C.} \bibnamefont{de~Melo}},
  \bibinfo{author}{\bibfnamefont{T.}~\bibnamefont{Frederico}},
  \bibinfo{author}{\bibfnamefont{E.}~\bibnamefont{Pace}},
  \bibinfo{author}{\bibfnamefont{S.}~\bibnamefont{Pisano}}, \bibnamefont{and}
  \bibinfo{author}{\bibfnamefont{G.}~\bibnamefont{Salme}},
  \bibinfo{journal}{Phys. Lett.} \textbf{\bibinfo{volume}{B671}},
  \bibinfo{pages}{153} (\bibinfo{year}{2009}).

\bibitem[{\citenamefont{Huang and Polyzou}(2009)}]{Huang:2008jd}
\bibinfo{author}{\bibfnamefont{Y.}~\bibnamefont{Huang}} \bibnamefont{and}
  \bibinfo{author}{\bibfnamefont{W.~N.} \bibnamefont{Polyzou}},
  \bibinfo{journal}{Phys. Rev.} \textbf{\bibinfo{volume}{C80}},
  \bibinfo{pages}{025503} (\bibinfo{year}{2009}).

\bibitem[{\citenamefont{Lin et~al.}(2008{\natexlab{b}})\citenamefont{Lin,
  Elster, Polyzou, Witala, and Glockle}}]{Lin:2008sy}
\bibinfo{author}{\bibfnamefont{T.}~\bibnamefont{Lin}},
  \bibinfo{author}{\bibfnamefont{C.}~\bibnamefont{Elster}},
  \bibinfo{author}{\bibfnamefont{W.~N.} \bibnamefont{Polyzou}},
  \bibinfo{author}{\bibfnamefont{H.}~\bibnamefont{Witala}}, \bibnamefont{and}
  \bibinfo{author}{\bibfnamefont{W.}~\bibnamefont{Glockle}},
  \bibinfo{journal}{Phys. Rev.} \textbf{\bibinfo{volume}{C78}},
  \bibinfo{pages}{024002} (\bibinfo{year}{2008}{\natexlab{b}}),
  \eprint{0801.3210}.

\bibitem[{\citenamefont{Biernat et~al.}(2009)\citenamefont{Biernat, Schweiger,
  Fuchsberger, and Klink}}]{Biernat:2009my}
\bibinfo{author}{\bibfnamefont{E.~P.} \bibnamefont{Biernat}},
  \bibinfo{author}{\bibfnamefont{W.}~\bibnamefont{Schweiger}},
  \bibinfo{author}{\bibfnamefont{K.}~\bibnamefont{Fuchsberger}},
  \bibnamefont{and} \bibinfo{author}{\bibfnamefont{W.~H.} \bibnamefont{Klink}},
  \bibinfo{journal}{Phys. Rev.} \textbf{\bibinfo{volume}{C79}},
  \bibinfo{pages}{055203} (\bibinfo{year}{2009}).

\bibitem[{\citenamefont{Desplanques}(2009)}]{Desplanques:2009kj}
\bibinfo{author}{\bibfnamefont{B.}~\bibnamefont{Desplanques}},
  \bibinfo{journal}{Eur. Phys. J.} \textbf{\bibinfo{volume}{A42}},
  \bibinfo{pages}{219} (\bibinfo{year}{2009}).

\bibitem[{\citenamefont{Frederico et~al.}(2009)\citenamefont{Frederico, Pace,
  Pasquini, and Salme}}]{Frederico:2009fk}
\bibinfo{author}{\bibfnamefont{T.}~\bibnamefont{Frederico}},
  \bibinfo{author}{\bibfnamefont{E.}~\bibnamefont{Pace}},
  \bibinfo{author}{\bibfnamefont{B.}~\bibnamefont{Pasquini}}, \bibnamefont{and}
  \bibinfo{author}{\bibfnamefont{G.}~\bibnamefont{Salme}},
  \bibinfo{journal}{Phys. Rev.} \textbf{\bibinfo{volume}{D80}},
  \bibinfo{pages}{054021} (\bibinfo{year}{2009}).

\bibitem[{\citenamefont{Fuda and Bulut}(2009)}]{Fuda:2009zz}
\bibinfo{author}{\bibfnamefont{M.~G.} \bibnamefont{Fuda}} \bibnamefont{and}
  \bibinfo{author}{\bibfnamefont{F.}~\bibnamefont{Bulut}},
  \bibinfo{journal}{Phys. Rev.} \textbf{\bibinfo{volume}{C80}},
  \bibinfo{pages}{024002} (\bibinfo{year}{2009}).

\bibitem[{\citenamefont{Whitney and Crater}(2014)}]{Whitney:2011aa}
\bibinfo{author}{\bibfnamefont{J.~F.} \bibnamefont{Whitney}} \bibnamefont{and}
  \bibinfo{author}{\bibfnamefont{H.~W.} \bibnamefont{Crater}},
  \bibinfo{journal}{Phys. Rev.} \textbf{\bibinfo{volume}{D89}},
  \bibinfo{pages}{014023} (\bibinfo{year}{2014}).

\bibitem[{\citenamefont{Witala et~al.}(2011)\citenamefont{Witala, Golak,
  Skibinski, Glockle, Kamada, and Polyzou}}]{Witala:2011yq}
\bibinfo{author}{\bibfnamefont{H.}~\bibnamefont{Witala}},
  \bibinfo{author}{\bibfnamefont{J.}~\bibnamefont{Golak}},
  \bibinfo{author}{\bibfnamefont{R.}~\bibnamefont{Skibinski}},
  \bibinfo{author}{\bibfnamefont{W.}~\bibnamefont{Glockle}},
  \bibinfo{author}{\bibfnamefont{H.}~\bibnamefont{Kamada}}, \bibnamefont{and}
  \bibinfo{author}{\bibfnamefont{W.~N.} \bibnamefont{Polyzou}},
  \bibinfo{journal}{Phys. Rev.} \textbf{\bibinfo{volume}{C83}},
  \bibinfo{pages}{044001} (\bibinfo{year}{2011}), \bibinfo{note}{[Erratum:
  Phys. Rev.C88,no.6,069904(2013)]}.

\bibitem[{\citenamefont{Salme et~al.}(2013)\citenamefont{Salme, Pace, and
  Romanelli}}]{Salme:2012tnh}
\bibinfo{author}{\bibfnamefont{G.}~\bibnamefont{Salme}},
  \bibinfo{author}{\bibfnamefont{E.}~\bibnamefont{Pace}}, \bibnamefont{and}
  \bibinfo{author}{\bibfnamefont{G.}~\bibnamefont{Romanelli}},
  \bibinfo{journal}{Few Body Syst.} \textbf{\bibinfo{volume}{54}},
  \bibinfo{pages}{769} (\bibinfo{year}{2013}).

\bibitem[{\citenamefont{Carbonell and Karmanov}(2014)}]{Carbonell:2014dwa}
\bibinfo{author}{\bibfnamefont{J.}~\bibnamefont{Carbonell}} \bibnamefont{and}
  \bibinfo{author}{\bibfnamefont{V.~A.} \bibnamefont{Karmanov}},
  \bibinfo{journal}{Phys. Rev.} \textbf{\bibinfo{volume}{D90}},
  \bibinfo{pages}{056002} (\bibinfo{year}{2014}).

\bibitem[{\citenamefont{Gross}(2014)}]{Gross:2014zra}
\bibinfo{author}{\bibfnamefont{F.}~\bibnamefont{Gross}},
  \bibinfo{journal}{Phys. Rev.} \textbf{\bibinfo{volume}{C89}},
  \bibinfo{pages}{064001} (\bibinfo{year}{2014}).

\bibitem[{\citenamefont{Hadizadeh et~al.}(2014)\citenamefont{Hadizadeh, Elster,
  and Polyzou}}]{Hadizadeh:2014yua}
\bibinfo{author}{\bibfnamefont{M.~R.} \bibnamefont{Hadizadeh}},
  \bibinfo{author}{\bibfnamefont{C.}~\bibnamefont{Elster}}, \bibnamefont{and}
  \bibinfo{author}{\bibfnamefont{W.~N.} \bibnamefont{Polyzou}},
  \bibinfo{journal}{Phys. Rev.} \textbf{\bibinfo{volume}{C90}},
  \bibinfo{pages}{054002} (\bibinfo{year}{2014}).

\bibitem[{\citenamefont{Santopinto and Ferretti}(2015)}]{Santopinto:2014opa}
\bibinfo{author}{\bibfnamefont{E.}~\bibnamefont{Santopinto}} \bibnamefont{and}
  \bibinfo{author}{\bibfnamefont{J.}~\bibnamefont{Ferretti}},
  \bibinfo{journal}{Phys. Rev.} \textbf{\bibinfo{volume}{C92}},
  \bibinfo{pages}{025202} (\bibinfo{year}{2015}).

\bibitem[{\citenamefont{Faustov and O.~Galkin}(2015)}]{Faustov:2015}
\bibinfo{author}{\bibfnamefont{R.}~\bibnamefont{Faustov}} \bibnamefont{and}
  \bibinfo{author}{\bibfnamefont{V.}~\bibnamefont{O.~Galkin}},
  \bibinfo{journal}{Physical Review D. 92. 10.1103/PhysRevD.92.054005.}
  \textbf{\bibinfo{volume}{92}}, \bibinfo{pages}{054005}
  (\bibinfo{year}{2015}).

\bibitem[{\citenamefont{Arslanaliev et~al.}(2018)\citenamefont{Arslanaliev,
  Kamada, Shebeko, Stepanova, Witala, and Yakovlev}}]{Arslanaliev:2018wkl}
\bibinfo{author}{\bibfnamefont{A.}~\bibnamefont{Arslanaliev}},
  \bibinfo{author}{\bibfnamefont{H.}~\bibnamefont{Kamada}},
  \bibinfo{author}{\bibfnamefont{A.}~\bibnamefont{Shebeko}},
  \bibinfo{author}{\bibfnamefont{M.}~\bibnamefont{Stepanova}},
  \bibinfo{author}{\bibfnamefont{H.}~\bibnamefont{Witala}}, \bibnamefont{and}
  \bibinfo{author}{\bibfnamefont{S.}~\bibnamefont{Yakovlev}},
  \bibinfo{journal}{Prob. Atomic Sci. Technol.}
  \textbf{\bibinfo{volume}{2018}}, \bibinfo{pages}{3} (\bibinfo{year}{2018}).

\bibitem[{\citenamefont{Carbonell and Karmanov}(2016)}]{Carbonell:2016udz}
\bibinfo{author}{\bibfnamefont{J.}~\bibnamefont{Carbonell}} \bibnamefont{and}
  \bibinfo{author}{\bibfnamefont{V.~A.} \bibnamefont{Karmanov}},
  \bibinfo{journal}{Few Body Syst.} \textbf{\bibinfo{volume}{57}},
  \bibinfo{pages}{533} (\bibinfo{year}{2016}).

\bibitem[{\citenamefont{de~Paula et~al.}(2016)\citenamefont{de~Paula,
  Frederico, Salme, and Viviani}}]{dePaula:2016oct}
\bibinfo{author}{\bibfnamefont{W.}~\bibnamefont{de~Paula}},
  \bibinfo{author}{\bibfnamefont{T.}~\bibnamefont{Frederico}},
  \bibinfo{author}{\bibfnamefont{G.}~\bibnamefont{Salme}}, \bibnamefont{and}
  \bibinfo{author}{\bibfnamefont{M.}~\bibnamefont{Viviani}},
  \bibinfo{journal}{Phys. Rev.} \textbf{\bibinfo{volume}{D94}},
  \bibinfo{pages}{071901} (\bibinfo{year}{2016}).

\bibitem[{\citenamefont{Del~Dotto et~al.}(2017)\citenamefont{Del~Dotto, Pace,
  Salme, and Scopetta}}]{DelDotto:2016vkh}
\bibinfo{author}{\bibfnamefont{A.}~\bibnamefont{Del~Dotto}},
  \bibinfo{author}{\bibfnamefont{E.}~\bibnamefont{Pace}},
  \bibinfo{author}{\bibfnamefont{G.}~\bibnamefont{Salme}}, \bibnamefont{and}
  \bibinfo{author}{\bibfnamefont{S.}~\bibnamefont{Scopetta}},
  \bibinfo{journal}{Phys. Rev.} \textbf{\bibinfo{volume}{C95}},
  \bibinfo{pages}{014001} (\bibinfo{year}{2017}).

\bibitem[{\citenamefont{Eichmann et~al.}(2016)\citenamefont{Eichmann,
  Sanchis-Alepuz, Williams, Alkofer, and Fischer}}]{Eichmann:2016yit}
\bibinfo{author}{\bibfnamefont{G.}~\bibnamefont{Eichmann}},
  \bibinfo{author}{\bibfnamefont{H.}~\bibnamefont{Sanchis-Alepuz}},
  \bibinfo{author}{\bibfnamefont{R.}~\bibnamefont{Williams}},
  \bibinfo{author}{\bibfnamefont{R.}~\bibnamefont{Alkofer}}, \bibnamefont{and}
  \bibinfo{author}{\bibfnamefont{C.~S.} \bibnamefont{Fischer}},
  \bibinfo{journal}{Prog. Part. Nucl. Phys.} \textbf{\bibinfo{volume}{91}},
  \bibinfo{pages}{1} (\bibinfo{year}{2016}).

\bibitem[{\citenamefont{Leitao et~al.}(2017)\citenamefont{Leitao, Stadler,
  Pena, and Biernat}}]{Leitao:2017mlx}
\bibinfo{author}{\bibfnamefont{S.}~\bibnamefont{Leitao}},
  \bibinfo{author}{\bibfnamefont{A.}~\bibnamefont{Stadler}},
  \bibinfo{author}{\bibfnamefont{M.~T.} \bibnamefont{Pena}}, \bibnamefont{and}
  \bibinfo{author}{\bibfnamefont{E.~P.} \bibnamefont{Biernat}},
  \bibinfo{journal}{Phys. Rev.} \textbf{\bibinfo{volume}{D96}},
  \bibinfo{pages}{074007} (\bibinfo{year}{2017}).

\bibitem[{\citenamefont{Sokolov}(1977)}]{Sokolov:1977}
\bibinfo{author}{\bibfnamefont{S.~N.} \bibnamefont{Sokolov}},
  \bibinfo{journal}{Dokl. Akad. Nauk SSSR} \textbf{\bibinfo{volume}{233}},
  \bibinfo{pages}{575} (\bibinfo{year}{1977}).

\bibitem[{\citenamefont{Keister and Polyzou}(1991)}]{Keister:1991sb}
\bibinfo{author}{\bibfnamefont{B.~D.} \bibnamefont{Keister}} \bibnamefont{and}
  \bibinfo{author}{\bibfnamefont{W.~N.} \bibnamefont{Polyzou}},
  \bibinfo{journal}{Adv. Nucl. Phys.} \textbf{\bibinfo{volume}{20}},
  \bibinfo{pages}{225} (\bibinfo{year}{1991}).

\bibitem[{\citenamefont{Schwinger}(1958)}]{Schwinger:1958qau}
\bibinfo{author}{\bibfnamefont{J.}~\bibnamefont{Schwinger}},
  \bibinfo{journal}{Proc. Nat. Acad. Sci.} \textbf{\bibinfo{volume}{44}},
  \bibinfo{pages}{956} (\bibinfo{year}{1958}).

\bibitem[{\citenamefont{Coester}(1965)}]{Coester:1965zz}
\bibinfo{author}{\bibfnamefont{F.}~\bibnamefont{Coester}},
  \bibinfo{journal}{Helv. Phys. Acta} \textbf{\bibinfo{volume}{38}},
  \bibinfo{pages}{7} (\bibinfo{year}{1965}).

\bibitem[{\citenamefont{Polyzou}(1989)}]{Polyzou:19892b}
\bibinfo{author}{\bibfnamefont{W.~N.} \bibnamefont{Polyzou}},
  \bibinfo{journal}{Ann. Phys.} \textbf{\bibinfo{volume}{193}},
  \bibinfo{pages}{367} (\bibinfo{year}{1989}).

\bibitem[{\citenamefont{Wiringa et~al.}(1995)\citenamefont{Wiringa, Stoks, and
  Schiavilla}}]{Wiringa:1994wb}
\bibinfo{author}{\bibfnamefont{R.~B.} \bibnamefont{Wiringa}},
  \bibinfo{author}{\bibfnamefont{V.~G.~J.} \bibnamefont{Stoks}},
  \bibnamefont{and}
  \bibinfo{author}{\bibfnamefont{R.}~\bibnamefont{Schiavilla}},
  \bibinfo{journal}{Phys. Rev.} \textbf{\bibinfo{volume}{C51}},
  \bibinfo{pages}{38} (\bibinfo{year}{1995}).

\bibitem[{\citenamefont{Entem and Machleidt}(2003)}]{Entem:2003ft}
\bibinfo{author}{\bibfnamefont{D.~R.} \bibnamefont{Entem}} \bibnamefont{and}
  \bibinfo{author}{\bibfnamefont{R.}~\bibnamefont{Machleidt}},
  \bibinfo{journal}{Phys. Rev.} \textbf{\bibinfo{volume}{C68}},
  \bibinfo{pages}{041001} (\bibinfo{year}{2003}).

\bibitem[{\citenamefont{Coester et~al.}(1975)\citenamefont{Coester, Pieper, and
  Serduke}}]{Coester:1975zz}
\bibinfo{author}{\bibfnamefont{F.}~\bibnamefont{Coester}},
  \bibinfo{author}{\bibfnamefont{S.~C.} \bibnamefont{Pieper}},
  \bibnamefont{and} \bibinfo{author}{\bibfnamefont{F.~J.~D.}
  \bibnamefont{Serduke}}, \bibinfo{journal}{Phys. Rev.}
  \textbf{\bibinfo{volume}{C11}}, \bibinfo{pages}{1} (\bibinfo{year}{1975}).

\bibitem[{\citenamefont{Chung and Coester}(1991)}]{Chung:1991st}
\bibinfo{author}{\bibfnamefont{P.~L.} \bibnamefont{Chung}} \bibnamefont{and}
  \bibinfo{author}{\bibfnamefont{F.}~\bibnamefont{Coester}},
  \bibinfo{journal}{Phys. Rev.} \textbf{\bibinfo{volume}{D44}},
  \bibinfo{pages}{229} (\bibinfo{year}{1991}).

\bibitem[{\citenamefont{Coester and Riska}(1994)}]{Coester:1994wp}
\bibinfo{author}{\bibfnamefont{F.}~\bibnamefont{Coester}} \bibnamefont{and}
  \bibinfo{author}{\bibfnamefont{D.~O.} \bibnamefont{Riska}},
  \bibinfo{journal}{Annals Phys.} \textbf{\bibinfo{volume}{234}},
  \bibinfo{pages}{141} (\bibinfo{year}{1994}).

\bibitem[{\citenamefont{Krutov and Troitsky}(2009)}]{Krutov:2009zz}
\bibinfo{author}{\bibfnamefont{A.~F.} \bibnamefont{Krutov}} \bibnamefont{and}
  \bibinfo{author}{\bibfnamefont{V.~E.} \bibnamefont{Troitsky}},
  \bibinfo{journal}{Phys. Part. Nucl.} \textbf{\bibinfo{volume}{40}},
  \bibinfo{pages}{136} (\bibinfo{year}{2009}).

\bibitem[{\citenamefont{Polyzou}(2002)}]{Polyzou:2002id}
\bibinfo{author}{\bibfnamefont{W.~N.} \bibnamefont{Polyzou}}
  (\bibinfo{year}{2002}).

\bibitem[{\citenamefont{Chung et~al.}(1989)\citenamefont{Chung, Keister, and
  Coester}}]{Chung:1989zz}
\bibinfo{author}{\bibfnamefont{P.~L.} \bibnamefont{Chung}},
  \bibinfo{author}{\bibfnamefont{B.~D.} \bibnamefont{Keister}},
  \bibnamefont{and} \bibinfo{author}{\bibfnamefont{F.}~\bibnamefont{Coester}},
  \bibinfo{journal}{Phys. Rev.} \textbf{\bibinfo{volume}{C39}},
  \bibinfo{pages}{1544} (\bibinfo{year}{1989}).

\bibitem[{\citenamefont{Coester and Polyzou}(2004)}]{Coester:2004qu}
\bibinfo{author}{\bibfnamefont{F.}~\bibnamefont{Coester}} \bibnamefont{and}
  \bibinfo{author}{\bibfnamefont{W.~N.} \bibnamefont{Polyzou}}
  (\bibinfo{year}{2004}).

\bibitem[{\citenamefont{Dirac}(1951)}]{Dirac:1951zz}
\bibinfo{author}{\bibfnamefont{P.~A.~M.} \bibnamefont{Dirac}},
  \bibinfo{journal}{Can. J. Math.} \textbf{\bibinfo{volume}{3}},
  \bibinfo{pages}{1} (\bibinfo{year}{1951}).

\bibitem[{\citenamefont{Greenberg}(2011)}]{Greenberg:2010wx}
\bibinfo{author}{\bibfnamefont{O.~W.} \bibnamefont{Greenberg}},
  \bibinfo{journal}{Int. J. Mod. Phys.} \textbf{\bibinfo{volume}{A26}},
  \bibinfo{pages}{935} (\bibinfo{year}{2011}).

\bibitem[{\citenamefont{Glaser et~al.}(1957)\citenamefont{Glaser, Lehmann, and
  Zimmermann}}]{Glaser:1957vrs}
\bibinfo{author}{\bibfnamefont{V.}~\bibnamefont{Glaser}},
  \bibinfo{author}{\bibfnamefont{H.}~\bibnamefont{Lehmann}}, \bibnamefont{and}
  \bibinfo{author}{\bibfnamefont{W.}~\bibnamefont{Zimmermann}},
  \bibinfo{journal}{Nuovo Cim.} \textbf{\bibinfo{volume}{6}},
  \bibinfo{pages}{1122} (\bibinfo{year}{1957}).

\bibitem[{\citenamefont{Dyson}(1949)}]{Dyson:1949ha}
\bibinfo{author}{\bibfnamefont{F.~J.} \bibnamefont{Dyson}},
  \bibinfo{journal}{Phys. Rev.} \textbf{\bibinfo{volume}{75}},
  \bibinfo{pages}{1736} (\bibinfo{year}{1949}).

\bibitem[{\citenamefont{Schwinger}(1951)}]{Schwinger:1951xk}
\bibinfo{author}{\bibfnamefont{J.~S.} \bibnamefont{Schwinger}},
  \bibinfo{journal}{Phys. Rev.} \textbf{\bibinfo{volume}{82}},
  \bibinfo{pages}{914} (\bibinfo{year}{1951}).

\bibitem[{\citenamefont{Schwinger}(1953)}]{Schwinger:1953tb}
\bibinfo{author}{\bibfnamefont{J.~S.} \bibnamefont{Schwinger}},
  \bibinfo{journal}{Phys. Rev.} \textbf{\bibinfo{volume}{91}},
  \bibinfo{pages}{713} (\bibinfo{year}{1953}).

\bibitem[{\citenamefont{Huang and Weldon}(1975)}]{Huang:1974cd}
\bibinfo{author}{\bibfnamefont{K.}~\bibnamefont{Huang}} \bibnamefont{and}
  \bibinfo{author}{\bibfnamefont{H.~A.} \bibnamefont{Weldon}},
  \bibinfo{journal}{Phys. Rev.} \textbf{\bibinfo{volume}{D11}},
  \bibinfo{pages}{257} (\bibinfo{year}{1975}).

\bibitem[{\citenamefont{Biernat et~al.}(2015)\citenamefont{Biernat, Gross,
  Pena, and Stadler}}]{Biernat:2015xya}
\bibinfo{author}{\bibfnamefont{E.~P.} \bibnamefont{Biernat}},
  \bibinfo{author}{\bibfnamefont{F.}~\bibnamefont{Gross}},
  \bibinfo{author}{\bibfnamefont{M.~T.} \bibnamefont{Pena}}, \bibnamefont{and}
  \bibinfo{author}{\bibfnamefont{A.}~\bibnamefont{Stadler}},
  \bibinfo{journal}{Phys. Rev.} \textbf{\bibinfo{volume}{D92}},
  \bibinfo{pages}{076011} (\bibinfo{year}{2015}).

\bibitem[{\citenamefont{Polyzou}(2014)}]{Polyzou:2013nga}
\bibinfo{author}{\bibfnamefont{W.~N.} \bibnamefont{Polyzou}},
  \bibinfo{journal}{Phys. Rev.} \textbf{\bibinfo{volume}{D89}},
  \bibinfo{pages}{076008} (\bibinfo{year}{2014}).

\bibitem[{\citenamefont{Aiello and Polyzou}(2016)}]{Aiello:2015jgc}
\bibinfo{author}{\bibfnamefont{G.}~\bibnamefont{Aiello}} \bibnamefont{and}
  \bibinfo{author}{\bibfnamefont{W.}~\bibnamefont{Polyzou}},
  \bibinfo{journal}{Phys. Rev.} \textbf{\bibinfo{volume}{D93}},
  \bibinfo{pages}{056003} (\bibinfo{year}{2016}).

\bibitem[{\citenamefont{Kopp and Polyzou}(2012)}]{Kopp:2011vv}
\bibinfo{author}{\bibfnamefont{P.}~\bibnamefont{Kopp}} \bibnamefont{and}
  \bibinfo{author}{\bibfnamefont{W.}~\bibnamefont{Polyzou}},
  \bibinfo{journal}{Phys. Rev.} \textbf{\bibinfo{volume}{D85}},
  \bibinfo{pages}{016004} (\bibinfo{year}{2012}).

\bibitem[{\citenamefont{Schwinger}(2015)}]{schwinger}
\bibinfo{author}{\bibfnamefont{J.}~\bibnamefont{Schwinger}},
  \emph{\bibinfo{title}{On Angular Momentum}} (\bibinfo{publisher}{Dover, NY},
  \bibinfo{year}{2015}).

\end{thebibliography}
\end{document}